\documentclass[12pt,article]{JHEP3}

\usepackage{amscd,amsmath,amssymb,amsfonts,xspace,mathrsfs}
\usepackage{epsfig}

\numberwithin{equation}{section}

\def\be{\begin{equation}}
\def\ee{\end{equation}}
\def\bea{\begin{eqnarray}}
\def\eea{\end{eqnarray}}
\def\bes{\be\begin{split}}
\def\ees{\end{split}\ee}
\def\barr{\begin{array}}
\def\earr{\end{array}}
\def\ba{\begin{align}}
\def\ea{\end{align}}
\def\bse{\begin{subequations}}
\def\ese{\end{subequations}}

\def\lfig#1#2#3#4#5{
 \begin{figure}
 \refstepcounter{figure}
 \label{#4}
 \addtocounter{figure}{-1}
 \epsfxsize=#3
 \centerline{\epsfbox{#2}}
 \vspace{#5}
 {\bf \caption{{\rm #1}}}
 \end{figure}
}

\def\varpi{t}

\def\det{\,{\rm det}\, }

\def\rank{{\rm rank}}

\def\tr{\,{\rm tr} }
\def\Tr{\,{\rm Tr} }

\def\Im{\,{\rm Im}\,}
\def\Re{\,{\rm Re}\,}

\def\Vol{{\rm Vol}}

\def\({\left(}
\def\){\right)}
\def\[{\left[}
\def\]{\right]}
\def\<{\left\langle}
\def\>{\right\rangle}

\def\hf{{1\over 2}}

\newcommand{\nn}{\nonumber}
\newcommand{\p}{\partial}

\def\vph{\varphi}

\newcommand{\eps}{\epsilon}

\newcommand{\vth}{\vartheta}

\renewcommand{\d}{\mathrm{d}}
\newcommand{\de}{\mathrm{d}}

\newcommand{\I}{\mathrm{i}}

\newcommand{\cA}{\mathcal{A}}

\newcommand{\cC}{\mathcal{C}}
\newcommand{\cD}{\mathcal{D}}
\newcommand{\cE}{\mathcal{E}}
\newcommand{\cF}{\mathcal{F}}
\newcommand{\cG}{\mathcal{G}}

\newcommand{\cI}{\mathcal{I}}
\newcommand{\cJ}{\mathcal{J}}
\newcommand{\cK}{\mathcal{K}}
\newcommand{\cL}{\mathcal{L}}
\newcommand{\cM}{\mathcal{M}}
\newcommand{\cN}{\mathcal{N}}
\newcommand{\cO}{\mathcal{O}}

\newcommand{\cQ}{\mathcal{Q}}

\newcommand{\cS}{\mathcal{S}}

\newcommand{\cW}{\mathcal{W}}
\newcommand{\cZ}{\mathcal{Z}}
\newcommand{\cX}{\mathcal{X}}
\newcommand{\cV}{\mathcal{V}}
\newcommand{\cU}{\mathcal{U}}

\newcommand{\tD}{\mathsf{D}}
\newcommand{\tN}{\mathsf{N}}
\newcommand{\tM}{\mathsf{M}}

\newcommand{\IF}{\mathbb{F}}
\newcommand{\IR}{\mathbb{R}}
\newcommand{\IC}{\mathbb{C}}
\newcommand{\IZ}{\mathbb{Z}}

\newcommand{\IP}{\mathbb{P}}

\newcommand{\bfF}{{\boldsymbol F}}

\newcommand{\bfR}{{\boldsymbol R}}

\newcommand{\bfr}{{\boldsymbol r}}
\newcommand{\bfw}{{\boldsymbol w}}

\newcommand{\bfcl}{\tilde{\boldsymbol c}}

\newcommand{\tzeta}{\tilde\zeta}

\newcommand{\tc}{\tilde c}

\newcommand{\txi}{\tilde\xi}
\newcommand{\tgam}{\tilde\gamma}

\def\cla{\tilde c_a}
\def\cl0{\tilde c_0}

\def\ba{\bar a}

\def\bu{\bar u}
\def\bz{\bar z}
\def\btau{\bar \tau}

\def\bY{\bar Y}
\def\bV{\bar V}
\def\bZ{\bar Z}

\def\bF{\bar F}

\def\hw{\hat w}

\def\hz{\hat z}
\def\hatt{\hat t}

\def\hD{\hat D}
\def\hK{\hat K}

\def\hN{\hat N}
\def\hA{\hat A}
\def\hB{\hat B}
\def\hC{\hat C}
\def\hI{\hat I}
\def\hJ{\hat J}
\def\hL{\hat L}

\newcommand{\hmu}{{\hat\mu}}
\newcommand{\hnu}{{\hat\nu}}
\newcommand{\hlambda}{{\hat\lambda}}
\newcommand{\hrho}{{\hat\rho}}
\newcommand{\hsigma}{{\hat\sigma}}

\def\cij#1{c}
\def\ci#1{c}

\def\mui#1{\mu^{[#1]}}
\def\txii#1{{\tilde\xi}^{[#1]}}

\def\ai#1{{\alpha}^{[#1]}}
\def\alpi#1{\alpha^{[#1]}}

\def\etai#1{\eta_{[#1]}}
\def\xii#1{\xi_{[#1]}}

\def\Hij#1{H^{[#1]}}

\newcommand{\Li}{{\rm Li}}

\def\Thkl{\Theta_{\gamma}}

\def\Hij#1{H^{[#1]}}

\def\XXint#1#2#3{{\setbox0=\hbox{$#1{#2#3}{\int}$}
\vcenter{\hbox{$#2#3$}}\kern-.5\wd0}}

\def\gfinv{n^{(0)}}

\newcommand{\cY}{\mathcal{Y}}

\def\Zg{Z_{\gamma}}
\def\bZg{\bar Z_{\gamma}}

\def\Om#1{\Omega_{#1}}
\def\bOm#1{\bar\Omega_{#1}}

\def\bOmMSW{{\bar \Omega}^{\rm MSW}}

\def\ellg#1{\ell_{#1}}

\def\Ilg{\cJ^{(1)}}
\def\Ilog#1{\cJ^{(1,#1)}}
\def\Irt{\cJ^{(2)}}
\def\Irat#1{\cJ^{(2,#1)}}

\def\Igp{\Ilog{+}_{\gamma}}
\def\Igm{\Ilog{-}_{\gamma}}
\def\Igpm{\Ilog{\pm}_{\gamma}}
\def\Igamp#1{\Ilog{+}_{#1}}

\def\Igg{\Ilg_{\gamma}}

\def\Igam#1{\Ilg_{#1}}

\def\rIg{\Irt_{\gamma}}

\def\rIgpm{\Irat{\pm}_{\gamma}}

\def\Ingam#1#2{\cJ^{(#1)}_{#2}}
\def\Insgam#1#2#3{\cJ^{(#1,#2)}_{#3}}

\def\Dn#1{\cD^{(#1)}}

\def\cIo{\cI^{(0)}}
\def\cIp{\cI^{(+)}}
\def\cIm{\cI^{(-)}}
\def\cIpm{\cI^{(\pm)}}

\def\ccl{c_{\rm cl}}
\def\fcl{f^{\rm cl}}
\def\bfcl{\bar f^{\rm cl}}
\def\Fcl{F^{\rm cl}}

\def\Mcl{\cM_H^{\rm cl}}
\def\Mpcl{\cM_H'^{\rm cl}}
\def\Mcorcl{\cM_H^{\rm cl,cor}}

\def\Fg{\cF}

\def\Uin{\mathbf{U}}

\def\gamD#1{\tilde\gamma}

\def\CY{\mathfrak{Y}}
\def\CYm{\mathfrak{\hat Y}}
\def\base{\mathfrak{B}}

\def\cXsf{\cX^{\rm sf}}

\def\Geff{\Gamma_{\rm rig}}

\def\vp{\vec p}
\def\vmu{\vec \mu}

\def\qr{\sigma_\gamma}
\def\qrp{\sigma_{\gamma'}}
\def\qrg#1{\sigma_{#1}}

\def\vv{v}

\def\zzb{{\bf z}}

\def\vvp{\vv^{(+)}}
\def\vvm{\vv^{(-)}}
\def\vvpm{\vv^{(\pm)}}
\def\vvn#1{\vv^{(#1)}}
\def\vvpn#1{\vv^{(+,#1)}}
\def\vvmn#1{\vv^{(-,#1)}}
\def\vvpmn#1{\vv^{(\pm,#1)}}

\def\Min{\cM}

\def\Uin{\mathbf{U}}

\def\cCf{\cC}
\def\cCd{\mathbf{C}}
\def\cAd{\mathbf{A}}
\def\cBd{\mathbf{B}}

\def\Vcy{V}
\def\uz{z}
\def\buz{\bar z}

\newcommand{\under}[2]{\mathop{#1}\limits_{#2}}

\newcommand{\dP}{dP}
\newcommand{\CP}{\IC P^1}
\newcommand{\Mpl}{M_{\rm Pl}}
\newcommand{\Mcor}{\cM_H^{\rm cor}}
\newcommand{\kahler}{{K\"ahler}\xspace}
\newcommand{\qk}{{quaternion-K\"ahler}\xspace}
\newcommand{\hpk}{{hyperk\"ahler}\xspace}

\def\nA{n_\infty}
\def\nX{n_{\rm fr}}
\def\nI{n'}

\def\vt{v}
\def\vtA{\vt_{A}}
\def\vtI{\vt_{I}}
\def\vtX{\vt_{X}}
\def\vvt{\vec v}
\def\vvtA{\vec\vt_{A}}
\def\vvtI{\vec\vt_{I}}
\def\vvtX{\vec\vt_{X}}

\def\escl#1{\sim\Lambda^{#1}}
\def\scl#1{\Lambda^{#1}}
\def\escc{\sim 1}
\def\scc{1}

\def\gT{\varrho}
\def\thtau{\vartheta_\tau}
\def\bthtau{\bar\vartheta_\tau}

\title{Rigid limit for hypermultiplets \\ and five-dimensional gauge theories}

\preprint{L2C:17-128\\
CERN-TH-2017-224\\
MPIM-2017-68\\
UUITP-39/17
}

\author{Sergei Alexandrov$^{1,2}$, Sibasish Banerjee$^{3,4}$ and Pietro Longhi$^5$
\\
$^1$ {\it
Laboratoire Charles Coulomb (L2C), UMR 5221 CNRS-Universit\'e de
Montpellier, F-34095, Montpellier, France}\\

$^2$ {\it Theoretical Physics Department, CERN, Geneva, Switzerland}\\

$^3$ {\it IPhT, CEA, Saclay, Gif-sur-Yvette, F-91191, France}\\

$^4$ {\it Max-Planck-Institut f\"ur Mathematik, Vivatsgasse 7, 53111 Bonn, Germany}\\

$^5$ {\it Department of Physics and Astronomy, Uppsala University, Uppsala, Sweden}

\vspace*{2mm} {\tt e-mail:
\email{sergey.alexandrov@umontpellier.fr},
\email{sbanerje@mpim-bonn.mpg.de}
\email{pietro.longhi@physics.uu.se}
}

\vspace*{-3mm}

}

\abstract{We study the rigid limit of a class of hypermultiplet moduli spaces
appearing in Calabi-Yau compactifications of type IIB string theory, which is induced by a local limit of the Calabi-Yau.
We show that the resulting hyperk\"ahler manifold is obtained by performing a hyperk\"ahler quotient
of the Swann bundle over the moduli space, along the isometries arising in the limit.
Physically, this manifold appears as the target space of the non-linear sigma model obtained by compactification
of a five-dimensional gauge theory on a torus.
This allows to compute dyonic and stringy instantons of the gauge theory from the known results on D-instantons in string theory.
Besides, we formulate a simple condition on the existence
of a non-trivial local limit in terms of intersection numbers of the Calabi-Yau, and find an explicit form
for the hypermultiplet metric including corrections from all mutually non-local D-instantons, which can be of independent interest.
}

\begin{document}

\section{Introduction}

String theory plays a prominent role in extracting the non-perturbative dynamics of supersymmetric gauge theories.
Indeed, due to the existence of various dualities, sometimes it is easier to solve a problem in string theory and then to take
the so called {\it rigid limit}, in which gravity decouples and one recovers a gauge theory description \cite{Kachru:1995fv,Klemm:1996bj}.
A particularly fruitful playground for this are theories with 8 supercharges corresponding to $N=2$ supersymmetry in 4 dimensions.
In such case, the dynamics of compactified string theory is captured at low energies by effective supergravity which comprises,
besides the gravitational multiplet, also vector and hypermultiplets.
The kinetic couplings of the former are encoded in the vector multiplet moduli space $\cM_V$,
which is a {\it projective} (also called {\it local}) special K\"ahler manifold.
In the rigid limit it directly reduces to a simpler {\it rigid} special K\"ahler manifold,
whose prepotential contains all information about the solution of the corresponding gauge theory.
Due to this, previous works mostly concentrated on the vector multiplet sector of string compactifications
\cite{Andrianopoli:1996cm,Katz:1996fh,Katz:1997eq,Billo:1998yr},
and one can say that the procedure of extracting the rigid limit there is understood fairly well
(see \cite{Gunara:2013rca} for a recent discussion).

Let us recall that $\cM_V$ is only one component of the moduli space of $N=2$ supergravity.
The second one is the hypermultiplet (HM) moduli space $\cM_H$, and it is natural to ask what happens to this space
after decoupling gravity. The local supersymmetry restricts $\cM_H$ to be \qk (QK) \cite{Bagger:1983tt},
i.e. a $4n$ real dimensional manifold with holonomy group $SU(2)\times Sp(n)$. For such manifolds
the Riemann curvature tensor decomposes as
\be
R_{\mu\nu\rho\sigma}=\kappa^2\hat R_{\mu\nu\rho\sigma}+W_{\mu\nu\rho\sigma},
\ee
where $\kappa^2=8\pi \Mpl^{-2}$ is the gravitational coupling, $\hat R_{\mu\nu\rho\sigma}$ is the dimensionless $SU(2)$
part of the curvature, and $W_{\mu\nu\rho\sigma}$ is the Weyl tensor.
Thus, one can expect that in the rigid limit only the second contribution survives and
one ends up with a Ricci-flat manifold with holonomy group $Sp(n)$, i.e. a \hpk (HK) manifold.
This is indeed a very natural expectation
because such manifolds are known to play an important role in the low energy description of theories with global supersymmetry.
For instance, they appear as Higgs branches of 4d $N=2$ gauge theories.
However, the metric on these Higgs branches is classically exact.
For this reason, and since $\cM_H$ does receive quantum corrections, we do not expect them to be relevant in our context.
A more interesting and, as we will see, relevant example is provided by target spaces
of $N=4$ non-linear sigma models in 3 dimensions \cite{AlvarezGaume:1980vs},
some of which can also be viewed as circle compactifications of 4d $N=2$ gauge theories \cite{Seiberg:1996nz}.

Unfortunately, it turns out that the naive decoupling leads to a flat \hpk geometry, and to get a non-trivial limit it is necessary to introduce
an additional mass scale, which is kept finite as $\kappa\to 0$.
As a result, no general treatment of the rigid limit for QK spaces exists in the literature, and
a non-trivial limit was produced only in a number of particular cases
\cite{Galicki:1985qv,Galicki:1987jz,Ambrosetti:2010tu,Antoniadis:2015egy,Antoniadis:2016kde}.
At the same time, the rigid limit, used to extract information about gauge theories from the vector multiplet sector
of string compactifications, usually has a geometric realization as a {\it local limit} on the compactification manifold $\CY$ where
one zooms in on the region near some singularity in the moduli space \cite{Katz:1996fh,Katz:1997eq}.
Since the metric on both moduli spaces, $\cM_V$ and $\cM_H$, is completely determined by the geometry of $\CY$, it is natural
to ask whether this zooming procedure is sufficient to induce the rigid limit of the HM moduli space.
This is the question that we investigate in this paper for compactifications of type IIB string theory on a Calabi-Yau (CY) threefold $\CY$.

The advantage of considering this type of compactifications is that in recent years substantial progress has been made
towards understanding the complete non-perturbative description of the corresponding HM moduli space
(see \cite{Alexandrov:2011va,Alexandrov:2013yva} for reviews).
As a result, we now have access to the metric on $\cM_H$ which includes most of the non-perturbative corrections.
In the type IIB formulation, the latter include D$p$-brane instantons (with $p=-1,1,3$ and 5) and NS5-brane instantons.
Only contributions of five-branes remain not well understood (although some partial results can be found in
\cite{Alexandrov:2010ca,Alexandrov:2014mfa,Alexandrov:2014rca}), whereas all D-instantons have been incorporated
\cite{Alexandrov:2008gh,Alexandrov:2009zh}
using a twistorial description of QK geometry \cite{MR1327157,Alexandrov:2008nk}.
Fortunately, it turns out that in any local limit the unknown five-brane contributions always decouple
and one remains with a metric which is completely under our control.

The last statement however needs a refinement.
Although the twistorial description, used to obtain the cited results, is very powerful,
it is also somewhat implicit because it encodes the QK metric
into the contact structure on the twistor space $\cZ_\cM$, a $\CP$ bundle over the original manifold, and it is not so easy to extract it.
Recently this problem was solved \cite{Alexandrov:2014sya}
only for {\it mutually local} D-instantons, i.e. a subset of all D-instantons whose charges $\gamma$ have vanishing
symplectic products $\langle\gamma,\gamma'\rangle$. In this paper we seize on the opportunity to improve the situation and calculate
the explicit HM metric, which includes all {\it mutually non-local} D-instanton corrections
and is parametrized by topological data on the CY, such as its triple intersection numbers $\kappa_{abc}$,
Euler characteristic $\chi_\CY$ and generalized Donaldson-Thomas (DT) invariants $\Om{\gamma}$.

Having at hand the explicit metric, we can study its behavior in the local limit.
To define it, we fix a set of $\nA$ vectors $\vvtA$
belonging to the boundary of the K\"ahler cone of $\CY$. They correspond to the directions in the moduli space along which
some of the (dimensionless) K\"ahler moduli are sent to infinity,
thereby introducing a new scale $\Lambda$.
Geometrically, they fix a set of 2-cycles which shrink in the local limit and have vanishing intersection with the divisors defined by $\vvtA$.

Then, evaluating the HM metric in the so-defined limit, we show that, besides a non-trivial finite part, it also features a divergent part.
This leads to the freezing of some moduli, including those which are sent to infinity.
As a result, all moduli can be split into 3 groups:
\begin{itemize}
\item
moduli appearing only in the vanishing part of the metric and thus dropping out in the limit;
\item
frozen moduli;
\item
moduli appearing in the finite, but not in the divergent parts of the metric and thus remaining dynamical.
\end{itemize}
Only the latter moduli parametrize the limiting manifold $\cM'_H$, which therefore has always a smaller dimension than the original $\cM_H$.
More precisely, the dimension of $\cM'_H$ is given by $4\nI$ where $\nI$ is the number of K\"ahler moduli remaining dynamical.
We show that $\nI$ coincides with the dimension of the intersection of the kernels for the matrices $M_{A,ab}=\kappa_{abc}\vtA^c$.
Note that the possibility of having a non-empty common kernel is a very non-trivial condition on both the vectors
$\vvtA$ and the triple intersection numbers, so that far from any CY allows for a non-trivial rigid limit even with $\nA=1$.

Furthermore, we prove that $\cM'_H$ is an HK manifold and can be constructed from $\cM_H$ in a pure geometric way (see Fig. \ref{fig-quotient}).
To this end, one should first note that the local limit induces on $\cM_H$ a set of $n-\nI$
commuting isometries where
$n=h^{1,1}(\CY)+1$ is the quaternionic dimension of $\cM_H$.
These isometries are present in the perturbative metric, but are broken in general by instanton corrections.
However, the relevant corrections vanish exponentially fast in our limit and thus can be ignored.
Next, one constructs a canonical $\IC^2/\IZ_2$ bundle $\cS_{\cM}$, known as Swann bundle \cite{MR1096180} or \hpk cone
in the physics literature \cite{deWit:2001dj}. $\cS_{\cM}$ is an HK manifold, which immediately brings us in the realm of \hpk geometry
with all its available methods. Finally, $\cM'_H$ is obtained by performing $n-\nI+1$ \hpk quotients
along the set of commuting isometries, which include those mentioned above plus one additional isometry corresponding
to a $U(1)$ symmetry on the fiber of the Swann bundle.

\lfig{Geometric construction of the rigid limit via the Swann bundle and \hpk quotient.}{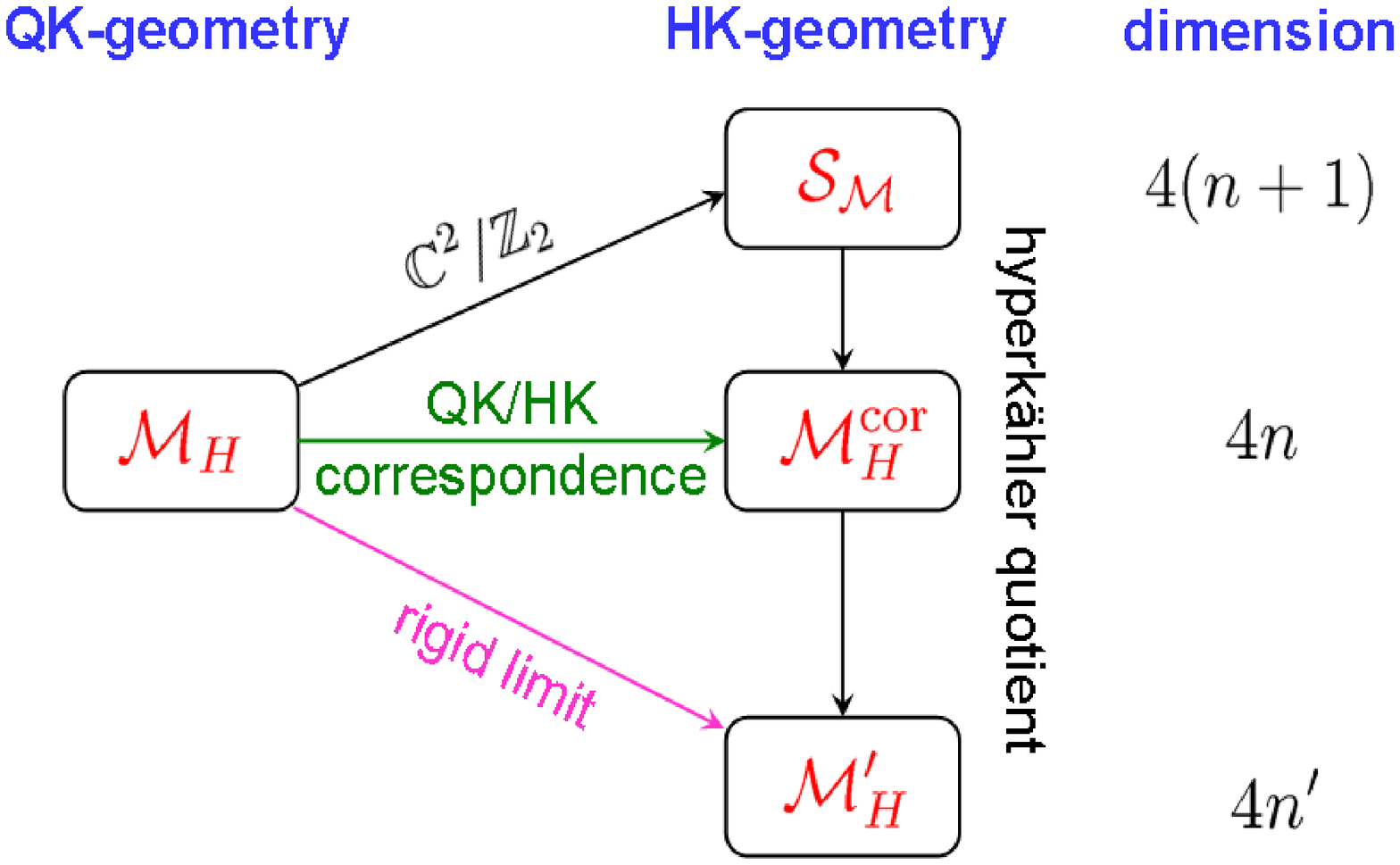}{11.7cm}{fig-quotient}{-0.9cm}

Interestingly, at an intermediate step of this quotient construction, one finds the HK manifold $\Mcor$
which is associated with $\cM_H$ by the so called QK/HK correspondence.
This correspondence establishes a one-to-one map between, on one hand, QK spaces with a quaternionic isometry
and, on the other hand, HK spaces of the {\it same} dimension with a rotational isometry,
equipped with a hyperholomorphic line bundle \cite{Haydys,Alexandrov:2011ac,Hitchin:2012}.
Its physical interpretation is in fact very close to the subject of this paper:
it translates into a formal correspondence between the D-instanton corrected HM moduli space $\cM_H$ and the moduli space
of a 4d $N=2$ gauge theory compactified on a circle, described by the same holomorphic prepotential as the CY.
In particular, the D-instantons are mapped into the gauge theory instantons
produced by BPS particles wrapping the circle.
In a sense, our rigid limit is a close analogue of this formal mathematical correspondence,
with the additional property that both sides realize concrete physical systems.

One should note that a similar geometric prescription for the rigid limit was already given in \cite{Haghighat:2011xx}
for a particular compactification on an elliptically fibered CY.
Here we extend it to the full non-perturbative level, prove it by carefully analyzing the metric, and generalize it to a generic CY.

The work \cite{Haghighat:2011xx} also suggests a physical interpretation of the HK manifold $\cM'_H$:
it is expected to describe the non-perturbative moduli space of a 5d $N=1$ gauge theory compactified on a torus,
where the complex structure of the torus is identified with the frozen axio-dilaton of compactified type IIB string theory.
Indeed, the chain of dualities, shown on Fig. \ref{fig-duality} and explained in detail in section \ref{subsec-duality},
demonstrates that $\cM_H$ is the same moduli space which is obtained by first compactifying M-theory on the same CY $\CY$
and then compactifying its vector multiplet sector on a torus. Since the torus compactification is expected to commute
with the rigid limit, the alternative way to get $\cM'_H$ is to start from 5d supergravity obtained from M-theory on $\CY$,
take the rigid limit in its vector multiplet sector, and only then compactify on $T^2$.
Then the above gauge theory interpretation immediately follows.

\lfig{Duality map and rigid limit of moduli spaces in string and gauge theories.}{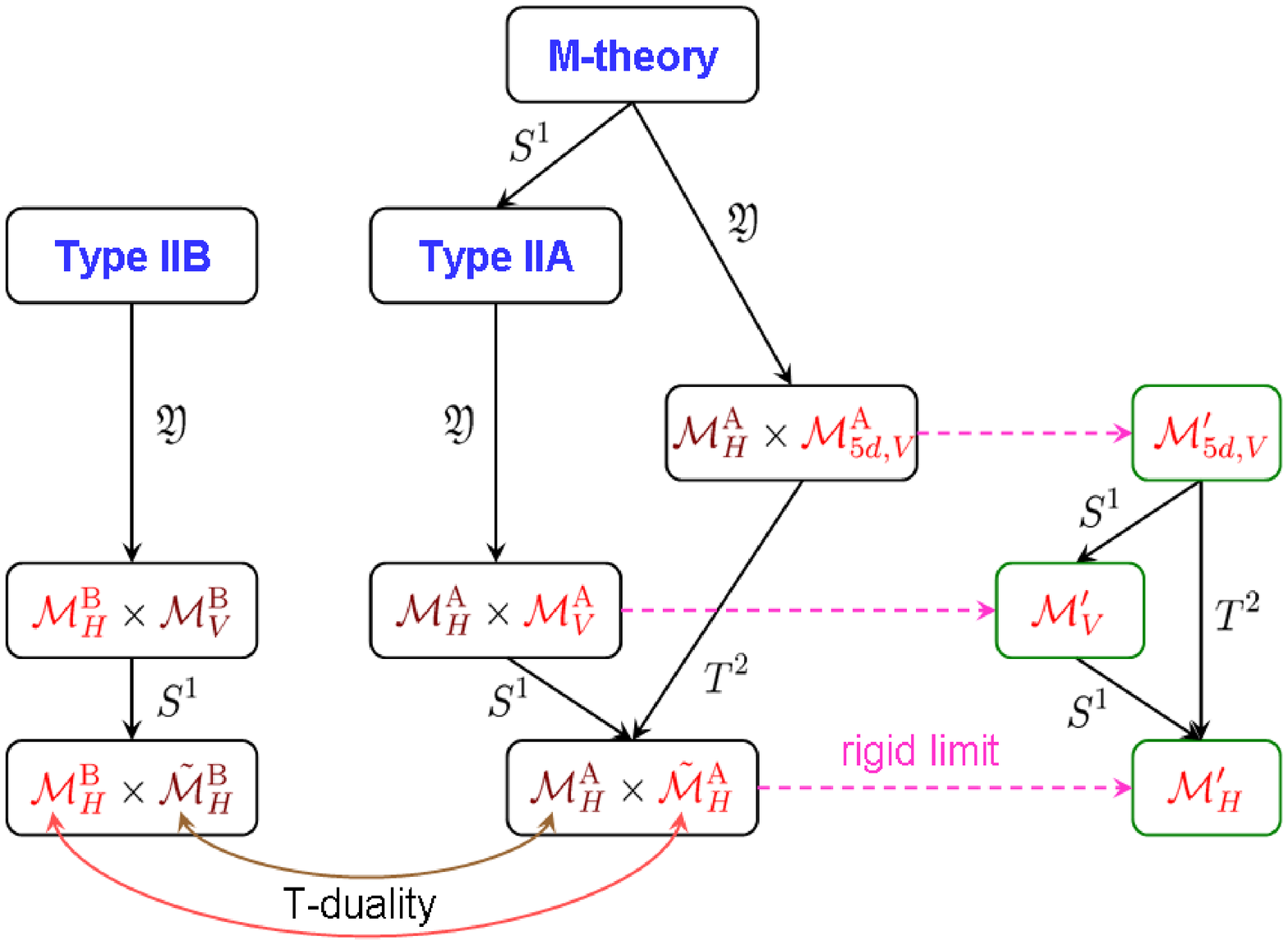}{14.8cm}{fig-duality}{-1cm}

This interpretation opens the possibility to derive non-perturbative effects in compactified 5d gauge theory,
such as dyonic and stringy instantons,
from the known results on D-instantons in CY string theory compactifications.
Although we leave the detailed study of this problem to a future research, here we discuss
various implications of this possibility.

The organization of the paper is as follows. In the next section we study the rigid limit of the HM moduli space $\cM_H$.
First, in \S\ref{subsec-def} we provide the definition of the limit. Then in \S\ref{subsec-cmap} we show how it works
on the example of the classical moduli space where the derivation is particularly explicit,
but contains all the features of the general construction. In \S\ref{subsec-metric} we present
the rigid limit for the full non-perturbative moduli space and in \S\ref{subsec-geometry} provide its geometric interpretation.
The physical interpretation is elaborated in section \ref{sec-5dtorus},
which starts from a discussion of string dualities suggesting the interpretation in terms of 5d gauge theories
(\S\ref{subsec-duality}), proceeds with a brief review of these theories (\S\ref{subsec-5d}),
their compactification on a torus (\S\ref{subsec-torus}),
and finishes with a discussion of
implications for dyonic and stringy instantons (\S\ref{subsec-phys}).
In section \ref{sec-examples} we provide several examples of our construction and
in section \ref{sec-discussion} discuss the results of the paper.
A few appendices contain details on special geometry (\S\ref{ap-special}),
calculations of the D-instanton corrected HM metric (\S\ref{ap-Dinst}),
of the rigid limit (\S\ref{ap-GMN}) and of compactification on a torus (\S\ref{ap-torus}),
and toric data for the examples presented in section \ref{sec-examples} (\S\ref{ap-toricdata}).

\section{Rigid limit}
\label{sec-limit}

In this section we study the rigid limit of the HM moduli space $\cM_H$ of type IIB string theory compactified on a CY threefold $\CY$.
We recall that the moduli space comprises
\begin{itemize}
\item
the axio-dilaton $\tau\equiv\tau_1+\I\tau_2=c^0+\I/g_s$;

\item
the K\"ahler moduli $z^a=b^a+\I t^a$ ($a=1,\dots,h^{1,1}(\CY)$) parametrizing the deformations of the complexified K\"ahler structure of $\CY$;

\item
the RR-fields $c^a,\cla,\cl0$, corresponding to periods of the RR
2-form, 4-form and 6-form on a basis of $H^{\rm even}(\CY,\IZ)$;

\item
and the NS-axion $\psi$ dual to the Kalb-Ramond two-form $B$ in four dimensions.
\end{itemize}
We will use $C^a$ and $D_a$ to denote a basis in the space of curves $H_2(\CY,\IZ)$ and divisors $H_4(\CY,\IZ)$, respectively,
and $\omega_a$ for the basis of harmonic 2-forms dual to $D_a$ so that the expansion of the K\"ahler form reads $J=t^a\omega_a$.
These objects satisfy
\be
C^a\cap D_b=\int_{C^a}\omega_b=\delta^a_b,
\qquad
D_a\cap D_b\cap D_c=\int_{\CY}\omega_a\wedge \omega_b\wedge \omega_c=\kappa_{abc}.
\label{normal-cond}
\ee
Finally, note that in this paper we work in terms of dimensionless moduli.
Therefore, the dimensionful volumes are obtained by dividing integrals of the K\"ahler form by a mass (squared) scale $\Lambda$.
For instance, for 2-cycles one has
\be
\Vol(C^a)=\Lambda^{-1}\int_{C^a}J=\Lambda^{-1} \,t^a.
\label{Volt}
\ee
In the rigid limit, this scale is sent to infinity together with $\Mpl$ so that the shrinking cycles correspond to the finite K\"ahler parameters,
whereas the cycles of finite volume correspond to the moduli scaling as $\Lambda$.

\subsection{Definition}
\label{subsec-def}

Our aim here is to provide a definition of a local limit of the CY manifold.
Usually, this is done by specifying either a set of shrinking 4-cycles or 2-cycles.
On the other hand, to apply it to the metric on the moduli space,
we need a workable definition in terms of the K\"ahler moduli.
Therefore, instead of shrinking cycles,
let us start from a set of $\nA$ linearly independent vectors $\vvtA$ belonging to the K\"ahler cone of $\CY$.
Given these vectors, we define a set of matrices
\be
M_{A,ab}=\kappa_{abc}\vtA^c
\label{matA}
\ee
which in turn allow to introduce another set of vectors $\vvtI$ --- a basis for the common kernel of $M_{A}$,
i.e. linearly independent vectors satisfying
\be
M_{A,ab}\,\vtI^b=0.
\label{deftI}
\ee
We denote their number (i.e. the number of values taken by index $I$) by $\nI$.
We assume that $\nI>0$ and that the two sets, $\vvtA$ and $\vvtI$, are linearly independent.
Already at this point it becomes clear that $\vvtA$ must belong to the boundary of the K\"ahler cone
because it is well known that for any vector inside the cone its contraction with the intersection numbers defines
a non-degenerate matrix of signature $(1,h^{1,1}(\CY)-1)$.
Thus, to have $\nI>0$, all vectors $\vvtA$ must belong to the boundary.\footnote{This has a simple physical explanation.
In a local limit one usually zooms in around a point in the moduli space where CY becomes singular,
and the vectors $\vvtA$ are supposed to point towards such singularity.
But CY can develop a singularity only when its moduli approach the boundary of the K\"ahler cone, which implies the condition on $\vvtA$.}
Finally, we complete these sets to a basis in $H_2(\CY,\IR)$,
which can be done by providing an additional set of $h^{1,1}-\nA-\nI\equiv\nX$ vectors $\vvtX$.
This allows to expand the K\"ahler moduli in the new basis
\be
t^a=\vtA^a\, \hatt^A+\vtX^a\, \hatt^X+\vtI^a\, \hatt^I\equiv \vt^a_b\, \hatt^b,
\label{changet}
\ee
where we combined three indices $A$, $X$ and $I$ into one index $b$.
Then our local limit is defined by taking the moduli $\hatt^A$ to scale as $\Lambda$, whereas $\hatt^X$ and $\hatt^I$
to stay finite (see the comment below \eqref{Volt}).
It is important that this definition of the limit does not depend on the choice of $\vvtX$.
Indeed, changing $\vvtX$ in \eqref{changet} can at most shift $\hatt^A$ and $\hatt^I$ by a combination of $\hatt^X$.
But this does not affect which variables grow with $\Lambda$ and which of them do not.

Let us show that the above definition is equivalent to the usual one in terms of shrinking cycles.
First, we define a rotated basis of divisors $\hD_a=v^b_a D_b$.
It is easy to see that $\hD_I$ are the divisors shrinking in the limit, whereas the divisors $\hD_{\hA}$,
where we introduced a combined index $\hA=(A,X)$, remain with a finite volume. Indeed,
\be
\begin{split}
\Vol(\hD_I)=&\frac{1}{2\Lambda^2}\int_{\hD_I}J\wedge J=\frac{1}{2\Lambda^2}\, \vtI^a\,\kappa_{abc}t^b t^c\escl{-2},
\\
\Vol(\hD_{\hA})=&\frac{1}{2\Lambda^2}\int_{\hD_{\hA}}J\wedge J=\frac{1}{2\Lambda^2}\, \vt_{\hA}^a\,\kappa_{abc}t^b t^c
\approx \frac{1}{2\Lambda^2}\,\vt_{\hA}^a \,M_{B,ab} \vt_{C}^b \hatt^B\hatt^C \escc,
\end{split}
\ee
where the first result follows from \eqref{deftI},
whereas the second is due to that none of vectors $\vec\vt_{\hA}$ belongs to the common kernel of $M_A$.\footnote{We consider
a generic point in the moduli space so that no accidental cancellations are possible due to contraction with $\hatt^A$.}

Second, we define a rotated basis of curves $\hC^a=(v^{-1})^a_b C^b$. Their volumes are given by $\Lambda^{-1}\hatt^a$
and therefore $\hC^A$ has a finite volume, whereas $C^{\hI}$, where we introduced another combined index $\hI=(I,X)$,
are shrinking.
It is important to note that all shrinking curves can be characterized by their orthogonality to the divisors $\hD_A$,
\be
\hC^{\hI}\cap \hD_{A}=0,
\label{orthCD}
\ee
since due to \eqref{normal-cond} the l.h.s is evaluated to $(\vt^{-1})^{\hI}_a\vtA^a =0$.

Thus, our definition of the local limit is equivalent to specifying either the set of shrinking divisors $\hD_I$
or the set of shrinking curves $\hC^{\hI}$. Both sets are in one-to-one correspondence with vectors $\vvtA$, and both
their definitions $\hD_I=\vtI^a D_a$ as well as the orthogonality relation \eqref{orthCD} do not depend on $\vvtX$.
Of course, to talk about a local limit, one must have at least one shrinking divisor, which gives the condition $\nI>0$.
Thus, the condition of having a non-trivial limit is that the common kernel of $M_A$ is non-empty.

Finally, we impose an additional condition on the vectors $\vvtA$ that
$\kappa_{abc}\vt_A^a\vt_B^b\vt_C^c$ is non-zero at least for some $A,B,C$.
It ensures that the volume of the CY, $\Vcy=\frac16\, \kappa_{abc}t^a t^b t^c$, scales as $\Lambda^3$ in the local limit.
As we will show below, under these conditions the three sets of moduli appearing in \eqref{changet} acquire in the limit
a very different status:
\begin{itemize}
\item
$\hatt^A$ become frozen and do not enter the finite part of the metric;
\item
$\hatt^X$ are also frozen, but appear in the finite part;
\item
$\hatt^I$ remain dynamical.
\end{itemize}
Correspondingly, their physical interpretation in the dual gauge theory will also be different:
while $\hatt^I$ are associated with the Coulomb branch moduli, $\hatt^X$ provide its physical parameters such as masses and
the gauge coupling.

In the following, to simplify notations, we assume that the rotation of the basis \eqref{changet} has already been done
and drop hats on the moduli adapted to the limit, i.e. consider $t^A$ to be of order $\Lambda$, whereas $t^X$ and $t^I$ as finite variables.
Then \eqref{deftI} implies that in this basis the intersection numbers possess the following property
\be
\kappa_{aAI}=0,
\label{restr-kap}
\ee
whereas the matrix $M_{\hA\hB}=\kappa_{\hA\hB C}t^C$ is non-degenerate.
In section \ref{sec-examples} we will return back to the original basis and
discuss in more detail the conditions for the existence of a non-trivial limit.

\subsection{Example: classical c-map}
\label{subsec-cmap}

Before attacking the problem of taking the rigid limit of the non-perturbative HM moduli space,
let us consider how it works at the classical level where all quantum corrections in $\alpha'$ and $g_s$ are ignored.
In this approximation the metric on $\cM_H$ is given by the {\it local c-map} \cite{Cecotti:1989qn,Ferrara:1989ik}
which is a QK manifold constructed in a canonical way from the holomorphic prepotential $F(X^\Lambda)$ ($\Lambda=0,\dots, h^{1,1}(\CY)$)
on the K\"ahler moduli space of $\CY$.
It has the simplest form in terms of the fields of type IIA string theory compactified on the mirror CY $\CYm$,
which comprise the four-dimensional dilaton $r=e^\phi$,
the complex structure moduli $z^a$, the RR-scalars $\zeta^\Lambda$, $\tzeta_\Lambda$
corresponding to periods of the RR 3-form on a basis of $H^3(\CYm,\IZ)$, and the NS-axion $\sigma$ dual to the $B$-field.
In these coordinates the metric reads
\be
\label{cmap}
\begin{split}
\de s^2_{\Mcl} =&\, \frac{\de r^2}{r^2}
- \frac{1}{2r}\, \Im\cN^{\Lambda\Sigma} \(\de\tzeta_\Lambda- \cN_{\Lambda\Lambda'}\de\zeta^{\Lambda'}\)
\(\de\tzeta_\Sigma- \bar{\cN}_{\Sigma\Sigma'}\de\zeta^{\Sigma'}\)
\\ &\,
+\frac{1}{16r^2}\(\de\sigma+\tzeta_\Lambda\de\zeta^\Lambda-\zeta^\Lambda\de\tzeta_\Lambda\)^2 + 4{\cK}_{a \bar{b}} \de z^a \de {\bz}^{\bar{b}},
\end{split}
\ee
where $\cK$ is the K\"ahler potential on the special K\"ahler space of complex structure deformations of $\CYm$ (we set $z^\Lambda=(1,z^a)$)
\be
\cK=- \log \[\I (\bz^\Lambda F_\Lambda - z^\Lambda \bF_\Lambda)\],
\label{Kahlerpot}
\ee
$F_\Lambda$, ${\cK}_{a \bar{b}}$, etc. denote derivatives of the corresponding quantities without indices,
and $\cN_{\Lambda\Sigma}$ is the matrix of the gauge couplings defined in \eqref{defcN}.
We refer to appendix \ref{ap-special} for the details on the special geometry encoded by the prepotential $F$.

To return to the type IIB fields, which we used to define the rigid limit, one should apply the {\it mirror map}.
In the classical approximation it was found in \cite{Bohm:1999uk} and identifies the complex structure moduli $z^a$ with
the complexified K\"ahler moduli as well as
\be
\begin{split}
& r = \frac{\tau_2^2}{2} \, \Vcy,
\qquad
\zeta^0 = \tau_1,
\qquad
\zeta^a = -c^a + \tau_1 b^a,
\\ &
\tzeta_a = \cla + \frac12\, \kappa_{abc} b^b (c^c -\tau_1 b^c),
\qquad
\tzeta_0 = \cl0 - \frac16\,\kappa_{abc} b^a b^b (c^c-\tau_1b^c),
\\ &
\sigma = -2\(\psi+\frac12\, \tau_1 \cl0\) + \cla(c^a-\tau_1 b^a)  - \frac16\, \kappa_{abc} b^a c^b (c^c-\tau_1 b^c).
\end{split}
\label{mirmap}
\ee
The classical prepotential to be used in \eqref{cmap} is completely determined by the triple intersection numbers of $\CY$
\be
\Fcl(X) = -\frac16\, \kappa_{abc} \, \frac{X^a X^b X^c}{X^0}.
\label{Fcl}
\ee
Let us now plug in this prepotential and the change of variables \eqref{mirmap} into the c-map metric.
Then, using the expressions \eqref{mmIR} for the gauge coupling matrix and its inverse,
after straightforward, but a bit tedious manipulations the metric can be brought
to the following form
\be
\begin{split}
\de s^2_{\Mcl} = &\,
\frac{(\de r)^2}{r^2} +\frac{\de\tau_1^2}{\tau_2^2}
+ 4\cK_{a\bar{b}}\[ \de t^a \de t^b+ \frac{1}{\tau_2^2} \(\de c^a -\tau \de b^a)(\de c^b-\btau\de b^b\)\]
\\
&\,
+\frac{\cK^{a\bar{b}}}{4\tau_2^2\Vcy^2}
\(\de \cla + \frac12\, \kappa_{acd} \(c^c \de b^d - b^c \de c^d\)\)\(\de \tilde{c}_b + \frac12\,\kappa_{bfg}\(c^f \de b^g-b^f \de c^g\)\)
\\
&\,
+ \frac{1}{\tau_2^2\Vcy^2} \(\de \cl0 + b^a \de \cla + \frac16\, \kappa_{abc} b^a(c^b\de b^c- b^b\de c^c)\)^2
\\
&\,
+ \frac{1}{\tau_2^4 \Vcy^2} \[\de\psi+\tau_1\de \cl0 -(c^a-\tau_1 b^a) \(\de \cla -\frac16 \kappa_{abc} (b^b\de c^c - c^b\de b^c)\)\]^2.
\end{split}
\label{cmapIIB}
\ee
Using \eqref{mirmap} and \eqref{Kmetric-expl}, the first three terms can be rewritten as
\be
\frac{|\de\tau|^2}{\tau_2^2}+2\(\de\log(\Vcy\tau_2^{3/2})\)^2
-\frac{\kappa_{abc}t^c}{\tau_2 \Vcy}\,\de(\sqrt{\tau_2} t^a) \de(\sqrt{\tau_2} t^b),
\ee
whereas the last two terms can be reorganizied in the following way
\bea
&&
\frac{1}{\tau_2^4\Vcy^2} \[\left| \de\psi+\tau\de\cl0\right|^2
+ \left|(c^a-\btau b^a) \(\de \cla -\frac16 \kappa_{abc} (b^b\de c^c - c^b\de b^c)\)\right|^2 \]
\\
&&
- \frac{1}{\tau_2^4 \Vcy^2}\bigg[(\de\psi+\tau\de\cl0)(c^a-\btau b^a) + (\de\psi+\btau\de\cl0)(c^a-\tau b^a)  \bigg]
\(\de \cla -\frac16 \kappa_{abc} (b^b\de c^c - c^b\de b^c)\).
\nn
\eea
This rewriting makes it manifest that the whole metric is invariant under the $SL(2,\IR)$ isometry group acting on the type IIB fields as
\be\label{SL2Z}
\begin{array}{c}
\displaystyle{
\tau \mapsto \frac{a \tau +b}{c \tau + d} \, ,
\qquad
t^a \mapsto t^a |c\tau+d| \, ,
\qquad
\cla\mapsto \cla \,  ,}
\\
\displaystyle{
\begin{pmatrix} c^a \\ b^a \end{pmatrix} \mapsto
\begin{pmatrix} a & b \\ c & d  \end{pmatrix}
\begin{pmatrix} c^a \\ b^a \end{pmatrix} ,
\qquad
\begin{pmatrix} \cl0 \\ \psi \end{pmatrix} \mapsto
\begin{pmatrix} d & -c \\ -b & a  \end{pmatrix}
\begin{pmatrix} \cl0 \\ \psi \end{pmatrix}},
\end{array}
\ee
where $a,b,c,d$ are the parameters of the transformation
${\scriptsize \begin{pmatrix} a & b \\ c & d \end{pmatrix}} \in SL(2,\IR)$ with $ad-bc=1$.
This symmetry descends from the S-duality group of type IIB supergravity in 10 dimensions,
but is broken to the discrete subgroup $SL(2,\IZ)$ by quantum corrections \cite{RoblesLlana:2006is}.
It is this symmetry that fixed the form of the mirror map \eqref{mirmap}
and it will play an important role in the physical interpretation of the rigid limit.

To extract this limit from the metric \eqref{cmapIIB}, it is enough to understand
the behavior of the special K\"ahler metric $\cK_{a\bar{b}}$ and its inverse.
This can be done using the representation \eqref{Kmetric-expl} valid in the classical approximation.
It involves the matrix $\kappa_{ab}=\kappa_{abc} t^c$ and its inverse, so first we establish the scaling for them.
Using notations for indices from the previous subsection, the restriction on intersection numbers \eqref{restr-kap},
the matrix $M_{\hA\hB}$ introduced below it,
the matrix $g_{IJ}=-\kappa_{IJ\hK}t^{\hK}$ and their inverse $M^{\hA\hB}$ and $g^{IJ}$, one finds
\bea
\kappa_{ab}&\approx & \(\begin{array}{cc}
M_{\hA\hB} & \kappa_{\hA J \hK}t^{\hK}
\\
\kappa_{I\hB \hK}t^{\hK} & -g_{IJ}
\end{array}\)
\sim
\(\begin{array}{cc}
\scl{} & \scc
\\
\scc & \scc
\end{array}\),
\label{sc-kappa}
\\
\kappa^{ab}&\approx & \(\begin{array}{cc}
M^{\hA\hB} & M^{\hA X}\kappa_{X K \hL}t^{\hL}g^{KJ}
\\
g^{IK}\kappa_{X K \hL}t^{\hL} M^{X\hB} & -g^{IJ}
\end{array}\)
\sim
\(\begin{array}{cc}
\scl{-1} & \scl{-1}
\\
\scl{-1} & \scc
\end{array}\).
\label{sc-invkappa}
\eea
Plugging these results into \eqref{Kmetric-expl}, one obtains
\be\hspace{-0.5cm}
\begin{array}{rllcrll}
4\Vcy\cK_{\hA\bar\hB}\approx & -M_{\hA\hB}+\frac{1}{4\Vcy}\,M_{\hA A} t^A M_{\hB B} t^B&\escl{},
&\quad &
\frac{1}{4\Vcy}\,\cK^{\hA\bar\hB}\approx & -M^{\hA\hB}+\delta^{\hA}_{A}\delta^{\hB}_B\, \frac{t^A t^B}{2\Vcy}&\escl{-1}
\\
4\Vcy\cK_{I\bar B}\approx &\frac{1}{4\Vcy}\,M_{BC} t^C\kappa_{I\hK\hL} t^{\hK} t^{\hL}  &\escl{-1},
&\quad &
\frac{1}{4\Vcy}\,\cK^{I\bar\hB}\approx & -g^{IJ}\kappa_{JX\hK}t^{\hK} M^{X\hB} &\escl{-1},
\\
4\Vcy\cK_{I\bar \hJ}\approx & -\kappa_{I\hJ\hK}t^{\hK}&\escc,
&\quad &
\frac{1}{4\Vcy}\,\cK^{I\bar J}\approx & g^{IJ} &\escc.
\end{array}
\label{scaleK}
\ee

On the basis of these scaling results, the bosonic Lagrangian defined by the metric \eqref{cmapIIB}
can be split into three contributions\footnote{The overall minus sign comes from that we work in the `most plus' signature $(-,+,+,+)$.}
\be
\cL_{\rm bos} =-\frac{\sqrt{-g}}{2\kappa^2\Vcy\tau_2^{3/2}}\(\cL_+ +\cL_0+\cL_{-}\),
\label{fullL}
\ee
where
\begin{subequations}
\bea
\cL_+ &=&
\frac{\Vcy}{2\tau_2^{1/2}}\[\(2\p_\mu \tau_2+\frac{\tau_2}{2\Vcy}\,\kappa_{abc}t^at^b\p_\mu t^c\)^2 +(\p_\mu\tau_1)^2\]
\nn\\
&&+ 2\tau_2^{3/2}\Vcy\cK_{\hA\hB}\[ \p_\mu t^{\hA} \p^\mu t^{\hB}+
\frac{1}{\tau_2^2} \(\p_\mu c^{\hA} -\tau \p_\mu b^{\hA})(\p^\mu c^{\hB}-\btau\p^\mu b^{\hB}\)\],
\\
\cL_0 &=& 4 \tau_2^{3/2}\Vcy\cK_{I\hB}\[ \p_\mu t^I \p^\mu t^{\hB}+ \frac{1}{\tau_2^2} \(\p_\mu c^I -\tau \p_\mu b^I)(\p^\mu c^{\hB}-\btau\p^\mu b^{\hB}\)\]
\nn\\
&& +2 \tau_2^{3/2}\Vcy\cK_{IJ}\[ \p_\mu t^I \p^\mu t^J+ \frac{1}{\tau_2^2} \(\p_\mu c^I -\tau \p_\mu b^I)(\p^\mu c^J-\btau\p^\mu b^J\)\]
+\frac{\cK^{I\bar J}}{8\tau_2^{1/2}\Vcy}  \, y_{I\mu} {y_{J}}^\mu,
\\
\cL_{-} &=& \frac{1}{8\tau_2^{1/2}\Vcy} \left\{\cK^{\hA\hB} y_{\hA\mu} {y_{\hB}}^\mu
+2\cK^{I\hB} y_{I\mu} {y_{\hB}}^\mu
+4\(\p_\mu \cl0 + b^a \p_\mu\cla + \frac16\, \kappa_{abc}b^a (c^b\p_\mu b^c- b^b\p_\mu c^c)\)^2
\right.
\nn\\
&&\left.
+ \frac{4}{\tau_2^2} \[\p_\mu\psi+\tau_1\p_\mu \cl0 -(c^a-\tau_1 b^a)
\(\p_\mu \cla -\frac16 \kappa_{abc} (b^b\p_\mu c^c - c^b\p_\mu b^c)\)\]^2\right\},
\eea
\end{subequations}
and we denoted
\be
y_{a\mu}=\p_\mu \tc_a + \frac12\, \kappa_{abc} \(c^b \p_\mu b^c- b^b \p_\mu c^c\).
\ee
Let us take the gravitational coupling $\kappa^2$ scaling as $\Lambda^{-3}$ so that $\kappa^2 \Vcy$ remains constant.
Then, as the notations suggest, $\cL_+$ corresponds in our limit to the divergent part of the Lagrangian, $\cL_0$ stays finite, and $\cL_-$ vanishes.
As a result, the fields $\psi$, $\tc_0$ and $\tc_{\hA}$, appearing only in $\cL_-$, simply drop out from the theory, whereas
the divergent part imposes its equations of motion as strong constraints. These leads to the freezing of the moduli
$\tau$, $t^{\hA}$, $b^{\hA}$ and $c^{\hA}$, which means that their fluctuations vanish or at least scale as $\Lambda^{-1}$,
and thus these fields can be considered as constant.
Taking this into account in $\cL_0$, one obtains that its non-vanishing part is determined by the following metric
\bea
\de s^2_{\Mpcl} &= & \hf\,\tau_2^{3/2} g_{IJ}\[ \de t^I \de t^J+ \frac{1}{\tau_2^2} \(\de c^I -\tau \de b^I)(\de c^J-\btau\de b^J\)\]
\label{metricIIB-rigid}\\
&& +\frac{g^{IJ}}{2\tau_2^{1/2}}  \(\de \tc_I + \frac12\, \kappa_{IK\hL} \(c^{\hL} \de b^K - b^{\hL} \de c^K\)\)
\(\de\tc_J + \frac12\, \kappa_{JM\hN}\(c^{\hN}\de b^M-b^{\hN} \de c^M\)\).
\nn
\eea
Note that it is manifestly $SL(2,\IR)$ invariant.
It is to keep this invariance we included the factor $\tau_2^{3/2}$ into the rescaling of the Lagrangian in \eqref{fullL}.

The metric \eqref{metricIIB-rigid} describes the rigid limit of the classical HM moduli space.
The space $\Mpcl$ where it leaves on is parametrized by $4\nI$ coordinates $t^I$, $b^I$, $c^I$ and $\tc_I$, whereas
$\tau$, $t^X$, $b^X$ and $c^X$ also appearing in the metric play the role of fixed parameters.
The geometric meaning of this metric can be elucidated by going back to the analogue of the type IIA variables.
Using the inverse mirror map relations \eqref{mirmap},
the metric can be rewritten as
\be
\de s^2_{\Mpcl}=\frac{1}{2\sqrt{\tau_2}}\[ \tau_2^2 \Im \fcl_{IJ}\de \uz^I\de\buz^J
+(\Im {\fcl})^{IJ}\(\de\tzeta_I-\fcl_{JK}\de \zeta^K\)\(\de \tzeta_J-\bfcl_{JL}\de \zeta^L\)\],
\label{rigL}
\ee
where the new prepotential is
\be
\fcl(\uz^I)=-\frac{1}{6}\, \kappa_{\hI\hJ\hK} \uz^{\hI} \uz^{\hJ} \uz^{\hK}.
\label{fcl}
\ee
One recognizes in \eqref{rigL} the well known {\it rigid c-map} \cite{Cecotti:1989qn}, which
describes an HK space constructed as a canonical bundle over the rigid special K\"ahler base with the holomorphic prepotential $\fcl(\uz^I)$.
Typically, it arises as the classical target space of three-dimensional non-linear $\sigma$-models obtained
by compactifications of gauge theories with eight supercharges.
The parameter $\tau_2$ controls the radius of compactification, but can be absorbed by the redefinition $u^I=\frac{\tau_2}{2}\, z^I$.

Furthermore, it is easy to see that $\Mpcl$ can be obtained from a larger rigid c-map space, which we call $\Mcorcl$
and which is determined by the prepotential $\fcl(u^\Lambda)=\Fcl(u^\Lambda)$.
The space $\Mcorcl$ has quaternionic dimension $n=h^{1,1}(\CY)+1$, and
its metric is given by exactly the same metric \eqref{rigL} (after the rescaling mentioned above) where however
the indices $I,J,\dots$ should be replaced by $\Lambda,\Sigma,\dots$ running over $0,\dots,h^{1,1}(\CY)$.
As any rigid c-map, $\Mcorcl$ has a set of commuting isometries acting by shifts of $\tzeta_\Lambda$,
with the triplet of moment maps given in the chiral basis by
$(\rho_+^\Lambda,\rho_-^\Lambda,\rho_3^\Lambda)=(u^\Lambda,\bu^\Lambda,\zeta^\Lambda)$.
Then performing $n-\nI$ \hpk quotients along $\tzeta_0$ and $\tzeta_{\hA}$ fixes the moment maps $\vec\rho^{\,\,0}$ and $\vec\rho^{\,\hA}$
and gives us back the manifold $\Mpcl$. The decoupling of the variables fixed by $\vec\rho^{\,A}$ is ensured by the condition \eqref{restr-kap}.
In particular, the prepotential $\Fcl(u^\Lambda)$, up to an overall factor and an irrelevant constant contribution,
reduces to \eqref{fcl} after identifying the moment maps of the first isometry as
$(\frac{\tau_2}{2},\frac{\tau_2}{2},\tau_1)$.\footnote{$\Mcorcl$ has an isometry which acts by multiplying all $u^\Lambda$ by a phase.
It can be used to cancel the phase of the moment maps $\rho_\pm^0$, this is why it is always possible to choose them to be real.\label{foot-isometry}}

In turn, the rigid c-map $\Mcorcl$ is known to be related to the local c-map $\Mcl$ by the QK/HK correspondence \cite{Alexandrov:2011ac}.
It proceeds via construction of the Swann bundle $\cS_{\cM}$ over the QK space with an isometry and subsequent \hpk quotient along
the isometry inherited on $\cS_{\cM}$.
In the case of the local c-map \eqref{cmap}, the role of such isometry is played by shifts of the NS-axion $\sigma$.
As a result, we arrive at the precise realization of the geometric scheme shown on Fig. \ref{fig-quotient}.

\subsection{Rigid limit of the non-perturbative HM moduli space}
\label{subsec-metric}

\subsubsection{Quantum corrections}
\label{subsubsec-quantum}

To extract the rigid limit of the full non-perturbative moduli space $\cM_H$,
let us first recall what kinds of quantum corrections affect the classical c-map metric considered in the previous subsection.
There are two classes of such corrections: one comes from quantum effects on the string worldsheet and is weighted by $\alpha'$,
and the other comes from physics in the target space and is weighted by $g_s$.
All $\alpha'$-corrections are captured as corrections to the holomorphic prepotential, and therefore the $\alpha'$-corrected HM metric
still falls into the class of metrics given by the local c-map.
However, the prepotential is now a deformation of the simple classical function \eqref{Fcl},
which is known to have the following form \cite{Candelas:1990rm,Hosono:1993qy}\footnote{In fact,
the prepotential also has a quadratic contribution $\frac12\, A_{\Lambda\Sigma} X^\Lambda X^\Sigma$ where $A_{\Lambda\Sigma}$ is real so that
this term does not affect the K\"ahler potential $\cK$ and is often omitted. However, it becomes important when one extends mirror symmetry
to the non-perturbative level \cite{Alexandrov:2010ca}.
Nevertheless, it is still possible to remove this term by a symplectic transformation. One should just take into account
that this transformation affects the integrality of D-brane charges which become rational.
This is the symplectic frame that is accepted in this work.\label{foot-ALS}}
\be
\label{lve}
F(X)=\Fcl(X)
- \chi_{\CY}\frac{\zeta(3)(X^0)^2}{2(2\pi\I)^3}
-\frac{(X^0)^2}{(2\pi\I)^3}\sum_{k_aC^a\in H_2^+(\CY)} \gfinv_{k_a}\,
\Li_3\left( e^{2\pi \I  k_a X^a/X^0}\right),
\ee
where the second term describes a perturbative $\alpha'$-correction, whereas the third term, parametrized by genus zero
Gopakumar-Vafa invariants $\gfinv_{k_a}$, corresponds to the contribution of worldsheet instantons wrapping effective curves $k_aC^a$.

The situation with $g_s$-corrections is more complicated.
At the perturbative level, the corrections appear only at one-loop \cite{Antoniadis:2003sw,Robles-Llana:2006ez} and the corresponding metric,
which is already {\it not} in the c-map class, is explicitly known \cite{Alexandrov:2007ec}.
At the non-perturbative level, there are two sources of $g_s$-corrections:
D-branes wrapping non-trivial cycles on the CY and NS5-branes wrapping the whole CY.
How to include the contributions of the former, to {\it all} orders in the instanton expansion, has been understood
(in the type IIA formulation) in \cite{Alexandrov:2008gh,Alexandrov:2009zh}, but only partial results are accessible for the latter
\cite{Alexandrov:2010ca,Alexandrov:2014mfa,Alexandrov:2014rca}.

Given such incomplete understanding of the HM moduli space, it is natural to ask whether it is possible to find
the exact rigid limit of $\cM_H$ or only its approximation?
It turns out that the lack of knowledge of the exact description of NS5-brane instantons does not pose a problem for evaluating
the rigid limit because these instantons necessarily decouple.
Indeed, they are known to have the following leading contribution \cite{Becker:1995kb}
\be
\sim e^{-2 \pi |k| \Vcy /g_s^2-\I\pi  k \sigma}.
\label{couplNS5}
\ee
At the same time, in a any local limit the (dimensionless) volume of the CY $\Vcy$ diverges and thus
the NS5-instantons are exponentially suppressed and can be ignored.

Furthermore, some of D-instantons decouple too.
Let us look as above at their leading contribution, which in the type IIA variables
has the following form \cite{Becker:1995kb}
\be
\label{d2quali}
\sim e^{ -2\pi|Z_\gamma|/g_s
- 2\pi\I (q_\Lambda \zeta^\Lambda-p^\Lambda\tzeta_\Lambda)} ,
\ee
where
\be
\label{defZ}
Z_\gamma(z) = q_\Lambda z^\Lambda- p^\Lambda F_\Lambda(z)
\ee
is the central charge function determined by the prepotential and the charge vector $\gamma=(p^\Lambda, q_\Lambda)$.
In the type IIA formulation, $\gamma$ picks out an element of $H_3(\CY,\IZ)$ wrapped by a D2-brane,
whereas in type IIB it decomposes as $\gamma=(p^0, p^a, q_a, q_0)$
and defines an element\footnote{In fact, the charges are not integer due to two reasons.
First, they have rational shifts because of the symplectic rotation mentioned in footnote \ref{foot-ALS}. And second, our charge lattice
is already a result of rotation \eqref{changet} to the basis adapted for the rigid limit in which,
in particular, the intersection numbers satisfy the condition \eqref{restr-kap}.}
of $H_{\rm even}(\CY,\IZ)$ corresponding to a D5-D3-D1-D(-1) bound state.
Substituting the prepotential \eqref{lve} into the central charge, one finds that in the local limit
the leading part of the D-instanton action behaves as
\begin{itemize}
\item
D5-instantons ($p^0\ne 0$): \hspace{2.57cm} $\sim|p^0|\,\Vcy$;
\item
D3-instantons ($p^0=0,\ p^a\ne 0$): \hspace{1.1cm} $\sim|p^a \kappa_{abC}t^b t^C|=|M_{ab}p^at^b|$;
\item
D1-instantons ($p^0=p^a=0,\ q_a\ne 0$): $\quad\sim|q_A t^A|$.
\end{itemize}
Thus, D5-instantons are always exponentially suppressed, and the same is true for D3-instantons with charges having
at least one non-vanishing component $p^{\hA}$ and D1-instantons with charges having at least one non-vanishing component $q_{A}$.
On the other hand, it is easy to check that the D-instantons with charges $\gamma=(0,p^I,q_{\hI},q_0)$ have a finite instanton action and do not decouple.
We denote the lattice of the remaining charges by $\Geff$.
Note that these results are in perfect agreement with the discussion in \S\ref{subsec-def}
because $\Geff$ precisely corresponds to the set of shrinking cycles, whereas for large K\"ahler moduli
the instanton action coincides with the volume of the cycle wrapped by the brane.

Finally, it is clear that the worldsheet instantons wrapping curves $C^A$ also decouple since their instanton action is proportional to $|k_At^A|$.
As a result, to extract the rigid limit, it is enough to consider the HM metric corrected by worldsheet instantons with charges $k_{\hI}$
and D-instantons with charges $\gamma\in\Geff$.

\subsubsection{D-instanton corrected HM metric}

As explained above, all of the instantons needed for the rigid limit are in principle known.
But do we know them in practice? In fact, in the case of D-instantons we do not.
In \cite{Alexandrov:2008gh,Alexandrov:2009zh} these instanton effects have been implemented at the level of the twistor space $\cZ_\cM$,
a canonical $\CP$ bundle over $\cM_H$, as deformations of its contact structure. More precisely,
this contact structure can be encoded in a set of holomorphic Darboux coordinates $(\xi^\Lambda, \txi_\Lambda, \alpha)$
on $\cZ_\cM$ expressed as functions of coordinates on $\cM_H$
and a holomorphic coordinate on the $\CP$ fiber (see appendix \ref{ap-Dinst} for details).
The instantons modify these functions and, as a result, the Darboux coordinates
become determined by a system of integral equations which has the form
of thermodynamic Bethe ansatz. Not only these equations cannot be solved in full generality, but also the procedure
to get the metric out of the Darboux coordinates is quite complicated and involves several non-trivial steps.

Recently, the problem of deriving the explicit metric corrected by D-instantons has been solved
for a subset of them \cite{Alexandrov:2014sya},
which can be characterized as instantons with charges all having vanishing symplectic products
\be
\langle\gamma,\gamma'\rangle=q_\Lambda p'^\Lambda-q'_\Lambda p^\Lambda
\label{sympprod}
\ee
and called usually mutually local.
A crucial simplification arising in this case is that the above mentioned integral equations become solvable.
However, this result is not sufficient for our purposes because the effective charge lattice $\Geff$ does contain mutually non-local charges.
These are, for instance, D3-instantons with charges $p^I$ and D1-instantons with charges $q_I$.
Thus, we need a generalization of the result presented in \cite{Alexandrov:2014sya}.

In appendix \ref{ap-Dinst}, we solve this problem and derive the HM metric including {\it all} D-instanton corrections.
The result is given by
\bea
\de s^2_{\cM_H} &= & \frac{2}{r^2}\(1-\frac{8r}{\tau_2^2\Uin}\)(\de r)^2
-\frac{1}{r}\(N^{\Lambda\Sigma}-\frac{\tau_2^2}{8r}\,z^\Lambda\bz^\Sigma\)\cY_\Lambda\bar\cY_\Sigma
-\frac{2}{r}\sum_{\gamma,\gamma'}(\vv\Min^{-1})_{\gamma\gamma'}\cY_{\gamma}\bar\cY_{\gamma'}
\nn\\
&&
+\frac{1}{r\Uin}\left|\sum_{\gamma}\((\zzb\Min^{-1})_\gamma \cY_\gamma+\frac{\tau_2}{4\pi}\,\cW_{\gamma}\de Z_{\gamma}\)\right|^2
\nn\\
&&
+\frac{\tau_2}{r}\sum_{\gamma,\gamma',\gamma''}\Min^{-1}_{\gamma\gamma'}
\[\vvpn{1}_{\gamma\gamma''}\(\de Z_{\gamma''}-\Uin^{-1}Z_{\gamma''}\p e^{-\cK}\) \bar\cY_{\gamma'}
+\cY_{\gamma'}\vvmn{1}_{\gamma\gamma''}\(\de\bZ_{\gamma''}-\Uin^{-1}\bZ_{\gamma''}\bar\p e^{-\cK}\)\]
\nn\\
&&
+\frac{\tau_2^2}{4r}\[ \Uin^{-1}|\p e^{-\cK}|^2-N_{\Lambda\Sigma}\de z^\Lambda \de\bz^\Sigma
-\frac{1}{2\pi\Uin}\sum_{\gamma}\Bigl(\cW_{\gamma}\de Z_{\gamma} \bar\p e^{-\cK}+\p e^{-\cK} \bar\cW_{\gamma}\de\bZ_{\gamma}\Bigr) \]
\label{metric}\\
&&
+\frac{\tau_2^2}{2r}\sum_{\gamma,\gamma'}\vvp_{\gamma\gamma'}\de Z_{\gamma'}\de \bZ_{\gamma}
-\frac{\tau_2^2}{r}\sum_{\gamma,\gamma'}(\Min^{-1}\cQ)_{\gamma\gamma'}\sum_{\tgam}\vvpn{1}_{\gamma\tgam}\de Z_{\tgam}
\sum_{\tgam'}\vvmn{1}_{\gamma'\tgam'}\de \bZ_{\tgam'}
\nn\\
&&
+ \frac{1}{32 r^2\(1-\frac{8r}{\tau_2^2\Uin}\)}\(\de \sigma+\tzeta_\Lambda\de \zeta^\Lambda-\zeta^\Lambda\de \tzeta_\Lambda
+\frac{1}{64\pi^4}\sum\limits_{\gamma,\gamma'}\Om{\gamma} \Om{\gamma'}\langle\gamma,\gamma'\rangle \Igg{}\de\Igam{\gamma'}+\cV\)^2.
\nn
\eea
We refer to the appendix for the explanation of all the notations appearing in \eqref{metric}.
Here we just note that this result is only semi-explicit because all the functions appearing in the metric are defined by
a solution of the integral equations which is supposed to be found as a perturbative series in the number of instantons.
Besides, the result involves two other expansions. One is used to define the matrices \eqref{mat-cI} entering the definition
of other quantities such as $\vv_{\gamma\gamma'}$ and $\vvpm_{\gamma\gamma'}$.
The other is due to the inverse of matrix $\Min_{\gamma\gamma'}$ which also can be found only as a perturbative series.
However, to every given order, both series can be easily evaluated and the metric follows by a direct substitution.
More importantly, this does not represent any obstacle for finding the rigid limit.

\subsubsection{The limit}

The first step to be done for taking the rigid limit of the metric \eqref{metric} is to pass to the IIB fields.
However, at the non-perturbative level this becomes problematic because the mirror map itself gets quantum corrections.
Fortunately, as we argue now, this step is not really necessary and all calculations can be done in the type IIA variables.

Indeed, the limit is defined as $t^A\to \infty$ keeping all other {\it type IIB} fields finite. In the classical mirror map
\eqref{mirmap} $t^A$ appear only in the imaginary part of $z^A$ (and the four-dimensional dilaton $r$ which we assume to be always expressed
through $\tau_2$ as in \eqref{mirmap} or \eqref{phiinstmany}).
Thus, in the classical approximation the limit can equally be defined as $\Im z^A\to \infty$
keeping all other {\it type IIA fields} finite.
At quantum level, the mirror map relations acquire additional terms which make all type IIA fields $t^A$-dependent.
Nevertheless, we can still define the limit in terms of these fields if all such $t^A$-dependent terms are exponentially suppressed as $t^A\to \infty$.
In other words, it is possible if the $t^A$-dependence of the mirror map drops out when one restricts to worldsheet instantons
with charges $k_{\hI}$ and D-instantons with charges from $\Geff$.
In fact, the quantum corrected mirror map is known only in the presence of worldsheet and D1-instantons \cite{Alexandrov:2009qq,Alexandrov:2012bu}
and D3-instantons in the large volume limit \cite{Alexandrov:2012au,Alexandrov:2017qhn} (i.e. when all K\"ahler moduli are taken to be large).
Although these cases do not cover all what we need (because of the large volume approximation used for D3-instantons), the inspection
shows that all known corrections to the mirror map respect the above property. We assume that it continues to hold beyond the large
volume approximation for D3-instantons as well,
and thus the rigid limit can be evaluated using the type IIA variables.

We do this evaluation in appendix \ref{ap-GMN}. It is very similar to the one presented in \S\ref{subsec-cmap} for the classical c-map
because the leading behavior of the most important quantities, such as the K\"ahler potential and the gauge coupling matrix,
is correctly captured by the classical contributions.
As a result, we find that:
\begin{itemize}
\item
The divergent part of the metric leads to the freezing of $\tau$, $z^{\hA}$ and $\zeta^{\hA}$.
\item
The fields $\sigma$, $\tzeta_0$ and $\tzeta_{\hA}$ appear only in the vanishing part of the metric and drop out after taking the limit.
This becomes possible because the dependence of quantum corrections on $\sigma$, $\tzeta_\Lambda$ and $\zeta^\Lambda$
arises only through the axionic couplings in the instanton contributions
\eqref{couplNS5} and \eqref{d2quali}, but due to the decoupling of NS5-instantons and the restriction to $\Geff$
the dependence on $\sigma$, $\tzeta_0$ and $\tzeta_{\hA}$ disappears.
\item
The finite part of the metric describes a space $\cM'_H$ parametrized by $z^I$, $\zeta^I$ and $\tzeta_I$ and depends
on $\tau$, $z^{X}$ and $\zeta^{X}$ as fixed parameters.
\end{itemize}
Explicitly, the limiting metric is given by
\bea
\de s^2_{\cM'_H} & =&
\frac{1}{2\sqrt{\tau_2}}\[\tau_2^2 g_{IJ} \de \uz^I\de\buz^J +g^{IJ} \cY'_I \bar\cY'_J
-4\sum_{\gamma,\gamma'\in\Geff} (v\Min^{-1})_{\gamma\gamma'} \cY'_\gamma \bar\cY'_{\gamma'}\]
\nn\\ &&
+ \sqrt{\tau_2}\sum_{\gamma,\gamma',\gamma''\in\Geff}\Min^{-1}_{\gamma\gamma'}
\[\vvpn{1}_{\gamma\gamma''} \de' Z_{\gamma''}\bar\cY'_{\gamma'} + \vvmn{1}_{\gamma\gamma''} \de' \bZ_{\gamma''} \cY'_{\gamma'}\]
\label{finLag}\\ &&
+\frac{\tau_2^{3/2}}{2} \sum_{\gamma,\gamma'\in\Geff} \vvp_{\gamma\gamma'} \de' Z_{\gamma'} \de' \bZ_{\gamma}
-\tau_2^{3/2} \sum_{\gamma,\gamma'\in\Geff}(\Min^{-1}\cQ)_{\gamma\gamma'}\sum_{\tgam\in\Geff}\vvpn{1}_{\gamma\tgam}\de' Z_{\tgam}
\sum_{\tgam'\in\Geff}\vvmn{1}_{\gamma'\tgam'}\de' \bZ_{\tgam'}.
\nn
\eea
Here $g_{IJ}=\Im F_{IJ}$, $\de'$ denotes the differential on $\cM'_H$, i.e. acting only on the dynamical fields,
and we refer to the appendix for all other notations.

\subsection{Geometric interpretation}
\label{subsec-geometry}

It is important to understand what kind of manifold is described by the metric \eqref{finLag}.
In appendix \ref{subap-HK} we prove that $\cM'_H$ is an HK manifold.
This is done by showing that the metric \eqref{finLag} comes from a holomorphic symplectic structure on the trivial $\CP$ bundle over $\cM'_H$,
which thus gets interpretation of the associated twistor space.
This symplectic structure encodes the triplet of K\"ahler structures on $\cM'_H$ and, similarly to the contact structure on $\cZ_{\cM}$,
can itself be encoded in a set of holomorphic Darboux coordinates $(\eta^I, \mu_I)$ satisfying certain integral equations.
The equations which we find (see \eqref{eqTBA-HK})
turn out to be identical to the ones describing the non-perturbative moduli space of 4d $N=2$ gauge theories
compactified on a circle \cite{Gaiotto:2008cd}, for the specific choice of the charge lattice $\Geff$ labeling 4d BPS states,
with $q_0$ and $q_X$ playing the role of flavor charges, and the holomorphic prepotential given by
\be
f(\uz^I)=-\frac{1}{6}\, \kappa_{\hI\hJ\hK} \uz^{\hI} \uz^{\hJ} \uz^{\hK}
-\frac{1}{(2\pi\I)^3}\sum_{k_{\hI}C^{\hI}\in H_2^+(\CY)} \gfinv_{k_{\hI}}\,
\Li_3\left( e^{2\pi \I  k_{\hI} \uz^{\hI}}\right).
\label{ffull}
\ee
This already establishes a connection to gauge theories with eight supercharges.
A more precise relation will be discussed in the next section.

Note that the twistor formalism provides us with an extremely simple way of taking the rigid limit.
As explained above, the QK geometry of $\cM_H$ is encoded in the Darboux coordinates $\xi^\Lambda,\txi_\Lambda,\alpha$.
Due to the decoupling of some of the instantons, the non-trivial integral equations determining these coordinates
involve only $\xi^I$ and $\txi_I$, whereas other Darboux coordinates either have a simple classical form (as e.g. \eqref{simpleDc})
or can be obtained from the solution for this pair. Then to obtain $\cM'_H$, it is enough
\begin{enumerate}
\item
to declare that the Darboux coordinates on its twistor space, $\eta^I$ and $\mu_I$,
satisfy the same equations as $\xi^I$ and $\txi_I$;
\item
to replace the prepotential entering the classical parts of Darboux coordinates by \eqref{ffull}.
\end{enumerate}
One can check that these two steps lead directly to the twistorial construction of an HK space whose metric coincides with the rigid limit \eqref{finLag}.
Essentially, this is the way which we use to prove that $\cM'_H$ carries the HK structure.

Given the twistorial description of $\cM'_H$, it is easy to see that, as it was in the case of the classical c-map,
it can be obtained by a series of \hpk quotients from a larger HK space $\Mcor$ which is also of the type described by
\cite{Gaiotto:2008cd}. This larger space has quaternionic dimension $n$ and is defined by
the original prepotential $F$.
Although the space is larger, the BPS states are restricted to belong to the same charge lattice $\Geff$ as before.
As a result, the metric on $\Mcor$ has the same form as in \eqref{finLag} (after the rescaling of $z^I$ by $\frac{\tau_2}{2}$ to absorb
this factor except the overall $\tau_2^{-1/2}$) where indices $I,J,\dots$ taking $\nI$ values
are replaced by $\Lambda,\Sigma,\dots$ running over $n$ values, but the charges run over the same lattice $\Geff$.
Due to the restriction of charges to $\Geff$, the Darboux coordinates $\eta^0$, $\eta^{\hA}$ and $\mu_A$ do not receive instanton corrections
and are given by quadratic polynomials in the coordinate $t$ parametrizing the $\CP$ fiber of the twistor space, e.g.
\be
\eta^{\hA}=u^{\hA}t^{-1}+\zeta^{\hA}-\bu^{\hA}t.
\label{undef-eta}
\ee
Besides, it leads to the existence of $n-\nI$ commuting isometries acting by shifts of $\tzeta_0$ and $\tzeta_{\hA}$ for which
the Darboux coordinates $\eta^0$ and $\eta^{\hA}$ play the role of moment maps.
Whereas on the twistor space they are the usual moment maps with respect to the holomorphic symplectic structure,
on $\Mcor$ they encode the whole triplet of moment maps: their 3 coefficients in the $t$-expansion \eqref{undef-eta}
provide the moment maps with respect to the triplet of K\"ahler forms on $\Mcor$.
Performing the \hpk quotients along these isometries, one freezes their moment maps and gets back $\cM'_H$.

On the other hand, $\Mcor$ is the HK manifold related to the non-perturbative HM moduli space $\cM_H$
(where NS5-instantons have been dropped and the charges of worldsheet and D-instantons are restricted as above) by the QK/HK correspondence
\cite{Alexandrov:2011ac}.
The easiest way to see this is to compare the two sets of Darboux coordinates, $(\xi^\Lambda,\txi_\Lambda)$ and $(\eta^\Lambda,\mu_\Lambda)$,
and to note that they are related as (c.f. step 1 above or \eqref{ident-Dc})
\be
\eta^\Lambda(t)=\xi^\Lambda(t \,e^{-\I\theta'}),
\qquad
\mu_\Lambda(t)=\txi_\Lambda(t \,e^{-\I\theta'}),
\ee
provided $u^\Lambda=\frac{\tau_2}{2}\, e^{\I\theta'} z^\Lambda$, i.e. $\theta'$ is the phase\footnote{It parametrizes
the isometry direction mentioned in footnote \ref{foot-isometry}. After the \hpk quotient along $\tzeta_0$, it can be set to zero.
This is why it does not appear in \eqref{ident-Dc} and in the relation between $u^I$ and $z^I$ on $\cM'_H$.}
of the complex coordinate $u^0$.
The isometry needed for the correspondence is ensured by the absence of NS5-instantons
and is again realized by shifts of the NS-axion $\sigma$ in \eqref{metric}.
This proves the geometric scheme presented on Fig. \ref{fig-quotient} and, in particular,
allows to obtain the rigid limit of $\cM_H$ as $n-\nI+1$ \hpk quotients of its Swann bundle.

\section{Physical interpretation: 5d gauge theory on a torus}
\label{sec-5dtorus}

\subsection{String dualities and rigid limit}
\label{subsec-duality}

In the previous section, taking the rigid limit of the HM moduli space appearing in CY compactifications of type IIB string theory,
we arrived at an HK manifold $\cM'_H$. The HK structure is an indication that this manifold should play a role in a physical theory
with rigid supersymmetry. Indeed, quantum corrected HK manifolds typically arise as moduli spaces, or more precisely target spaces
of 3d $N=4$ non-linear $\sigma$ models. But what class of $\sigma$-models are we describing?
We already saw that the twistorial description of $\cM'_H$ makes it clear that it fits into
the mathematical framework of \cite{Gaiotto:2008cd} developed for describing the class of $\sigma$-models
arising as circle compactifications of 4d $N=2$ gauge theories.
However, we can still ask how to characterize the subclass corresponding to $\cM'_H$.

In this section we propose an answer to this question.
Our reasoning mainly follows the reverse of the one presented in \cite{Haghighat:2011xx} and is based on a chain of string dualities,
which allow to establish a connection between $\cM'_H$ and 5d $N=1$ gauge theories compactified on a torus.
The appearance of the torus compactification should not come as a surprise because $\cM'_H$ is expected to carry an isometric action of
the torus modular group $SL(2,\IZ)$. We saw this explicitly in the classical approximation in \S\ref{subsec-cmap},
where the symmetry group was enhanced to $SL(2,\IR)$, but this should remain true even in the presence of quantum corrections.
The reason for this expectation is that, on one hand, the initial HM moduli space $\cM_H$ does carry such an isometry
and, on the other hand, its action on the K\"ahler parameters used to define the limit is a simple rescaling (see \eqref{SL2Z}),
which implies that the rigid limit should commute with the $SL(2,\IZ)$ action.

To begin with, let us note that under compactification on a circle the HM sector does not change and the corresponding moduli space
carries the same metric in both dimensions. In contrast, each four-dimensional $N=2$ vector multiplet gives rise to a hypermultiplet in three dimensions.
Indeed, each vector gives rise to two scalars: one is the vector component along the circle and the second appears after dualization of the three-dimensional
vector field. Combining them with the complex scalar from the 4d multiplet, one finds four real scalars representing
the bosonic content of a hypermultiplet. As a result, if we consider type IIB string theory compactified down to three dimensions on $\CY\times S^1$,
its moduli space is a direct product of two QK manifolds $\cM_H^{\rm B}\times\tilde\cM_H^{\rm B}$:
one is identical to the HM moduli space in 4d and the second comes from the vector multiplet sector of
the intermediate 4d theory.

Now let us perform T-duality along $S^1$.
Then type IIB string theory on $\CY\times S^1_R$ is mapped to type IIA string theory on $\CY\times S^1_{1/R}$.
Hence the moduli spaces of the two theories should also be identical.
Since $\cM_H^{\rm B}$ and $\tilde\cM_H^{\rm B}$ involve K\"ahler and complex structure moduli of $\CY$, respectively,
whereas $\cM_H^{\rm A}$ and $\tilde\cM_H^{\rm A}$ involve them in the opposite way, T-duality
exchanges the two factors and we have
\be
\cM_H^{\rm B}=\tilde\cM_H^{\rm A},
\qquad
\tilde\cM_H^{\rm B}=\cM_H^{\rm A}.
\label{identMM}
\ee
Note that this fact is heavily used in the physical derivation of the c-map metric \cite{Cecotti:1989qn,Ferrara:1989ik}
and is responsible for the identification of the instanton degeneracies $\Om{\gamma}$
with degeneracies of BPS black holes \cite{Alexandrov:2008gh}.

Next, one realizes that since type IIA string theory can be viewed as compactification of M-theory on a circle,
the same moduli spaces arise by considering M-theory on $\CY\times T^2$.
But let us stop in five dimensions after compactification on the CY.
The corresponding 5d $N=1$ supergravity contains the HM sector with
the moduli space $\cM_H^{\rm A}$ and the vector multiplet sector.
Taking the rigid limit of the latter, one arrives at a 5d $N=1$ gauge theory.
Finally, assuming that the rigid limit commutes with compactification on a torus,
one concludes that the rigid limit of $\cM_H^{\rm B}=\tilde\cM_H^{\rm A}$
should be the same as the torus compactification of this five-dimensional gauge theory.
All these dualities and limits are shown in detail in Fig. \ref{fig-duality} in the introduction.

Below we review some basic aspects of 5d $N=1$ gauge theories, their torus compactifications and discuss
some implications of their relation with the non-perturbative HM moduli space of string theory.

\subsection{Low energy description of 5d gauge theories}
\label{subsec-5d}

A 5d supersymmetric gauge theory with the gauge group $G$ is specified by a coupling of the vector multiplet
with a number of hypermultiplets representing the matter fields. The on-shell vector multiplet includes
a vector field $A_{\hmu}$, a real scalar $\vph$ and a Dirac spinor $\psi$, all taking values in the Lie algebra of $G$,
where $\hmu=0,\dots,4$ will denote 5-dimensional spacetime indices.
On the Coulomb branch of the moduli space the real scalar field $\vph$ takes non-vanishing vacuum expectation
values in the Cartan subalgebra, and at a generic point of this branch the gauge group $G$
is broken to its maximal torus $U(1)^{r}$ where $r = \rank(G)$.
Thus, the fields from the Cartan subalgebra, $\vph^I$ and $A^I$ with $I=1,\dots, r$, remain massless,
whereas the fields associated with other generators of the Lie algebra form massive vector multiplets with masses
determined by the expectation values of $\vph^I$.

In the low energy limit the effective Lagrangian for the massless fields takes the following general form which includes,
in particular, the Chern-Simons (CS) coupling
\be
\cL^{5d}_{bos} =
-\frac{\Fg_{IJ}(\vph)}{2\pi}\(\frac{1}{4}\, F^I_{\hmu\hnu} F^{J\hmu\hnu}
+\frac{1}{2}\,  \p_\hmu\vph^I  \p^\hmu\vph^J \)
- \frac{\Fg_{IJK}}{48\pi}\,  \epsilon^{\hmu\hnu\hlambda\hrho\hsigma} A^I_\hmu F^J_{\hnu\hlambda} F^K_{\hrho\hsigma}
\label{bosL5d}
\ee
and is completely determined by the prepotential $\Fg(\vph)$, a real function on the Coulomb branch.
The prepotential gets one-loop contributions from all dynamical fields, but is at most cubic in $\vph^I$
\cite{Seiberg:1996bd,Intriligator:1997pq}
\be
\Fg(\vph)=\frac{\pi}{g_0^2}\, h_{IJ} \vph^I\vph^J+\frac{\ccl}{12\pi} \, d_{IJK}\vph^I\vph^J\vph^K
+\frac{1}{24\pi}\(\sum_{\bfr}|\bfr\cdot\vph|^3-\sum_{i=1}^{N_f}\sum_{\bfw_i}|\bfw_i\cdot\vph+m_i|^3\)
+\frac{c_I\vph^I}{2\pi},
\label{prep5d}
\ee
where $g_0$ is the bare gauge coupling, $\bfr$ are the roots of $G$, $\bfw_i$ are the weights of $G$ in the representation $\bfR_i$,
\be
\begin{split}
h_{IJ}=&\,\tr_{\bfF}\,(T_I T_J),
\\
d_{IJK}=&\, \hf\, \tr_{\bfF}\, T_I(T_J T_K+T_K T_J),
\end{split}
\ee
and $\tr_{\bfF}$ denotes the trace in the fundamental representation.\footnote{Comparing to \cite{Intriligator:1997pq},
we accept the same normalization for the generators $\tr_{\bfF}T_I^2=2$ and take $m_0=4\pi^2 g_0^{-2}$.
Besides, we divide the whole prepotential by $2\pi$ so that our normalizations are consistent with the quantization of
the Chern-Simons coupling in \eqref{bosL5d}.}
Note that $d_{IJK}$ are non-zero only for $SU(N)$ theories with $N>2$.
In such case, $\ccl$ is the CS level in the ultraviolet Lagrangian.
We also allow for a non-vanishing linear term specified by coefficients $c_I$.
Such term is not seen in the Lagrangian \eqref{bosL5d}, but contributes to the tension of magnetic strings discussed below.
The important feature of the quantum corrected prepotential \eqref{prep5d} is that it is not smooth at loci where $\bfw_i\cdot\vph+m_i=0$,
which physically correspond to some charged matter fields becoming massless.
As a result, the Coulomb branch is divided into several chambers where the
prepotential takes different forms.

For future reference, let us specialize \eqref{prep5d} for
the $SU(2)$ gauge theory with $N_f$ hypermultiplets in the fundamental representation,
in which case one has
\be
2\pi\Fg_{SU(2)}=\frac{4\pi^2}{g_0^2}\, \vph^2 +\frac{4}{3}\,\vph^3-\frac{1}{12}\sum_{i=1}^{N_f}|\vph-m_i|^3-\frac{1}{12}\sum_{i=1}^{N_f}|\vph+m_i|^3+ c\vph,
\label{prepSU2}
\ee
and for the pure $SU(3)$ theory, which gives
\be
2\pi\Fg_{SU(3)}=\frac{4\pi^2}{g_0^2}\( \vph_1^2-\vph_1\vph_2+\vph_2^2\) +\frac{\ccl}{2}\(\vph_1^2\vph_2-\vph_1\vph_2^2\)
+\frac16\(8\vph_1^3-3\vph_1^2\vph_2-3\vph_1\vph_2^2+8\vph_2^3\)+c_I\vph^I.
\label{prepSU3}
\ee

Although five-dimensional gauge theories are non-renormalizable, they can have non-trivial fixed points at strong coupling
and thus be ultraviolet complete \cite{Seiberg:1996bd}.
Conditions on the matter content which ensure the existence of such a fixed point were studied in detail in
\cite{Intriligator:1997pq} where they have been derived by requiring that the second derivatives of the prepotential form
a positive definite matrix in all chambers of the Coulomb branch.
Recently, it has been noticed that this excludes some of the gauge theories,
including in particular quiver gauge theories, which can be obtained from string or brane constructions and therefore have
to be ultraviolet complete \cite{Katz:1997eq,Aharony:1997ju,Bergman:2015dpa,DelZotto:2017pti}.
This led to a proposal to relax the criterion of \cite{Intriligator:1997pq} and to require only that
$\cF_{IJ}$ is positive definite in the regions of the Coulomb branch where all non-perturbative degrees
of freedom remain massive \cite{Jefferson:2017ahm}.

These non-perturbative degrees of freedom are given by BPS states which, besides the usual electrically charged particles
with masses determined by the central charge
\be
Z_{\vec e}=e_I \vph^I+e_f^i m_i
\label{elecmass}
\ee
where $e_I$, $e_f^i$ are gauge and flavor charges, respectively,
include dyonic instantons \cite{Lambert:1999ua} (see also \cite{Kim:2008kn,Collie:2008vc})
and magnetic monopole strings \cite{Boyarsky:2002ck}.
The former are four-dimensional instantons lifted to solitons in $4+1$ dimensions.
They are charged under both local gauge symmetry and an additional global $U(1)_I$ symmetry.
This symmetry has the current
\be
j =\star \tr(F\wedge F)
\label{U1cur}
\ee
which is always conserved in five dimensions and the corresponding charge is equal to the instanton winding number $k$ \cite{Seiberg:1996bd}.
The central charge of dyonic instantons is given by
\be
Z_{k,\vec e}=k\(\frac{8\pi^2}{g_0^2}+\beta_I\vph^I\)+Z_{\vec e}\,,
\label{dyonmass}
\ee
where the additional term $\beta_I\phi^I$ arises at quantum level due to a mixing between the gauge symmetries and the global $U(1)_I$ symmetry
which can be traced back to the presence of the CS coupling in the bosonic Lagrangian \eqref{bosL5d}.
The monopole strings are magnetic dual to the electric particles and have tensions determined by derivatives of the prepotential
\be
Z_{\vec p}=p^I \Fg_I(\vph).
\label{Zstr}
\ee
All these central charges are real functions which must be positive in the physical region of the Coulomb branch.

\subsection{Torus compactification}
\label{subsec-torus}

Let us now compactify the 5d gauge theory considered above on a torus.
To this end, we choose spacetime to have topology $\IR^3\times T^2$ and to carry the metric
\be
g_{\hmu\hnu}=\(\begin{array}{cc}
\eta_{\mu\nu} & 0
\\
0 & \gT_{mn}
\end{array}\),
\qquad
\gT_{mn}=\frac{\cV}{\tau_2}\(\begin{array}{cc}
|\tau|^2 & \tau_1
\\
\tau_1 & 1
\end{array}\),
\label{metric5d}
\ee
where Greek indices $\mu,\nu$ label coordinates on the flat three-dimensional Minkowski spacetime,
Latin indices $m,n$ correspond to directions along the torus, $\cV$ is its volume and $\tau$ is its complex structure.

At classical level the compactified theory is given by the Kaluza-Klein reduction of the Lagrangian \eqref{bosL5d}.
This reduction is straightforward and we perform it in appendix \ref{ap-torus} generalizing (and correcting a few sign errors)
the procedure presented in \cite{Haghighat:2011xx}.
The result \eqref{S3d} represents a 3d non-linear sigma model with the target space parametrized by the 5d real scalars $\vph^I$,
the components of the 5d vector fields along the torus $\vth_1^I$ and $\vth_2^I$ \eqref{eq:holonomies-norm-tilde},
which can be combined in complex fields $\thtau^I$ \eqref{defvart},
and scalars $\lambda_I$ dual to the 3d vector fields.
The metric on this target space obtained by the Kaluza-Klein reduction has the following form
\bea
\de s^2_{3d}
&=& \Fg_{IJ}\(  \frac{ \pi}{\tau_2}\, \de \thtau^I  \de \bthtau^J
+\frac{\cV}{4\pi}\, \de\vph^I\, \de\vph^J\)
\label{met3d}\\
&&
+\frac{4\pi^3}{\cV}\, \Fg^{IJ} \(\de\lambda_I+\hf\,\Fg_{IKL} \(\vth_2^K\de\vth_1^L - \vth_1^K \de\vth_2^L\)\)
\(\de\lambda_J+\hf\,\Fg_{JMN} \(\vth_2^M\de \vth_1^N - \vth_1^M \de\vth_2^N\)\).
\nn
\eea
It is immediate to see that the metric is invariant under the action of $SL(2,\IZ)$ group which simultaneously transforms
the torus modular parameter $\tau$ by the usual fractional transformation and the three-dimensional fields as
\be
\vph^I\mapsto \vph^I,
\qquad
\lambda_I\mapsto\lambda_I,
\qquad
\begin{pmatrix} \vth_1^I \\ \vth_2^I\end{pmatrix} \mapsto
\begin{pmatrix} a & b \\ c & d  \end{pmatrix}
\begin{pmatrix} \vth_1^I \\ \vth_2^I \end{pmatrix} .
\label{SL2Z5d}
\ee
Since any theory on a torus must possess such invariance,
it can be seen as a consistency check of the derived metric.

Furthermore, comparing this metric with the rigid c-map \eqref{metricIIB-rigid},
which we obtained as the rigid limit of the classical HM moduli space,
one finds that the two metrics coincide up to the multiplicative factor $2\pi\cV^{-1/2}$ provided
\be
\Fg_{IJ}=\sqrt{\frac{\tau_2}{\cV}}\,g_{IJ}
\label{indetFF}
\ee
and the two sets of coordinates are identified as follows
\be
\begin{array}{rclcrcl}
\vph^I&=&\displaystyle{2\pi\sqrt{\frac{\tau_2}{\cV}}\(t^I+\beta_X^I t^X\),}
&\qquad&
\vth_1^I&=&c^I,
\\
\lambda_I&=&\displaystyle{\tc_I+\hf\,\kappa_{IJX}\(b^J c^X-c^J b^X\),}
&\qquad&
\vth_2^I&=&b^I,
\end{array}
\label{identvar}
\ee
where $\beta^I_X$ are some constant coefficients. Note that these identifications are perfectly consistent
with the $SL(2,\IZ)$ transformations \eqref{SL2Z} and \eqref{SL2Z5d}. They imply $\Fg_{IJK}=-\frac{1}{2\pi}\,\kappa_{IJK}$
and that the gauge theory parameters $1/g_0^2$ and $m_i$ are given by linear combinations of the frozen K\"ahler parameters $t^X$.
The concrete form of these relations depends, on one hand, on the intersection numbers of the CY and, on the other hand,
on the gauge group and matter content of 5d theory. Matching these data allows to determine which particular 5d theory
is captured by the rigid limit of a given Calabi-Yau manifold. We consider several examples of this in section \ref{sec-examples}.

However, the metric \eqref{met3d} is only the classical approximation to an exact result which includes contributions
from instantons originating from BPS states wrapping the torus. The simplest type of BPS states are electrically charged particles.
In particular, in the case of pure $SU(2)$ theory, the contribution from the W-bosons to the quantum corrected metric of the 3d $\sigma$-model
was computed in \cite{Haghighat:2011xx}
by integrating out the tower of massive Kaluza-Klein states in the one-loop approximation.
But as we saw in the previous subsection, there are two other types of BPS states which can generate instantons: dyonic instantons and magnetic strings.
Their contributions are much more difficult to calculate, and only a few partial results on stringy instantons are available
at the moment \cite{Haghighat:2011xx,Haghighat:2012bm}.

On the other hand, the argument presented in \S\ref{subsec-duality} implies that
the full non-perturbative metric including contributions from all instantons should coincide with the metric \eqref{finLag}
describing the rigid limit of the non-perturbative HM moduli space.
In particular, the instantons on the string theory side should match those on the gauge theory side.
Let us now show that this is indeed the case.

First, we claim that the contributions from perturbative $\alpha'$ and $g_s$-corrections as well as from D(-1)-instantons,
which are known to correct the metric on $\cM_H$, do not appear on $\cM'_H$. The easiest way to see this is to
look at the twistorial formulation of the rigid limit. On $\cM_H$ these corrections are encoded by
the second term in the prepotential \eqref{lve}, the logarithmic term parametrized by coefficient $c$ in the Darboux coordinate $\alpha$ \eqref{alp},
and D-instantons with charges $\gamma=(q_0,0,0,0)$, respectively. In particular, the latter affect only the Darboux coordinates $\txi_0$ and $\alpha$,
as can be seen from the integral equations \eqref{eqTBA}. But going to $\cM'_H$, these Darboux coordinates drop out from the twistorial formulation
and the prepotential \eqref{ffull} does not contain the perturbative correction term anymore. Thus, the twistorial formulation of $\cM'_H$
does not contain all these contributions.\footnote{Heuristically, this can be understood as follows.
In the type IIB formulation, these quantum corrections can be resumed into modular functions represented typically by
$\tau$-dependent non-holomorphic Eisenstein series \cite{RoblesLlana:2006is}.
Since in our case $\tau$ is a fixed parameter, all such contributions are constant and can be absorbed into a redefinition of variables.
A similar phenomenon happens when one applies the QK/HK correspondence to the one-loop corrected local c-map: the resulting HK space
coincides with the standard rigid c-map and is independent of the parameter controlling the one-loop correction \cite{Alexandrov:2011ac}.}

Next, let us consider the contributions of worldsheet and D1-instantons.
Combining them together, one can perform a resummation which turns them into $(p,q)$-instantons
with the instanton action of the following form \cite{RoblesLlana:2006is,Alexandrov:2009qq}
\be
S_{\vec q,\,m,n}=2\pi|m\tau+n|\,|q_{\hI}t^{\hI}|-2\pi \I q_{\hI}(m c^{\hI}+n b^{\hI}),
\label{Sinst-pq}
\ee
where we took into account that due to the restriction to $\Geff$ the only non-vanishing components, which D1-instanton charge can have, are $q_{\hI}$.
We would like to identify these $(p,q)$-instantons with dyonic instantons wrapping one-dimensional cycles of the torus.
Expressing the real part of the instanton action in terms of the gauge theory variables, one finds
\be
\Re S_{\vec q,\,m,n}=\sqrt{\frac{\cV}{\tau_2}}\,|m\tau+n| \, |Z_{\vec q}|,
\qquad
Z_{\vec q}= q_I\(\vph^I+b^I_X m^X\)+q_X\( b^X_I\vph^I+b^X_Y m^Y\),
\label{instactdyon}
\ee
where we denoted $m^X=(8\pi^2 g_0^{-2}, m_i)$ and encoded the identification between the K\"ahler moduli $t^{\hI}$
and the gauge theory variables $\vph^I$ and $m^X$ in a matrix $b^{\hI}_{\hJ}$ with $b^{I}_{J}=\delta_I^J$.
The factor in front of $Z_{\vec q}$ has a clear interpretation: this is the volume of the one-dimensional closed
cycle on the torus, labeled by two integers $(m,n)$, which is wrapped by the instanton.
Then the second factor should be identical to the dyonic central charge \eqref{dyonmass}.
Setting $e_{\hI}=b^{\hJ}_{\hI}q_{\hJ}$, one obtains that
\be
Z_{\vec q}=e_{I}\vph^I+e_X m^X=\frac{8\pi^2 e_0}{g_0^{2}} +Z_{\vec e}.
\label{ZZe}
\ee
This coincides with \eqref{dyonmass} upon identifying $e_0$ with the instanton charge $k$,
up to the shift of the bare gauge coupling $g_0^{-2}$. Of course, for vanishing $e_0$ one reproduces the central charge
of the usual electrically charged BPS particles.

To reproduce the shift of the gauge coupling in \eqref{dyonmass}, one should note two facts.
First, only the rational part of the coefficients $\beta_I$ is
unambiguously defined since their integer part can be absorbed into a redefinition of the charge lattice
which can be done, for instance, by $e_I\mapsto e_I-k[\beta_I]$. Second, the rotation of the charge lattice induced by $b^{\hJ}_{\hI}$
generically does not preserve its integrality. Furthermore, the lattice of charges $q_{\hI}$ was already a result of the rotation
to the basis adapted for taking the rigid limit (see \S\ref{subsec-def}), which also can spoil the integrality.
Taking this into account, the naive identification of $e_{\hI}$ with the set of electric, flavor and instanton charges of gauge theory
suggested by \eqref{ZZe} may not be correct, and a more careful analysis is required.
We will see in section \ref{sec-examples} on a concrete example how such analysis allows to get a non-trivial shift of the gauge coupling
in the dyonic central charge.

It is worth also to note that the identification of $(p,q)$ and dyonic instantons implies that
the definition of a 5d gauge theory at the non-perturbative level involves new parameters in addition to masses and the gauge coupling.
These are $c^X$ and $b^X$ appearing as $\theta$-angle terms in \eqref{Sinst-pq}.
We obtained them as frozen periods of the RR 2-form and the $B$-field along curves $C^X$ on the CY.
What is their origin in gauge theory? To answer this question, let us recall that the gauge theory parameters
can be thought as background gauge superfields related to gauging global symmetries associated with these parameters \cite{Seiberg:1993vc}.
In particular, the flavor masses can be identified with the scalar components of the vector superfields gauging the flavor symmetry,
whereas the gauge coupling appears as the scalar component of the superfield for the $U(1)_I$ symmetry discussed around \eqref{U1cur}.
Once the theory is put on a torus, each background vector field gives rise to two new parameters given by holonomies around the basis of
one-dimensional cycles on the torus, which are precisely $c^X$ and $b^X$.\footnote{It is amusing to note that background fields
can be thought of as dynamical fields whose kinetic terms have infinite coefficients \cite{Seiberg:1993vc}.
This remark closes the circle of ideas since it returns us back to the origin of the additional parameters in the rigid limit.}

The last type of the instanton effects contributing to the metric on $\cM'_H$ comes from D3-branes wrapping divisors $D_I$.
Their instanton action is given by
\be
S_{\vec p}=2\pi \tau_2|p^I \fcl_I|-2\pi\I p^I \(\tc_I+ \frac12\, \kappa_{I\hJ\hK} b^{\hJ} (c^{\hK} -\tau_1 b^{\hK})\).
\ee
Let us set for simplicity $b^{\hI}=0$.
Then if the relation \eqref{indetFF} can be integrated to
\be
\fcl_I|_{b^{\hI}=0}=-\frac{\cV}{2\pi\tau_2}\,\Fg_{I},
\label{indetF}
\ee
which can always be achieved by tuning the coefficients $c_I$ of the linear term in \eqref{prep5d},
then the instanton action takes the simple form
\be
S_{\vec p}|_{b^{\hI}=0}=\cV \, Z_{\vec p}-2\pi\I p^I\lambda_I.
\ee
Its real part coincides with the instanton action of a stringy instanton given by the volume of the torus wrapped by a magnetic string
multiplied by its tension. The imaginary part is also natural since $\lambda_I$ are the scalars dual to the vector fields of the gauge theory
and therefore should be sourced by magnetic objects.

Thus, all non-perturbative effects surviving in the rigid limit find their interpretation in the supersymmetric gauge theory
compactified on a torus.

\subsection{BPS spectrum and modular invariance}
\label{subsec-phys}

In the previous subsections we argued that the non-perturbative moduli space of a 5d gauge theory compactified on a torus
is captured by the metric on $\cM'_H$, the rigid limit of the HM moduli space of type IIB string theory on the appropriate CY threefold.
Expanding this metric around the classical rigid c-map allows to read off various instanton corrections which, as we saw, can all be identified
either with instantons from electrically charged BPS particles, or dyonic, or stringy instantons.
Thus, string theory provides us with concrete predictions for the instanton contributions in compactified 5d gauge theory.

In practice, all that we need in order to get these contributions is to know the BPS spectrum and
a relation between the frozen moduli and the gauge theory parameters.
The latter can be found by matching the classical prepotentials.
The BPS spectrum, however, represents a much more serious problem.
On the gauge theory side, only some partial results about the spectrum of dyonic instantons are available \cite{Kol:1998cf,Bergman:1998gs},
which have been obtained using the technique of string web diagrams \cite{Aharony:1997bh},
and even less is known about the spectrum of magnetic strings.
On the string theory side, the former spectrum is captured by Gopakumar-Vafa invariants of the CY, whereas the latter spectrum
is encoded in more complicated generalized DT invariants.

Here we would like to bring attention to unexpected constraints on the spectrum of bound states of magnetic strings and dyonic instantons
arising as a consequence of the $SL(2,\IZ)$ modular invariance. On the gauge theory side, this symmetry appears to be just an artefact
of compactification on a torus, and it is not clear how it can constrain the spectrum in five-dimensions.
But in string theory, it is a duality playing a fundamental role.
In particular, imposing it as an isometry of the HM moduli space, one arrives at the condition that the D3-D1-D(-1) bound states
form an $SL(2,\IZ)$ invariant subsector. To put this condition in a clear mathematical form, let us consider the four-dimensional dilaton $r=e^\phi$
which is known to transform under $SL(2,\IZ)$ as
\be
r\mapsto \frac{r}{|c\tau+d|}.
\label{trans-r}
\ee
Classically, $r$ has a simple expression thorough the volume of the CY given in \eqref{mirmap}, but at quantum level it gets
various corrections and can be expressed as in \eqref{phiinstmany}. Then the consistency with \eqref{trans-r} requires that
the D3-instanton contribution to $r$ transforms as a non-holomorphic modular form of weight $(-\tfrac12,-\tfrac12)$.
This turns out be a non-trivial requirement which leads to certain constraints on the spectrum of these instantons,
i.e. DT invariants $\Om{\gamma}$ with charges $\gamma=(0,p^a,q_a,q_0)$, some of which descend in the rigid limit
to the BPS degeneracies of bound states of magnetic strings and dyonic instantons.

Such constraints are typically formulated in terms of modular properties of a generating function of these invariants.
More precisely, let us introduce the so called MSW invariants \cite{Maldacena:1997de}, $\bOmMSW_\gamma=\bOm{\gamma}(z^a_\infty(\gamma))$,
given by the rational DT invariants
\be
\label{defntilde}
\bOm{\gamma}= \sum_{d|\gamma}  \frac{1}{d^2}\,
\Om{\gamma/d}
\ee
evaluated at the `large volume attractor point',
$z^a_\infty(\gamma)= \mathop{\lim}\limits_{\lambda\to +\infty}\(-q^a+\I\lambda  p^a\)$.
We recall that the specification of the complex moduli $z^a$ is important because the DT invariants are only piecewise constant
on the moduli space due to the wall-crossing phenomenon \cite{Denef:2000nb}.
An important property of the MSW invariants is that they do not change under the spectral flow symmetry
acting on charges as \cite{Manschot:2009ia,Manschot:2010xp}
\be
\label{flow}
q_a \mapsto q_a - \kappa_{abc}p^b \epsilon^c,
\qquad
q_0 \mapsto q_0 - \epsilon^a q_a + \frac12\, \kappa_{abc}p^a \epsilon^b \epsilon^c,
\qquad
\epsilon^a\in \IZ.
\ee
As a result, they only depend on $p^a$, the charge $\hat q_{0}=q_0 -\frac12\, \kappa^{ab} q_a q_b$ invariant under the spectral flow,
and a residue class $\mu_a$ which takes into account that not all integer charges $q_a$ can be obtained by varying $\epsilon^a$ in \eqref{flow}.
This allows to write $\bOmMSW_\gamma=\bOm{\vp,\vmu}( \hat q_0)$.
Furthermore, since the invariant charge $\hat q_0$ is bounded from above by $\hat q_0^{\rm max}=\tfrac{1}{24}(p^3+c_{2,a}p^a)$,
it is possible to define the generating function of the MSW invariants
\be
h_{\vp,\vmu}(\tau) = \sum_{\hat q_0 \leq \hat q_0^{\rm max}}
\bOm{\vp,\vmu}(\hat q_0)\,e^{-2\pi\I \hat q_0 \tau }
\label{defhDT}
\ee
with fixed magnetic charge and residue class.
It is this function that must have an appropriate  modular behavior. In particular, in the one-instanton approximation, i.e.
when a D3-brane wraps an {\it irreducible} divisor of $\CY$, it must be a vector-valued holomorphic modular form
of negative weight $-\(\tfrac{1}{2}\,h^{1,1}(\CY)+1\)$ \cite{Gaiotto:2006wm,deBoer:2006vg,Denef:2007vg}.
Even more interesting behavior appears if one goes beyond the one-instanton approximation, i.e. considers
branes on {\it reducible} divisors, in which case $h_{\vp,\vmu}$ was shown to be a vector-valued holomorphic {\it mock} modular form \cite{Alexandrov:2016tnf}.

Very similar constraints should arise in the gauge theory setup. These constraints can be derived either by taking the rigid limit of
the above construction or by studying the constraints of modular invariance directly for the metric \eqref{finLag}.
In either case, one expects to find non-trivial restrictions on the modular behavior of a generating function of BPS degeneracies
of magnetic strings so that their spectrum will be severely constrained.

Note that the mock modularity of the generating function \eqref{defhDT} evaluated for reducible divisors takes its roots
in the wall crossing of the DT invariants.
This raises the question about the wall crossing in 5d gauge theories.
The reality of the moduli and the central charges, \eqref{dyonmass} and \eqref{Zstr}, represents an essential difference
from the more familiar four-dimensional case.
Nevertheless, the relation to string theory indicates that the central charge of a bound state is complex and given by
\be
Z_{\vec p,k,\vec e}=Z_{\vec p}+\I Z_{k,\vec e}.
\ee
Thus, one might have a non-trivial wall crossing even if the moduli space is real.
Here we restrict ourselves just to these comments and postpone
the study of five-dimensional wall crossing and modularity constraints to future research.

\section{Examples}
\label{sec-examples}

In this section we present several examples illustrating our rigid limit
for different types of Calabi-Yau manifolds and its relation to five-dimensional $N=1$ gauge theories.

\subsection{Elliptic fibrations and $SU(2)$ gauge theory}
\label{subsec-elliptic}

We start with the most studied example of a family of elliptically fibered CYs which are well known to be related to $SU(2)$ gauge theories
with $N_f<8$ flavors \cite{Morrison:1996xf}.
The elliptic fibrations $\pi\ :\ \CY\to\base$, where $\base$ is a complex two-dimensional base, can locally be described by a Weierstrass form
\be
y^2=4x^3- g_2(u_1,u_2)x w^4-g_3(u_1,u_2) w^6,
\label{Weier}
\ee
where $u_1,u_2$ are coordinates on the base.
We assume that the fibration is smooth with a single section $\sigma$ represented by the base $\base$.
This implies that singularities on the fiber can only be of Kodaira type $I_1$, which means that
the discriminant $\Delta=g_2^3-27 g_3^2$ of \eqref{Weier} has only simple zeros on $\base$.
This restricts the possible choice of $\base$ to smooth almost Fano twofolds which include
the Hirzebruch surfaces $\IF_m$, $m=0,1,2$, del Pezzo surfaces $\dP_m$, $m=0,\dots,8$, and the toric surfaces described
by the 16 reflexive two-dimensional polytopes. Here we consider only the first two possibilities, $\IF_m$ and $\dP_m$.
Their geometric description can be found, for instance, in \cite{Klemm:2012sx,Haghighat:2012bm,Cvetic:2014gia}.

For all smooth elliptic fibrations a basis of $H^{1,1}(\CY)$ generating the K\"ahler cone is given by
$\{\omega_e,\pi^\star \omega_{\alpha}\}$, $\alpha=1,\dots,h^{1,1}(\base)$,
where
\be
\omega_e=\sigma+\pi^\star c_1(\base)
\label{section}
\ee
and $\omega_\alpha$ are the generators of the K\"ahler cone on the base.
We denote the corresponding basis of dual divisors by $\{D_e,D_\alpha\}$. The divisor $D_e$ is dual to the elliptic fiber curve $\cE$
in the sense that it does not intersect any curve in $\base$ and obeys $D_e\cap \cE=1$.

Let us expand the first Chern class of the base in the basis of $\omega_\alpha$
\be
c_1(\base)=c_1^\alpha\, \omega_\alpha,
\ee
and denote by $C_{\alpha\beta}$ the intersection matrix on $\base$
\be
\int_\base \omega_\alpha\wedge \omega_\beta=C_{\alpha\beta},
\ee
which has signature $(1, h^{1,1}(\base)-1)$.
Then, using the adjunction formula which leads to the relation $\sigma^2=-c_1\sigma$,
the triple intersection numbers of $\CY$ can be shown to be
\be
\kappa_{\alpha\beta\gamma}=0,
\qquad
\kappa_{e\,\alpha\beta}=C_{\alpha\beta},
\qquad
\kappa_{ee\,\alpha}=C_{\alpha\beta}c_1^\beta,
\qquad
\kappa_{eee}=C_{\alpha\beta}c_1^\alpha c_1^\beta.
\label{kappa-ell}
\ee
Thus, all intersection numbers are determined by the intersection numbers of the base and its first Chern class.

A crucial property of the intersection numbers \eqref{kappa-ell} is that the matrix $M_{ab}=\kappa_{eab}$ is degenerate, i.e. $\det M=0$.
This suggests that the vector $\vt_1^a=\delta^a_e$, playing the role of $\vvtA$ of \S\ref{subsec-def}, defines a non-trivial local limit.
Indeed, it belongs to the K\"ahler cone, the kernel of the matrix \eqref{matA}, coinciding with $M_{ab}$ defined above, is non-empty,
and its self-intersection number $\kappa_{eee}$ given in \eqref{kappa-ell} is non-vanishing for all the bases under consideration.
The kernel of $M_{ab}$ is one-dimensional and described by the vector\footnote{In this section we accept the convention that
the indices $A$ corresponding to the large moduli run over $1,\dots, \nA$, the indices $I$ corresponding to the moduli remaining dynamical
run over $n-\nI,\dots,n-1$, and the indices $X$ labeling the frozen moduli run in-between. We recall that $n=h^{1,1}(\CY)+1$.}
\be
\vvt_{n-1}=(-1,c_1^\alpha),
\label{vecvI}
\ee
playing the role of $\vvtI$ of \S\ref{subsec-def}.
Remarkably, the corresponding shrinking divisor $\hD_{n-1}=\vvt_{n-1}^a D_a$ can be expressed using \eqref{section} as
\be
\hD_{n-1}=-D_e+c_1^a D_a=-\base
\ee
and thus it coincides with the base of the elliptic fibration.

Finally, one can complete the two vectors $\vvt_1$ and $\vvt_{n-1}$ to a basis in $H_2(\CY,\IR)$ by choosing $\vvtX$ with $X=2,\dots,n-2$.
This structure indicates that in the local limit one K\"ahler modulus grows, one remains dynamical and $\nX=n-3=h^{1,1}(\base)-1$ moduli become frozen.
This is consistent with the expectation that such limit produces an $SU(2)$ 5d gauge theory with $\nX-1$ flavors since one of the frozen moduli should
play the role of the gauge coupling, whereas others can be associated with flavor masses.
To verify this claim and establish a precise relation between the moduli and the gauge theory variables,
we need to specify the choice of the base $\base$ and to analyze its homology lattice.

\subsubsection{Hirzebruch surfaces}

First, we choose $\base=\IF_m$.
The Hirzebruch surface $\IF_m$ is a $\IP^1$ bundle over $\IP^1$ of the form $\IP(\cO\oplus\cO(m))$ for $m\geq 0$.
The Mori cone, dual to the K\"ahler cone, is generated by two effective curves, the isolated section $S$ of the bundle and the fiber $F$.
These curves have the following intersections
\be
S\cap S= -m,
\qquad
S\cap F = 1 ,
\qquad
F\cap F = 0.
\ee
The dual generators of the K\"ahler cone, $D_\alpha$, are given by
\be
D_1 = F ,
\qquad
D_2 = S+m F
\label{basisHirz}
\ee
and have the following intersection matrix
\be
C_{\alpha\beta}=\( \begin{array}{cc} 0 & 1 \\ 1 & m \end{array}  \).
\ee
Finally, the first Chern class is known to be
\be
c_1(\IF_m) = 2S +(2+m)F = (2-m)D_1 + 2 D_2,
\label{c1F}
\ee
where we used Poincar\'e duality to write it in terms of divisors.

Computing the intersection numbers of a smooth elliptic fibration over $\IF_m$ using \eqref{kappa-ell}, one obtains
that the classical cubic prepotential \eqref{Fcl} is given by
\be
\Fcl=-\[\frac43\, (z^e)^3+(z^e)^2 z^1+\(1+\frac{m}{2}\)(z^e)^2 z^2+z^e z^1 z^2+\frac{m}{2}\,z^e (z^2)^2\].
\ee
Next, we perform the rotation \eqref{changet} to the basis adapted for taking the local limit.
As explained above, the rotation is generated by the vectors $\vvt_A,\vvtX,\vvtI$ which in this case are taken as
\be
\begin{split}
\vvt_1=&\,(1,0,0),
\\
\vvt_2=&\,(0,1,0),
\\
\vvt_3=&\,(-1,2-m,2).
\label{vectorF}
\end{split}
\ee
It brings the prepotential to the form
\be
\Fcl=-\frac43\, (\hz^1)^3-(\hz^1)^2 \hz^2+\hz^2(\hz^3)^2 +\frac{4}{3}\,(\hz^3)^3.
\ee
Note that the prepotential in the new basis does not depend on $m$ and
that the moduli $\hz^1$ and $\hz^3$ are decoupled, which makes possible to define the local limit as $\hatt^1\to \infty$.
Then from our general discussion it follows that the limit is described by the prepotential
\be
\fcl=\hz^2(\hz^3)^2 +\frac{4}{3}\,(\hz^3)^3,
\label{prep-limitFk}
\ee
where $\hz^3$ is dynamical, whereas $\hz^2$ is fixed to be constant.

Since we expect that the local limit of the elliptic fibration over $\IF_m$ corresponds to the pure $SU(2)$ gauge theory,
the prepotential \eqref{prep-limitFk} is to be compared to \eqref{prepSU2} with $N_f=0$, i.e.
\be
2\pi \Fg_{SU(2)}^{N_f=0}=\frac{4\pi^2}{g_0^2}\, \vph^2 +\frac{4}{3}\,\vph^3+ c\vph.
\label{prepSU20}
\ee
It is immediate to see that this implies $c=0$ and leads to the following identification of the moduli and the gauge theory variables
\be
\hatt^2=\frac{4\pi^2 a}{g_0^2}\,,
\qquad
\hatt^3=a\vph,
\ee
where $a$ is a proportionality coefficient which, according to \eqref{identvar}, is given by $a=\frac{1}{2\pi}\,\sqrt{\frac{\cV}{\tau_2}}$.

Furthermore, computing the central charge introduced in \eqref{instactdyon}, which is supposed to encode the mass of dyonic instantons,
in terms of the charges defined with respect to the original basis \eqref{basisHirz},
one finds
\be
Z_{\vec q}=q_e(a^{-1}\hatt^1-\vph) + q_1\( \frac{4\pi^2}{g_0^2}+(2-m)\vph\)+2 q_2 \vph.
\label{Z-Hirz}
\ee
Here for completeness we included also the charge $q_e$ associated with the elliptic fiber, which is set to zero in the rigid limit.
Comparing with the dyonic central charge \eqref{dyonmass}, we see that the instanton winding number $k$ can be identified with charge $q_1$,
whereas the function multiplying it receives a $\vph$-dependent contribution, which can be traced back to
the rotation of the basis induced by \eqref{vectorF}. In particular, we deduce that $\beta=2(2-m)$.
However, this coefficient is not uniquely defined as it can be changed by shifting the electric charge $e$ by a multiple of $k$,
which shows that the identification of $e$ with $q_2$ suggested by \eqref{Z-Hirz} is also ambiguous.

\subsubsection{Del Pezzo surfaces}

Our second choice of the base is $\base=\dP_m$.
The del Pezzo surface $\dP_m$ is a blowup of $\IP^2$ at $m$ points. It can be viewed as a fibration over $\IP^1$
where the generic fiber is also $\IP^1$, but degenerates over $m-1$ points into two $\IP^1$'s intersecting at a point.
The number of blow-up points can vary from 0 to 9, but we restrict to $1\le m\le 8$ to present a uniform
description.\footnote{For instance, $\dP_9$ is qualitatively different since it is
a rational elliptic surface with infinite dimensional Mori and K\"ahler cones.}
A standard choice of basis for $H_{2}(\dP_m,\IZ)$ is given by the hyperplane class $H$ of $\IP^2$
and by the exceptional divisors $E_i$, $i=1,\dots, m$, of the blow-ups.
Their intersections are
\be
H\cap H = 1,
\qquad
E_i\cap E_j = -\delta_{i,j},
\qquad
H\cap E_i = 0.
\ee
However, this basis is not a basis of the K\"ahler cone. The latter can be obtained by choosing
\be
\label{eq:del-pezzo-kahler-gen}
D_i=H-E_i,
\qquad
D_{m+1}=H.
\ee
In fact, for $m>2$ this choice is not unique because the K\"ahler cone is non-simplicial and the number of its generators exceeds the dimension
of $H_2(\dP_m,\IZ)$. All K\"ahler generators can be found in \cite{Cvetic:2014gia} and different choices of the basis
correspond to different sub-cones.
In the basis \eqref{eq:del-pezzo-kahler-gen}, the first Chern class is given by
\be
c_1(\dP_m) = 3 H - \sum_{i=1}^{m} E_i =\sum_{i=1}^{m} D_i +(3-m)D_{m+1} .
\ee

Substituting these data into equations \eqref{kappa-ell} for the intersection numbers of the CY constructed over $\dP_m$, one obtains
the following prepotential
\be
\Fcl=-\[\frac{9-m}{6}\, (z^e)^3+(z^e)^2 \sum_{i=1}^m z^i+\frac32(z^e)^2 z^{m+1}+z^e \sum_{i,j=1\atop i<j}^{m+1}z^i z^j+\hf\,z^e (z^{m+1})^2\].
\ee
The vectors performing the rotation \eqref{changet} to the basis adapted to the local limit can be chosen as
\be
\begin{split}
\vvt_1=&\,(1,0,\cdots,0),
\\
\vvt_{i}=&\,(0,\under{\underbrace{0,\cdots,0}}{i-1},-1,\under{\underbrace{0,\cdots,0}}{m-i},1),
\qquad i=2,\dots, m,
\\
\vvt_{m+1}=&\,(0,-1,0,\cdots,0,2),
\\
\vvt_{m+2}=&\,(-1,1,\cdots,1,3-m).
\end{split}
\label{vectorsdP}
\ee
After the rotation, the prepotential becomes
\be
\begin{split}
\Fcl =&\, \frac{9-m}{6}\, \((\hz^{m+2})^3 - (\hz^1)^3\)   +\hf \((\hz^{m+2})^2 -(\hz^1)^2\) \(\sum_{i=2}^{m} \hz^i +4 \hz^{m+1}\)
\\
&\,
- \hf\, (\hz^{m+2} - \hz^1) \, \sum_{i=2}^{m} (\hz^i)^2 .
\end{split}
\ee
Again the moduli $\hz^1$ and $\hz^{m+2}$ are decoupled and in the limit $\hatt^1\to \infty$
the relevant part of the prepotential is given by
\be
\fcl = \frac{9-m}{6}\, (\hz^{m+2})^3  +\hf (\hz^{m+2})^2 \(\sum_{i=2}^{m} \hz^i +4 \hz^{m+1}\)
- \hf\,\hz^{m+2} \sum_{i=2}^{m} (\hz^i)^2 ,
\label{prep-limitdP}
\ee
where $\hz^{m+2}$ is dynamical and all other moduli are frozen.

We compare the prepotential \eqref{prep-limitdP} with the one of the $SU(2)$ gauge theory with $N_f=m-1$ flavors \eqref{prepSU2}
where we consider the chamber of the moduli space with $\vph\pm m_i>0$. In this chamber the gauge theory prepotential takes the form
\be
2\pi\Fg_{SU(2)}^{N_f=m-1}=\frac{9-m}{6}\,\vph^3+\frac{4\pi^2}{g_0^2}\, \vph^2 +\[ c-\frac{1}{2} \( \sum_{i=1}^{m-1} {m_i^2}\)\]\vph.
\label{prepSU2m}
\ee
Then again we should set $c=0$, whereas the other variables are identified as follows
\be
\hatt^i=a m_{i-1},
\qquad
\hatt^{m+1}=a\(\frac{2\pi^2}{g_0^2}-\frac14\sum_{i=1}^{m-1}m_i\) ,
\qquad
\hatt^{m+2}=a\vph.
\ee

\subsection{Two large moduli}
\label{subsec-two-moduli}

In the elliptic fibrations considered so far, the local limit was obtained by sending only one K\"ahler modulus to infinity.
However, one may expect that this is a very restricted set of examples because shrinking some of the divisors
is a local procedure, which should not affect the cycles which are ``far away" from them.
Thus, in general, in the local limit several 2-cycles can stay finite and therefore several moduli are taken to infinity.

In our language, this will happen whenever there exist two or more linearly independent vectors $\vvtA$ such
that the intersection of the respective kernels (\ref{deftI}) is non-empty.
To give a concrete example of such situation, we consider one of the toric hypersurfaces constructed
from the Kreuzer-Skarke list of reflexive polytopes \cite{Kreuzer:2000xy, Kreuzer:2000qv}.
A useful database containing information about these CY varieties can be found
in \cite{Altman:2014bfa}, and we will make extensive use of the data analyzed by these authors.
We provide some details about these data and toric geometry in appendix \ref{ap-toricdata}, whereas
some relevant background can also be found in  \cite{Aspinwall:1993nu, Altman:2014bfa}.

Let us consider CY which corresponds to geometry 1 of polytope 337 in the database \cite{Altman:2014bfa}.
It is defined by the data given in \eqref{polytop1} and \eqref{triang1} and has $h^{1,1}=4$.
We choose the following four generators as a basis of $H_4(\CY,\IZ)$
\be
D_1 = \tD_5,
\qquad
D_2 = \tD_6,
\qquad
D_3 = \tD_7,
\qquad
D_4 = \tD_8,
\label{basis2large}
\ee
where we expressed them in terms of divisors $\tD_i$ of the ambient toric space.
Extracting the triple intersection numbers in the basis from the database, the classical cubic prepotential can be written as
\be
\begin{split}
\Fcl= &\, -\frac16\,\Bigl[-(z^1)^3+3 (z^1)^2 z^2 +3(z^1)^2 z^4-3 z^1(z^2)^2+12 z^1 z^2 z^4-6 z^1(z^3)^2
\Bigr.
\\
& +18 z^1 z^3 z^4-3 z^1(z^4)^2-5 (z^2)^3 -18 (z^2)^2 z^3+ 12 (z^2)^2 z^4-18 z^2 (z^3)^2
\\
&\, \Bigl.
+36 z^2 z^3 z^4-6 z^2 (z^4)^2-13 (z^3)^3+27 (z^3)^2 z^4-9 z^3 (z^4)^2+(z^4)^3\Bigr].
\end{split}
\label{Fcl-2large}
\ee
The K\"ahler moduli here
are constrained by the requirement that the volumes of all Mori generators $C^i$ must be positive
\be\label{eq:KC-volume-condition}
	\int_{C^i} J =t^a D_a\cap C^i  \geq  0\,.
\ee
The matrix of intersections of Mori generators with the basis divisors is as follows
\be
	C^i \cap D_a =
\left(
\begin{array}{cccc}
 0 & -1 & -1 & 1 \\
 0 & 0 & 1 & 0 \\
 0 & 1 & 0 & 0 \\
1 & 2 & 3 & -1 \\
1 & 0 & 0 & 0 \\
-1 & 1 & 0 & 1
\end{array}
\right)
\ee
and leads to the following inequalities
\be
\begin{split}
-t^2-t^3+t^4\geq 0\,, &
\qquad
t^3\geq 0 \,,
\qquad
t^2\geq 0 \,,
\\
t^1+2 t^2 + 3t^3 -t^4 \geq 0 \,,  &
\qquad
t^1\geq0\,,
\qquad
-t^1+t^2+t^4\geq0 .
\end{split}
\label{condKc}
\ee

The intersection numbers encoded by the prepotential \eqref{Fcl-2large} and the K\"ahler cone conditions
provide the starting point for defining the local limit. Let us choose the following two vectors
\be
\vvt_1 = (1,0,0,1),
\qquad
\vvt_2 = (0,1,0,2).
\label{vA-large2}
\ee
It is easy to check that their components satisfy the inequalities \eqref{condKc}, saturating some of them, so that
both vectors belong to the boundary of the K\"ahler cone.
They give rise to the following intersection matrices \eqref{matA}
\be
M_{1,ab} =
\left(
\begin{array}{cccc}
 0 & 3 & 3 & 0 \\
 3 & 3 & 6 & 0 \\
 3 & 6 & 7 & 0 \\
 0 & 0 & 0 & 0
\end{array}
\right),
\qquad
M_{2,ab} =
\left(
\begin{array}{cccc}
 3 & 3 & 6 & 0 \\
 3 & 3 & 6 & 0 \\
 6 & 6 & 12 & 0 \\
 0 & 0 & 0 & 0
\end{array}
\right),
\ee
whose kernels have dimensions 1 and 2, respectively.
The two kernels overlap along the real line generated by
\be
\vvt_4 = (0,0,0,-1),
\ee
which is linearly independent from the vectors \eqref{vA-large2}.
We complete all three vectors to a basis by taking
\be
\vvt_3 = (0,0,1,2).
\ee

Changing the basis as in \eqref{changet}, one arrives at the new form of K\"ahler cone conditions
\be
\begin{split}
\hatt^{1} + \hatt^{2} + \hatt^3 -\hatt^4\geq 0 ,
&\qquad
\hatt^3 \geq 0 ,
\qquad\qquad
\hatt^{2} \geq 0 ,
\\
\hatt^3 +\hatt^4 \geq 0 ,\qquad
&
\qquad
\hatt^1 \geq 0,
\qquad
3\hatt^2  + 2\hatt^3 - \hatt^4\geq 0
\end{split}
\label{Kalercone-large2n}
\ee
and the classical prepotential
\be
\begin{split}
\Fcl =&\,  -\frac16\, \Bigl[
9 (\hz^{1})^2 \hz^{2}
+9 \hz^{1} (\hz^{2})^2
+3 (\hz^{2})^3
+9 (\hz^{1})^2 \hz^3
\Bigr.\\
&\,
+36 \hz^{1} \hz^{2} \hz^3
+18 (\hz^{2})^2 \hz^3
+21 \hz^{1} (\hz^3)^2
+36 \hz^{2} (\hz^3)^2
\\
&\, \Bigl.
+13 (\hz^3)^3-3(\hz^3)^2 \hz^4-3 \hz^3 (\hz^4)^2
-(\hz^4)^3
\Bigr].
\end{split}
\ee
As it should be, $\hz^4$ is decoupled from $\hz^1$ and $\hz^2$.
Thus, in the local limit $\hatt^1$ and $\hatt^2$ are sent to infinity, $\hatt^3$ becomes frozen, and $\hatt^4$ remains dynamical.
As a result, the prepotential reduces to
\be
\fcl=-\frac{13}{6}\, (\hz^3)^3+\frac12\,(\hz^3)^2 \hz^4+\frac12\, \hz^3 (\hz^4)^2+\frac16\, (\hz^4)^3.
\ee
Note that the K\"ahler cone conditions \eqref{Kalercone-large2n} ensure that the effective gauge coupling
\be
	\Im\frac{\p^2\fcl}{(\p \hz^4)^2} =
	\hatt^3 + \hatt^4 \geq 0
\ee
is positive definite. Comparing with the prepotential \eqref{prepSU20} of the pure $SU(2)$ gauge theory, one finds the following identifications
\be
\hatt^3 =\frac{2\pi^2 a}{g_0^2},
\qquad
\hatt^4 = 2a\vph,
\qquad
c=\frac{4\pi^4a^2}{g_0^4}.
\ee
Note that in this model we obtain a non-vanishing coefficient of the linear term which contributes to the tension of magnetic strings.
This contribution however still vanishes at the SCFT point where the gauge coupling is sent to infinity.

\subsection{SU(3) gauge theory}
\label{subsec-su3-gt}

Another variation on the models considered in \S\ref{subsec-elliptic}
are geometries whose local limits give rise to higher-rank gauge theories.
To illustrate this possibility, in this subsection we explore a Calabi-Yau
admitting a local limit which leads to the pure $5d$ $N=1$ $SU(3)$ gauge theory.

Let us consider a toric hypersurface described by geometry 2 of polytope 1439 in \cite{Altman:2014bfa}.
Its defining data can be found in \eqref{polytop2} and \eqref{triang2}, and it has $h^{1,1}=4$.
In terms of toric divisors $\tD_i$, the basis of $H_4(\CY,\IZ)$ is chosen as
\be
D_1 = \tD_4,
\qquad
D_2 = \tD_5,
\qquad
D_3 = \tD_6,
\qquad
D_4 = \tD_7.
\label{basisDSU3}
\ee
In this basis, the classical prepotential encoding the triple intersection numbers is given by the polynomial
\be
\begin{split}
\Fcl=&\, -\frac16\, \Bigl[
3 (z^1)^2 z^2 +6 (z^1)^2 z^3+3 (z^1)^2 z^4 -9 z^1 (z^2)^2
\Bigr.
\\
&\,\Bigl.
+18 z^1 (z^3)^2-9 z^1 (z^4)^2+8(z^2)^3+16 (z^3)^3+9 (z^4)^3
\Bigr].
\end{split}
\ee
The intersections of Mori generators with the basis divisors are given by the matrix
\be
	C^i \cap D_a =
	\left(
\begin{array}{cccc}
1 & 0 & 0 & -3 \\
0 & -1 & 0 & 0 \\
0 & 1 & 1 & 0 \\
0 & 0 & 0 & 1
\end{array}
\right),
\ee
so that the K\"ahler moduli are subject to the following constraints
\be
t^1-3t^4\geq 0,
\qquad
t^2\leq 0,
\qquad
t^2+t^3\geq 0,
\qquad
t^4\geq 0.
\label{KahlerconeSU3}
\ee

To take the local limit, we choose
\be
\vvt_1 =  (3,0,0,1),
\ee
which trivially satisfies the K\"ahler cone conditions \eqref{KahlerconeSU3}.
Its intersection matrix \eqref{matA}
\be
	M_{ab} =
\left(
\begin{array}{cccc}
 1 & 3 & 6 & 0 \\
 3 & -9 & 0 & 0 \\
 6 & 0 & 18 & 0 \\
 0 & 0 & 0 & 0
\end{array}
\right)
\ee
has a two-dimensional kernel spanned by the following basis vectors
\be
\vvt_3 = (-6,-2,2,-2),
\qquad
\vvt_4= (3,1,-1,0).
\ee
The three vectors are linearly independent and can be completed to a basis by\footnote{In fact,
the vectors $\vvt_a$, $a=2,3,4$, may be chosen in a simpler form. Our choice instead
allows to have simpler relations to the gauge theory variables, which are found below in \eqref{identSU3}.}
\be
\vvt_2=(1,1,0,0).
\ee

The rotation of the basis \eqref{changet} modifies the K\"ahler cone conditions to
\be
\hatt^2+3\hatt^4\geq 0,
\qquad
2\hatt^3-\hatt^2-\hatt^4\geq 0,
\qquad
\hatt^2\geq 0,
\qquad
\hatt^1- 2\hatt^3\geq 0,
\label{KahlerconeSU3r}
\ee
and brings the prepotential to the following form
\be
\begin{split}
\Fcl =&\,
-\frac32\,(\hz^{1})^3
-6 (\hz^{1})^2 \hz^{2}
+\hz^{1} (\hz^{2})^2
\\
&\,
-\frac13\, (\hz^{2})^3
- (\hz^{2})^2 \hz^3
+ (\hz^{2})^2 \hz^4
+2\hz^2(\hz^3)^2
-2 \hz^{2} \hz^3\hz^{4}
+2\hz^2(\hz^{4})^2
\\
&\,
+\frac43\, (\hz^3)^3-2(\hz^3)^2 \hz^4+\hz^3 (\hz^4)^2
+\frac43\,(\hz^4)^3.
\end{split}
\ee
As is expected, $\hz^3$ and $\hz^4$ are decoupled from $\hz^1$.
Thus, in the local limit where $\hatt^1$ is large, we find two dynamical moduli, $\hatt^3$ and $\hatt^4$, whereas $\hatt^2$ is frozen.
The effective prepotential is
\be
\begin{split}
\fcl =&\,
-\frac13\, (\hz^{2})^3
- (\hz^{2})^2 \hz^3
+ (\hz^{2})^2 \hz^4
+2\hz^2(\hz^3)^2
-2 \hz^{2} \hz^3\hz^{4}
+2\hz^2(\hz^{4})^2
\\
&\,
+\frac43\, (\hz^3)^3-2(\hz^3)^2 \hz^4+\hz^3 (\hz^4)^2
+\frac43\,(\hz^4)^3
\end{split}
\label{effprepSU3}
\ee
and gives rise to the following matrix of the effective gauge couplings
\be
g_{IJ}=	\Im\frac{\p^2\fcl}{\p \hz^I\, \p \hz^J}  =
2 \left(
\begin{array}{ccc}
2(\hatt^2+ 2\hatt^3 -\hatt^4) & & -\hatt^2-2 \hatt^3+ \hatt^4  \\
-\hatt^2 -2 \hatt^3+ \hatt^4 &  & 2\hatt^2 + \hatt^3+4 \hatt^4
\end{array}
\right).
\label{effcouplSU3}
\ee
The trace and the determinant of this matrix are given by
\be
\begin{split}
\Tr\, g=&\, 2\(4\hatt^2+5\hatt^3+2\hatt^4\)
=5 (2\hatt^3-\hatt^2-\hatt^4)+3(\hatt^2+3\hatt^4)+10\hatt^2,
\\
\det g =&\, 12\(\hatt^2+2\hatt^3-\hatt^4\)\(\hatt^2+3\hatt^4\).
\end{split}
\ee
It follows immediately from the K\"ahler cone conditions \eqref{KahlerconeSU3r}
that both of them are positive,
which ensures the positive definiteness of \eqref{effcouplSU3}.
Comparing \eqref{effprepSU3} with the prepotential \eqref{prepSU3} of the pure $SU(3)$ gauge theory, one obtains the following dictionary
\be
\begin{split}
&
\hatt^2=\frac{2\pi^2 a}{g_0^2},
\qquad\
\hatt^3 =a\vph_1,
\qquad\quad
\hatt^4 = a\vph_2,
\\
&
\ccl=-3,
\qquad
c_1=-\frac{4\pi^4 a^2}{g_0^4},
\qquad
c_2=\frac{4\pi^4 a^2}{g_0^4}.
\end{split}
\label{identSU3}
\ee

\subsection{No local limit}
\label{subsec-no-limit}

The previous examples could make an impression that most of Calabi-Yau manifolds allow a non-trivial local limit in the sense of \S\ref{subsec-def}.
However, this is not so. It is easy to find examples which do not allow any such limit.
For instance, let us consider a complete intersection Calabi-Yau (CICY) manifold defined by the following configuration matrix
\be
	\left[  \begin{array}{c||ccc}
	2 & 2 & 1 & 0\\
	2 & 1 & 0 & 2 \\
	2 & 0 & 2 & 1
	\end{array} \right]_{-48}\, .
\label{confCY}
\ee
It appears first in the list of CICYs studied in \cite{Anderson:2017aux}, which simultaneously has $h^{1,1} = 3$
and the property of being \emph{K\"ahler favourable} (model 5299 in this database).
The latter property means that its K\"ahler cone descends from the one of the ambient projective space.
Thus, choosing a basis of divisors given by a subset of the divisors of the ambient space,
the K\"ahler cone admits a particularly simple description as the positive orthant $t^a \geq 0, \ a=1,2,3$.
The intersection numbers can be computed from the configuration matrix \eqref{confCY}
using the standard technique (see e.g. \cite{Hubsch:1992nu}) and give rise to the following cubic prepotential
\be
\begin{split}
\Fcl=&\,
-\Bigl[(z^1)^2 z^2+2(z^1)^2 z^3+2 z^1 (z^2)^2+9 z^1 z^2 z^3
+ z^1(z^3)^2+(z^2)^2 z^3+2 z^2 (z^3)^2\Bigr].
\end{split}
\ee

To have a non-trivial local limit, we have to find at least one vector $\vvt$, belonging to the K\"ahler cone,
such that the matrix $M_{ab}=\kappa_{abc}\vt^c$ is degenerate.
In particular, this implies that it must have vanishing determinant.
Calculating the determinant for a generic vector, one finds
\be\label{eq:detM-CICY}
\begin{split}
\det M=	&\,18\Bigl[ 4 \((\vt^1)^3+ (\vt^2)^3+ (\vt^3)^3\)+9 \((\vt^1)^2\vt^2+ \vt^1(\vt^3)^2+ (\vt^2)^2 \vt^3\)
\Bigr.\\
&\, \Bigl. +18 \(\vt^1(\vt^2)^2+(\vt^1)^2\vt^3+\vt^2 (\vt^3)^2\)+69\vt^1 \vt^2 \vt^3\Bigr].
\end{split}
\ee
Note that all coefficients are positive. Thus, the determinant can vanish only if some components $\vt^a$ have opposite signs.
An example of such vector is provided by
\be
\vvt = (1,-4,0).
\ee
But any such vector does not belong to the K\"ahler cone which requires the positivity of all coefficients.
Therefore, we conclude that this Calabi-Yau does not admit a  non-trivial local limit.

\section{Conclusions}
\label{sec-discussion}

In this paper we analyzed the rigid limit of the HM moduli space $\cM_H$ of type IIB string theory
compactified on a CY threefold $\CY$. Whereas for generic QK manifolds the rigid limit is not well defined,
for the HM moduli space we suggested to induce it by a local limit of the CY.
When such local limit exists, we showed that the original manifold reduces to a manifold $\cM'_H$ of real dimension $4\nI$,
where $\nI$ is the number of shrinking divisors on $\CY$, and computed the {\it exact} non-perturbative metric on it.
To accomplish this, we significantly improved the understanding of the D-instanton corrected metric on $\cM_H$ by
computing explicitly its exact expression for {\it all mutually non-local} charges.

Furthermore, we proved that $\cM'_H$ is an HK manifold and can be obtained by a series of HK quotients of the Swann bundle over $\cM_H$.
An intermediate step of this quotient procedure coincides with the HK manifold $\Mcor$ related to $\cM_H$ by the QK/HK correspondence.
All these relations become particularly simple in the twistor formalism where
the metric on a quaternionic manifold is encoded in a set of Darboux coordinates on its twistor space.
Then it turns out that the rigid limit simply reduces one system of Darboux coordinates to another by restricting
to the charge lattice of shrinking cycles, whereas the HK quotient along an isometry just removes
a symplectic pair of Darboux coordinates, one of which plays the role of the moment map.

We would like to point out that our rigid limit is essentially different from the one considered, for instance,
in \cite{Ambrosetti:2010tu,Antoniadis:2015egy,Antoniadis:2016kde}. In these papers the limiting HK manifold
has the same dimension as the original QK manifold and the procedure heavily relies on the existence of continuous isometries.
In our case, the dimension is always reduced because of the decoupling of the ``universal hypermultiplet"
containing the dilaton. Such decoupling is very natural since this multiplet has a gravitational origin and should not contribute to
the gauge theory physics recovered in the limit. Besides, although some isometries do appear at intermediate steps
of our procedure, the original manifold is taken to be fully non-perturbative where all classical isometries are broken
by instanton corrections.

Our limit is also different from the rigid limit suggested in \cite{Gunara:2013rca} which relies on a simple rescaling.
Although this allows to decouple some multiplets, including the universal hypermultiplet, and thus to reduce the effective dimension,
the decoupled fields do not disappear, but just support the flat metric.
In contrast, in our limit some fields do drop out and others become frozen.
Besides, the procedure of \cite{Gunara:2013rca} was performed only for the classical metric described by the c-map,
and an inspection shows that its direct generalization to the instanton corrected metric does not appear to produce sensible results.
At the same time, the instanton corrections to the metric on $\cM'_H$ all turn out to have a physical interpretation.

This interpretation comes from a general relation of $\cM'_H$ to the physics of five-dimensional $N=1$ gauge theories.
Following \cite{Haghighat:2011xx}, we argued that this manifold coincides with the non-perturbative target space of the $\sigma$-model
obtained by compactifying a 5d gauge theory on a torus. Which 5d gauge theory is recovered in the limit can be established
by matching the classical prepotentials. We demonstrated this matching procedure on several examples, including
a family of elliptically fibered CYs and a few toric hypersurfaces.

Note that in the usual notion of local limit, one zooms in around a point in the moduli space where the CY develops a singularity.
The study of the relation between five-dimensional gauge theories and singularities in CY threefolds
has a long history (see, for instance, \cite{Morrison:1996xf,Douglas:1996xp,Intriligator:1997pq} and \cite{Xie:2017pfl} for a recent work).
We hope that this paper can make at least two contributions to this subject.
First, we suggest a very simple condition in terms of intersection numbers of $\CY$ for the existence of a local limit.
It simply requires that there exists a set of vectors belonging to the boundary of the K\"ahler cone
such that the intersection of kernels of certain matrices constructed from them and the triple intersections
is non-empty.
It would be interesting to understand the precise relation of this criterion to the mathematical
conditions for the existence of CY singularities \cite{Reid}.

Second, our work extends the discussion to the setting of torus compactifications
where BPS states of 5d gauge theory generate non-perturbative effects.
Our results provide precise predictions from string theory for their contributions to the metric on the moduli space.
In particular, some of $(p,q)$-instantons are identified with the dyonic instantons of gauge theory and D3-instantons
correspond to the instantons generated by magnetic strings.
None of them has been computed exactly, and this work fills in this essential gap.

The compactification on a torus gives rise to the modular invariance of the effective three-dimensional theory,
which can be identified with the $SL(2,\IZ)$ symmetry of type IIB string theory surviving compactification on CY and the rigid limit.
This symmetry severely restricts both the form of the metric and the BPS spectrum, which remains in our results as a necessary input data.
We suggested how such constraints on the spectrum can be derived along the lines of \cite{Alexandrov:2012au,Alexandrov:2016tnf}
which should result in specific modular properties of a generating function of BPS degeneracies of magnetic strings bound to dyonic instantons.
This function is also expected to have a relation to the modular partition function studied in \cite{Haghighat:2012bm} in the same context.

Another interesting and related problem is to understand the wall crossing in 5d gauge theories.
Although there are some important differences with wall crossing in four dimensions, it is natural to expect
some relation between the two, as the theories can be related by compactification on a circle.
Moreover, while 5d $N=1$ gauge theories on a torus are richer than 4d $N=2$ gauge theories on a circle
considered in \cite{Gaiotto:2008cd}, their effective low energy descriptions are captured by the same mathematical framework,
so that both of them appear to be just particular cases of a general structure which is built on the wall crossing formula
discovered by Kontsevich and Soibelman \cite{ks}. It would be interesting to understand the role of this structure directly in five dimensions.

\acknowledgments

We would like to thank Benjamin Assel, Seung-Joo Lee, Joe Minahan, Boris Pioline, Stefan Vandoren and Timo Weigand
for correspondence and valuable discussions.
The work of P.L. is supported by the grants ``Geometry and Physics'' and
``Exact Results in Gauge and String Theories'' from the Knut and Alice Wallenberg Foundation.

\appendix

\section{Special geometry in the classical approximation}
\label{ap-special}

The local special geometry is determined by a prepotential $F(X)$, a holomorphic function homogeneous of degree 2 in coordinates $X^\Lambda$.
It defines the two main quantities of interest: the K\"ahler potential $\cK$ \eqref{Kahlerpot} and the matrix of the gauge couplings
\be
\cN_{\Lambda\Sigma} =  \bF_{\Lambda\Sigma} - \I \frac{(Nz)_\Lambda (Nz)_\Sigma}{(zNz)},
\label{defcN}
\ee
where $N_{\Lambda\Sigma} = -2\Im F_{\Lambda\Sigma}$. The imaginary part of $\cN_{\Lambda\Sigma}$ plays a particularly important role.
It is a negative definite matrix and for its inverse one can establish the following general result
\be
\Im\cN^{\Lambda\Sigma} =  2  N^{\Lambda\Sigma} - 2\,e^{\cK} \(z^\Lambda \bz^\Sigma + \bz^\Lambda z^\Sigma\),
\label{invcN}
\ee
where $N^{\Lambda\Sigma}$ is the inverse of $N_{\Lambda\Sigma}$.

In the particular case of the classical prepotential \eqref{Fcl}, it is possible to find more concrete representations for the above objects.
First, it is straightforward to compute
\be
N_{\Lambda\Sigma}
=  \begin{pmatrix} \frac23\,\kappa_{abc} \(3b^ab^bt^c-t^at^bt^c\) && -2\kappa_{bcd} b^ct^d \\  -2\kappa_{acd} b^ct^d && 2\kappa_{abc} t^c\end{pmatrix}.
\label{Ncl}
\ee
This implies
\be
\begin{split}
(zNz) =&\,  (\bz N\bz) = - \frac83\, \kappa_{abc} t^a t^b t^c= -16 \Vcy,
\\
e^{-\cK} =&\, \frac43\, \kappa_{abc} t^a t^b t^c= 8 \Vcy.
\end{split}
\ee
The inverse matrix $N^{\Lambda\Sigma}$ can be found in terms of $\kappa^{ab}$, the inverse of $\kappa_{ab}=\kappa_{abc} t^c $.
The result reads
\be
N^{\Lambda\Sigma}
= -\frac{1}{4\Vcy} \begin{pmatrix} 1 && b^b \\ b^a &&  - 2\Vcy\kappa^{ab} + b^a b^b\end{pmatrix}.
\label{Nclinv}
\ee
Using these results and notation $(vu)_a=\kappa_{abc}v^b u^c$, one then finds
the K\"ahler metric and its inverse
\be
\begin{split}
\cK_{a\bar{b}} =&\, -\frac{1}{4\Vcy} \(\kappa_{ab}- \frac{1}{4\Vcy}\,(tt)_a(tt)_b\),
\\
\cK^{a\bar{b}} = &\, -4\Vcy\kappa^{ab} +2\,t^at^b,
\end{split}
\label{Kmetric-expl}
\ee
and the real and imaginary parts of the gauge coupling matrix
\bea
\Re\cN_{\Lambda\Sigma} &=&
\begin{pmatrix}
-\frac13 (bbb) && \frac12 (bb)_b \\ \frac12 (bb)_a && - \kappa_{abc}b^c
\end{pmatrix},
\nn\\
\Im\cN_{\Lambda\Sigma} &=& - \Vcy
\begin{pmatrix}
1+4\cK_{a\bar{b}}b^a b^b&& -4 \cK_{a\bar{b}} b^b \\
 -4\cK_{a\bar{b}} b^a && 4\cK_{a\bar{b}}
\end{pmatrix},
\label{mmIR}\\
\Im\cN^{\Lambda\Sigma} &=&- \Vcy^{-1}
\begin{pmatrix}
1 && b^b \\ b^a && b^a b^b + \frac14\,\cK^{a\bar{b}}
\end{pmatrix}.
\nn
\eea

\section{Derivation of the D-instanton corrected HM metric}
\label{ap-Dinst}

\subsection{Twistorial description of QK manifolds}
\label{subap-twistors}

QK manifolds represent a very complicated type of geometry. Although they carry a quaternionic structure
given by the triplet of almost complex structures $\vec J$, all these almost complex structures are non-integrable
so that QK manifolds are not even complex \cite{MR664330}. A very efficient way to deal with such manifolds is
to work with their twistor spaces $\cZ_\cM$ whose $\CP$ fiber describes normalized linear combinations of $J_i$, $i=1,2,3$.
In contrast to the original manifold $\cM$, its twistor space is a \kahler manifold and, most importantly, it carries
a {\it holomorphic contact structure} \cite{MR1327157} defined as the kernel of the canonical (1,0)-form on $\cZ_\cM$
\be
D t = \de t + p_+ - \I p_3 t + p_- t^2,
\label{canform}
\ee
where $t$ is the standard stereographic coordinate parametrizing $\CP$, $\vec p$ is the $SU(2)$ part of the Levi-Civita connection on $\cM$,
and we used the chiral components defined as $p_\pm=-\hf\(p_1\mp \I p_2\)$.
Rescaling $Dt$, one can make from it a {\it holomorphic} one-form\footnote{In general, the rescaling factor may depend
holomorphically on the fiber coordinate $t$ and is different in different patches of an open covering of the twistor space,
which implies that the contact one-form is not globally defined and has different local realizations $\cX^{[i]}$.
However, we will not need such generic construction which becomes relevant only after inclusion of NS5-brane instantons.}
\be
\cX = \frac{4}{\I t} \,e^{\phi}\, D t
\label{relcontform}
\ee
such that $\cX\wedge \(\de \cX\)^n$ is the non-vanishing holomorphic top form.
The rescaling function $\phi$ is called the {\it contact potential}. The properties of $\cX$ imply that locally,
by a proper choice of coordinates, it can always be trivialized as
\be
\cX = \de \ai{i} + \xii{i}^\Lambda \de \txii{i}_\Lambda,
\label{contform}
\ee
where the index $\scriptstyle{[i]}$ labels open patches of an atlas, $\cZ=\cup\, \cU_i$, and
$(\xii{i}^\Lambda,\txii{i}_\Lambda,\ai{i})$ is the set of {\it Darboux coordinates} in $\cU_i$.
These coordinates is the central element of this construction because
knowing them as functions on the base $\cM$ and of the fiber coordinate $t$ is, in principle, equivalent to knowing the metric on $\cM$.
Let us spell out the steps necessary to compute it \cite{Alexandrov:2008nk}:
\begin{enumerate}
\item
First, one finds the Laurent expansion of the Darboux coordinates near $t=0$.
Denoting by $\scriptstyle{[+]}$ the patch surrounding the north pole of $\CP$,
we assume that the expansion has the following general form
\be
\begin{split}
\xii{+}^\Lambda=&\,\xii{+}^{\Lambda,-1}\varpi^{-1}+\xii{+}^{\Lambda,0}+O(\varpi),
\\
\txii{+}_\Lambda=&\,\txii{+}_{\Lambda,0}+O(\varpi),
\\
\ai{+}=&\,4\I c\log \varpi+\ai{+}_{0}+O(\varpi),
\end{split}
\label{expDc}
\ee
which is consistent with the form of Darboux coordinates in the case of the D-instanton corrected HM moduli space (see the next subsection).

\item
One specifies the almost complex structure $J_3$ by providing a basis of (1,0) forms on $\cM$.
Such a basis was found in \cite{Alexandrov:2008nk} and, after some simplifications, it takes the following form
\be
\label{defPi}
\pi^a =\de \(\xii{+}^{a,-1}/\xii{+}^{0,-1}\) ,
\qquad
\tilde\pi_\Lambda= \de\txii{+}_{\Lambda,0}  ,
\qquad
\tilde\pi_\alpha = \frac{1}{2\I}\,\de\ai{+}_0 +2c \,\de\log\xii{+}^{0,-1}.
\ee

\item
Substituting the expansions \eqref{expDc} into the contact one-form $\cX$ \eqref{contform}
and comparing it with the canonical form $Dt$ \eqref{canform} using \eqref{relcontform}, one finds the contact potential $\phi$ and
the components of the SU(2) connection
\be
\begin{split}
p_+ &=\frac{\I}{4}\, e^{-\phi}
\, \xi^{\Lambda,-1}_{[+]}  \de\txi^{[+]}_{\Lambda,0},
\\
p_3 &= -\frac{1}{4}\, e^{-\phi} \left( \de\alpha^{[+]}_0 +
\xi^{\Lambda,0}_{[+]}  \de\txi^{[+]}_{\Lambda,0} +
\xi^{\Lambda,-1}_{[+]}  \de\txi^{[+]}_{\Lambda,1}  \right) .
\end{split}
\label{connection}
\ee

\item
The $SU(2)$ connection $\vec p$ can then be used to compute the triplet of quaternionic two-forms $\vec \omega$
which are defined by the metric and the triplet of almost complex structures as $\vec \omega(X,Y)=g(\vec J X,Y)$,
but are known to be proportional to the curvature of the $SU(2)$ connection \cite{MR664330}.
In particular, for $\omega_3$ the formula reads
\be
\omega_3 = -2{\rm d} p_3+ 4\I  p_+ \wedge p_-.
\label{Kform}
\ee

\item
Finally, the metric is recovered as $g(X,Y) = \omega_3(X,J_3 Y)$.
To do this in practice, one should rewrite $\omega_3$, computed by \eqref{Kform} in terms of differentials of (generically real)
coordinates on $\cM$, in the form which makes explicit that it is of (1,1) Dolbeault type.
Using for this purpose the basis $\pi^X=(\pi^a,\tilde\pi_\Lambda,\tilde\pi_\alpha)$ given in \eqref{defPi}, the final result should look like
\be
\omega_3=\I g_{X\bY} \pi^X\wedge \bar\pi^{Y},
\label{metom}
\ee
from which the metric readily follows as $\de s^2 =2 g_{X\bY} \pi^X \otimes \bar\pi^{Y}$.
Technically, this is the most non-trivial step, which we realize for the D-instanton corrected HM moduli space in \S\ref{subap-metric}.

\end{enumerate}

\subsection{D-instantons in twistor space}
\label{subap-Dinst}

As we saw above, a QK manifold can be specified by a system of Darboux coordinates on its twistor space.
For the D-instanton corrected HM moduli space this was done in \cite{Alexandrov:2008gh,Alexandrov:2009zh}
where it was shown that Darboux coordinates $\xi^\Lambda$ and $\txi_\Lambda$ are determined by a system of integral equations.
To write it explicitly, it is convenient to introduce the exponentiated version of the Darboux coordinates labeled by charge
\be
\cX_\gamma =\qr\, e^{-2\pi \I \(q_\Lambda \xi^\Lambda-p^\Lambda\txi_\Lambda\)},
\ee
where $\qr$ is a sign function, known as quadratic refinement, which satisfies $\qr\qrp=(-1)^{\langle \gamma,\gamma'\rangle}\qrg{\gamma+\gamma'}$
and can be chosen as $\qr=(-1)^{q_\Lambda p^\Lambda}$.
Then the equations read
\be
\cX_\gamma(t) = \cXsf_\gamma(t)\, \exp\[
\frac{1}{4\pi\I}
\sum_{\gamma'} \Om{\gamma'}\, \langle \gamma ,\gamma'\rangle
\int_{\ell_{\gamma'} }\frac{\de t'}{t'}\, \frac{t+t'}{t-t'}\,
\log\(1-\cX_{\gamma'}(t')\)\],
\label{eqTBA}
\ee
where
\be
\cXsf_\gamma(t)=\exp\[-2\pi\I\(\Thkl+\frac{\tau_2}{2}\(Z_\gamma \,t^{-1}-\bZ_\gamma\,t\)\)\],
\label{defXsf}
\ee
$\Thkl= q_\Lambda \zeta^\Lambda - p^\Lambda\tzeta_\Lambda$ is a combination of RR-fields, $Z_\gamma$ is the central charge \eqref{defZ},
$\Om{\gamma'}$ is the generalized DT invariant,
$\langle\gamma,\gamma'\rangle$ is the skew-symmetric product \eqref{sympprod}, and
$\ell_\gamma$ is the so-called BPS ray on $\CP$ joining $t=0$ and $t=\infty$
along the direction determined by the phase of the central charge
\be
\ell_\gamma=\{ t\, :\ Z_\gamma(z)/t\in \I \IR^-\}.
\ee
In the perturbative approximation where the D-instantons are ignored, the Darboux coordinates are given by $\cXsf_\gamma$,
whereas the D-instantons are incorporated by the integral contribution in \eqref{eqTBA} weighted by DT invariants.
Given a solution of these integral equations, the remaining Darboux coordinate $\alpha$ can be found by simple integration
\be
\alpha(t)=4\I c \log t-\hf\, \sigma
-\frac{\tau_2}{4}\(\varpi^{-1} \cW-\varpi \bar \cW\)
- \frac{\I}{16\pi^3}\sum_{\gamma} \Om{\gamma}
\int_{\ellg{\gamma}}\frac{\d t'}{t'}\,\frac{t+t'}{t-t'}\,
L_{\gamma}(t')-\hf\, \xi^\Lambda\, \txi_\Lambda,
\label{alp}
\ee
where the parameter $c = \frac{\chi_\CY}{192\pi}$ encodes the one-loop $g_s$-correction,
\be
L_\gamma(t)= \Li_2\(\cX_\gamma\)
+\hf\,\log \(\qr^{-1}\cX_\gamma\)\log\(1-\cX_\gamma\)
\label{defrogers}
\ee
is a variant of the Rogers dilogarithm and
\be
\cW=F_\Lambda\zeta^\Lambda-z^\Lambda\tzeta_\Lambda
+\frac{1}{8\pi^2}\sum_{\gamma} \Om{\gamma} \, Z_{\gamma} \,\int_{\ellg{\gamma}}\frac{\d t}{t}\,\log\(1-\cX_\gamma\).
\ee

\subsection{Computation of the metric}
\label{ap-om3first}

Now we will follow the procedure outlined in \S\ref{subap-twistors} towards evaluation of the metric
corresponding to the twistorial construction of the previous subsection.
All equations given below are a direct generalization of the ones which can be found in \cite{Alexandrov:2014sya}
where the additional restriction $\langle\gamma,\gamma'\rangle=0$ has been imposed.
Under this restriction the integral equations \eqref{eqTBA} are trivially solved and
the one-instanton approximation to the Darboux coordinates becomes exact, which simplifies the derivation of the metric.
However, as we will show below, this derivation can be done even avoiding the assumption of mutual locality.

Before we start, let us introduce a few useful notations:
two measures
\be
\begin{split}
\Dn{1}_\gamma[t] =&\, \frac{\d t}{t}\,\log\(1-\cX_\gamma(t)\),
\\
\Dn{2}_\gamma[t] =&\, \frac{\d t}{t}\,\frac{\cX_\gamma(t)}{1-\cX_\gamma(t)},
\end{split}
\label{defmeasure}
\ee
and integrals
\be
\begin{array}{rclrcl}
\Igg{}& = &\displaystyle
\int_{\ellg{\gamma}}\Dn{1}_{\gamma}[t],
\quad &
\rIg&=&
\displaystyle \int_{\ellg{\gamma}}\Dn{2}_{\gamma}[t] ,
\\
\Igpm& = &\displaystyle
\pm\int_{\ellg{\gamma}}t^{\mp 1}\Dn{1}_{\gamma}[t],
\quad &
\rIgpm&=&
\displaystyle \pm \int_{\ellg{\gamma}}t^{\mp 1}\Dn{2}_{\gamma}[t],
\end{array}
\label{newfun-expand}
\ee
which appear in the expansion around $t=0$ of the $t$-dependent integrals in \eqref{eqTBA} and similar equations.
Note that they satisfy the reality properties
\be
\overline{\Ingam{n}{\gamma}}=\Ingam{n}{-\gamma},
\qquad
\overline{\Insgam{n}{+}{\gamma}}=\Insgam{n}{-}{-\gamma}.
\ee
Besides, by partial integration one can find the following identity
\be
\Zg{}\Insgam{n}{+}{\gamma}-\bZg{}\Insgam{n}{-}{\gamma}
+\frac{1}{8\pi^2}\sum_{\gamma'}\Om{\gamma'}\langle\gamma,\gamma'\rangle
\int_{\ellg{\gamma}}\Dn{n}_{\gamma}[t]\int_{\ellg{\gamma'}}\Dn{1}_{\gamma'}[t']\,\frac{tt'}{(t-t')^2}=0,
\label{mod-ident}
\ee
which takes a very simple form for $n=1$ and after summing over charges
\be
\sum_{\gamma} \Om{\gamma}\(\Zg{}\Igp-\bZg{}\Igm\)=0.
\label{indent-cJ}
\ee

These notations become already useful when one writes the result for the contact potential, the rescaling factor appearing in \eqref{relcontform}
\be
e^\phi= \frac{\tau_2^2}{16}\,e^{-\cK}-c
-\frac{ \I\tau_2}{64\pi^2}\sum\limits_{\gamma}\Om{\gamma} \(\Zg{}\Igp+\bZg{}\Igm\) .
\label{phiinstmany}
\ee
Comparing the first term with the first relation in \eqref{mirmap}, one observes that the contact potential provides
a generalization of the four-dimensional dilaton to quantum level \cite{Alexandrov:2008nk}.
This partially explains the important role played by this function both in physics and mathematics \cite{Alexandrov:2014wca}.
As for the dilaton, we will also use for it the notation $r=e^\phi$.

\begin{enumerate}
\item
The first step is to find the expansion of the Darboux coordinates around $t=0$.
However, the Darboux coordinates defined by \eqref{eqTBA} and \eqref{alp} live in a patch of $\CP$
which does not include its north and south poles.
This is seen from the presence of additional poles compared to \eqref{expDc}
in the perturbative part and an essential singularity in the instanton part at $t=0$.
The additional singularities can be removed by performing a holomorphic contact transformation,
i.e. a change of Darboux coordinates preserving the contact one-form \eqref{contform}.
Such contact transformation is given by
\be
\begin{split}
\xii{+}^\Lambda &=  \xi^\Lambda +\p_{\txi_\Lambda }\Hij{+},
\\
\txii{+}_\Lambda &=  \txi_\Lambda - \p_{\xi^\Lambda } \Hij{+},
\\
\ai{+} &=  \alpha-\Hij{+}+ \xii{+}^\Lambda \p_{\xi^\Lambda}\Hij{+},
\end{split}
\label{QKgluing}
\ee
where the holomorphic function $\Hij{+}$ was found in \cite{Alexandrov:2009zh} to have the following form
\be
\label{gensymp}
\Hij{+}=  F(\xii{+}) +\cG(\xii{+},\txi).
\ee
Here the second term is a complicated, but irrelevant function for us because, as was shown in \cite{Alexandrov:2009zh},
it affects only $O(t^2)$ terms in the Laurent expansion of the Darboux coordinates.
Thus, we can safely ignore it for our purposes, and this allows to replace $\xii{+}^\Lambda$ on the r.h.s. of \eqref{QKgluing} by
$\xi^\Lambda$. Then one finds the following coefficients of the Laurent expansion of the Darboux coordinates:
\bea
\xii{+}^{\Lambda,-1} &=& \frac{\tau_2}{2}\, z^\Lambda,
\nn\\
\xii{+}^{\Lambda,0}&=&\zeta^\Lambda-\frac{1}{8\pi^2}\sum\limits_{\gamma} \Om{\gamma} p^\Lambda \Igg{},
\nn\\
\txii{+}_{\Lambda,0}&=&\tzeta_\Lambda-F_{\Lambda\Sigma}\zeta^\Sigma
-\frac{1}{8\pi^2}\sum\limits_{\gamma} \Om{\gamma} V_{\gamma\Lambda}\Igg{},
\nn\\
\txii{+}_{\Lambda,1}&=&-\frac{\I\tau_2}{2}\,\bz^\Sigma N_{\Lambda\Sigma}
-\frac{1}{\tau_2}\, F_{\Lambda\Sigma\Theta}\zeta^\Sigma\zeta^\Theta
-\frac{1}{4\pi^2}\sum\limits_{\gamma} \Om{\gamma} \bigg[V_{\gamma\Lambda}\Igamp{\gamma}
\nn\\
&&
-\frac{1}{\tau_2}\, F_{\Lambda\Sigma\Theta}p^\Sigma\zeta^\Theta\Igg{}
+\frac{1}{16\pi^2\tau_2}\, F_{\Lambda\Sigma\Theta}p^\Sigma\Igg{}\sum_{\gamma'} \Om{\gamma'} p'^\Theta\Igam{\gamma'}\bigg],
\label{coeff-expD}\\
\alpi{+}_{0}&=&
-\hf\(\sigma+\zeta^\Lambda\tzeta_\Lambda-F_{\Lambda\Sigma}\zeta^\Lambda\zeta^\Sigma\)+2\I\(r+c\)
\nn\\
&&
-\frac{1}{8\pi^2}\sum\limits_{\gamma} \Om{\gamma} \[
\frac{1}{2\pi\I }\int_{\ellg{\gamma}}\frac{\d t}{t}\,
\Li_2\(\cX_\gamma\)
- V_{\gamma\Lambda}\zeta^\Lambda \Igg{}
-\frac{\tau_2}{2}\, Z_\gamma \Igp
\right.
\nn\\
&& \left.
+\frac{1}{16\pi^2}\,\sum_{\gamma'}\Om{\gamma'}\( p^\Lambda\Igg{} V_{\gamma'\Lambda}\Igam{\gamma'}
-\langle\gamma,\gamma'\rangle\int_{\ellg{\gamma}}\Dn{1}_{\gamma}[t]
\int_{\ellg{\gamma'}}\Dn{1}_{\gamma'}[t'] \,\frac{t+t'}{t-t'}
\)\],
\nn
\eea
where we introduced a useful shorthand notation
\be
V_{\gamma\Lambda}= q_\Lambda -F_{\Lambda\Sigma}p^\Sigma.
\label{defVgam}
\ee

\item
Using these coefficients, it is straightforward to compute the basis of (1,0) forms \eqref{defPi}.
However, it can be further simplified since one can drop all terms proportional to $\pi^a=\de z^a$
in other basis elements. Furthermore, it turns out to be convenient to add to $\tilde\pi_\alpha$ the term
$-\frac{\I}{2}\,\xii{+}^{\Lambda,0} \tilde\pi_\Lambda$.
As a result, one arrives at the following basis
\bea
\de z^a, &&
\nn\\
\cY_\Lambda &= &
\de\tzeta_\Lambda-F_{\Lambda\Sigma}\de\zeta^\Sigma
-\frac{1}{8\pi^2}\sum\limits_{\gamma} \Om{\gamma} \(q_{\Lambda}-p^\Sigma F_{\Lambda\Sigma}\)\de \Igg{},
\label{holforms}\\
\Sigma&=& \de r +2c\,\de \log\frac{\tau_2}{2}
+\frac{\I}{4}\(\de \sigma+\tzeta_\Lambda\de \zeta^\Lambda-\zeta^\Lambda\de \tzeta_\Lambda\)
\nn\\
&&
+\frac{\I}{32\pi^2}\sum\limits_{\gamma} \Om{\gamma} \[\Igm \de\(\tau_2\bZg{}\)-\tau_2\Zg{} \de\Igp
+\frac{1}{8\pi^2}\sum\limits_{\gamma'} \Om{\gamma'}\langle\gamma,\gamma'\rangle \Igg{}\de\Igam{\gamma'}
\].
\nn
\eea

\item
Substituting the Laurent coefficients \eqref{coeff-expD} into \eqref{connection}, one obtains
the explicit expression for the components of the $SU(2)$ connection:
\bea
p_+&=&
\frac{\I\tau_2}{8r}\, z^\Lambda\cY_\Lambda
=\frac{\I\tau_2 }{8r}\[z^\Lambda\(\de\tzeta_\Lambda - F_{\Lambda\Sigma} \de\zeta^\Sigma \)- \frac{1}{8\pi^2}\sum_\gamma \Om{\gamma} Z_{\gamma} \de\Igg{}\],
\nn\\
p_3 &= & \frac{1}{8r} \left[ \de\sigma
+\tzeta_\Lambda\de \zeta^\Lambda-\zeta^\Lambda\de \tzeta_\Lambda
+\frac{\I\tau_2^2}{2}\, e^{-\cK}\cA_K
-\frac{\tau_2}{8\pi^2}\sum\limits_{\gamma} \Om{\gamma} \(\Igp \de\Zg{}-\Igm \de\bZg{}\) \right.
\nn\\
&&
\qquad\left.
+ \frac{1}{64\pi^4} \sum\limits_{\gamma,\gamma'} \Om{\gamma}\Om{\gamma'}  \langle\gamma,\gamma'\rangle \cJ_\gamma^{(1)} \de \cJ_{\gamma'}^{(1)}\],
\eea
where we introduced the K\"ahler connection on the complex structure moduli space
\be
\cA_K = \frac{\I}{2}\, (\cK_a \de z^a - \cK_{\bar{a}} \de\bz^a)
= \frac{\I}{2}\,e^\cK N_{\Lambda\Sigma}\(z^\Lambda\de\bz^\Sigma-\bz^\Sigma \de z^\Lambda \).
\ee

\item
The $SU(2)$ connection allows to find the quaternionic 2-form \eqref{Kform}:
\be
\begin{split}
\omega_3 = &\, \frac{1}{4r^2}\, \de r\wedge \[ \de\sigma
+\tzeta_\Lambda\de \zeta^\Lambda-\zeta^\Lambda\de \tzeta_\Lambda
-\frac{\tau_2}{8\pi^2}\sum\limits_{\gamma} \Om{\gamma} \(\Igp \de\Zg{}-\Igm \de\bZg{}\) \right.
\\ &\, \qquad
\left.
+ \frac{1}{64\pi^4} \sum_{\gamma,\gamma'} \Om{\gamma}\Om{\gamma'}  \langle\gamma,\gamma'\rangle\cJ_\gamma^{(1)} \de \cJ_{\gamma'}^{(1)}
\]
+\frac{\tau_2^2}{8 r}\, e^{-\cK}\de\log\frac{r}{\tau_2^2}\wedge\cA_K
\\
&\,
+\frac{1}{2r}\(\de\zeta^\Lambda\wedge\de\tzeta_\Lambda
-\frac{\I\tau_2^2}{4} N_{\Lambda\Sigma}\de z^\Lambda \wedge\de\bz^\Sigma
+\frac{\I\tau_2^2}{8r}\,z^\Lambda\bz^\Sigma\cY_\Lambda\wedge\bar\cY_\Sigma\)
\\
&\, +\frac{1}{32\pi^2 r}\sum_\gamma\Om{\gamma}\(\de \Igp \wedge\de(\tau_2\Zg{})-\de\Igm\wedge \de(\tau_2\bZg{}) \)
\\ &\,
-\frac{1}{256\pi^4 r} \sum_{\gamma,\gamma'} \Om{\gamma}\Om{\gamma'} \langle\gamma,\gamma'\rangle\,  \de\cJ_{\gamma}^{(1)} \wedge \de\cJ_{\gamma'}^{(1)}.
\end{split}
\label{om3-start}
\ee

\end{enumerate}
The last step, which is supposed to bring $\omega_3$ into the form \eqref{metom}, is technically very complicated.
Therefore, we relegate it into the next separate subsection.

\subsection{The last step}
\label{subap-metric}

The main complication arising due to mutual non-locality is that it is impossible to get $\cX_\gamma$ in a closed form.
However, what is crucial for the derivation of the metric is not $\cX_\gamma$ itself, but its differential.
From \eqref{eqTBA} one can derive an integral equation which it satisfies. This equation is simpler than the original equation on $\cX_\gamma$
because it is linear, and its solution can be given in terms certain $t$-dependent matrices on the charge lattice, i.e.
matrices acting on the (infinite-dimensional) space of vectors whose components are enumerated by charges.
More precisely, we define them
as the following infinite series of nested integrals
\bea
\cIo_{\gamma\gamma'}(t_0)&=& \delta_{\gamma\gamma'}+\sum_{n=1}^\infty\( \frac{\I}{4\pi}\)^n
\sum_{\gamma_1,\dots,\gamma_{n-1}\atop \gamma_0=\gamma, \ \gamma_n=\gamma'}
\prod_{k=1}^{n}\[\Om{\gamma_k}\langle\gamma_{k-1},\gamma_k\rangle \int_{\ellg{\gamma_k}}\Dn{2}_{\gamma_k}[t_k] \,\frac{t_{k-1}+t_k}{t_{k-1}-t_k}\],
\label{mat-cI}\\
\cIpm_{\gamma\gamma'}(t_0)&=& \delta_{\gamma\gamma'}+\sum_{n=1}^\infty \( \frac{\I}{4\pi}\)^n
\sum_{\gamma_1,\dots,\gamma_{n-1}\atop \gamma_0=\gamma, \ \gamma_n=\gamma'}
\prod_{k=1}^{n}\[\Om{\gamma_k}\langle\gamma_{k-1},\gamma_k\rangle \int_{\ellg{\gamma_k}}\Dn{2}_{\gamma_k}[t_k] \,\frac{t_{k-1}+t_k}{t_{k-1}-t_k}\]
\(\frac{t_0}{t_n}\)^{\pm 1}.
\nn
\eea
They can be checked to satisfy the simple conjugation properties
\be
\overline{\cIo_{\gamma\gamma'}(1/\bar t)}=\cIo_{-\gamma,-\gamma'}(t),
\qquad
\overline{\cIp_{\gamma\gamma'}(1/\bar t)}=\cIm_{-\gamma,-\gamma'}(t).
\ee
Applying the differential operator to the integral equation \eqref{eqTBA} and performing iterations, one finds
that these matrices encode the differential of the Darboux coordinates:
\be
\begin{split}
&\,
-\frac{1}{2\pi\I}\,\de\log\cX_\gamma(t)=\sum_{\gamma'}\[\cIo_{\gamma\gamma'}(t)\de\Theta_{\gamma'}
+\frac{\tau_2}{2}\(t^{-1}\cIp_{\gamma\gamma'}(t)\de Z_{\gamma'}-t\cIm_{\gamma\gamma'}(t)\de \bZ_{\gamma'}\)
\right.\\
&\, \left.\qquad
+\frac{1}{2}\(t^{-1}\cIp_{\gamma\gamma'}(t)Z_{\gamma'}-t\cIm_{\gamma\gamma'}(t)\bZ_{\gamma'}\)\de\tau_2\].
\end{split}
\ee

Note that all the matrices \eqref{mat-cI} collapse to $\delta_{\gamma\gamma'}$ under the condition of mutual locality.
This implies that most of our general results can be obtained from the equations in \cite{Alexandrov:2014sya}
by insertion of these matrices in proper places.
Due to this reason we will not repeat the calculation which is similar to the one done in \cite[appenidix B.3]{Alexandrov:2014sya}
and showing how to arrive at the representation \eqref{metom} for $\omega_3$. Instead, we just give the final result which,
upon substitution of all definitions, can be checked to reproduce the initial expression \eqref{om3-start}.
But first we need to introduce several notations which allow to write the result in a readable form.

\begin{itemize}
\item
First, we define a simple matrix constructed from the vector \eqref{defVgam}
\be
\begin{split}
\cQ_{\gamma\gamma'}=&\, V_{\gamma\Lambda} N^{\Lambda\Sigma}\bV_{\gamma'\Sigma}+\frac{\I}{2}\,\langle\gamma,\gamma'\rangle
\\
=&\, N^{\Lambda\Sigma}\Re V_{\gamma\Lambda}\Re V_{\gamma'\Sigma}+\frac14\, N_{\Lambda\Sigma} p^\Lambda p'^\Sigma.
\end{split}
\label{defmatQ}
\ee

\item
Next, we introduce an integrated version of the matrices \eqref{mat-cI}
\be
\begin{split}
\vv_{\gamma\gamma'}=&\, \frac{\Om{\gamma}}{4\pi}\int_{\ellg{\gamma}}\Dn{2}_{\gamma}[t]\,\cIo_{\gamma\gamma'}(t),
\qquad
\vvn{\pm n}_{\gamma\gamma'}= (\pm 1)^n\,\frac{\Om{\gamma}}{4\pi}\int_{\ellg{\gamma}}\Dn{2}_{\gamma}[t]\, t^{\mp n}\cIo_{\gamma\gamma'}(t),
\\
\vvpm_{\gamma\gamma'}=&\,\frac{\Om{\gamma}}{4\pi}\int_{\ellg{\gamma}}\Dn{2}_{\gamma}[t]\,\cIpm_{\gamma\gamma'}(t),
\qquad
\vvpmn{n}_{\gamma\gamma'}= (\pm 1)^n\,\frac{\Om{\gamma}}{4\pi}\int_{\ellg{\gamma}}\Dn{2}_{\gamma}[t]\, t^{\mp n}\cIpm_{\gamma\gamma'}(t).
\end{split}
\label{notvec}
\ee
Not all of them are actually independent since it is easy to check that they satisfy the following properties:
\begin{itemize}
\item
the matrices $\vv_{\gamma\gamma'}$, $\vvpn{2}_{\gamma\gamma'}$ and $\vvmn{2}_{\gamma\gamma'}$ are symmetric;
\item
identities under transposition
\be
\vvp_{\gamma\gamma'}=\vvm_{\gamma'\gamma},
\qquad
\vvpn{1}_{\gamma\gamma'}=\vvn{1}_{\gamma'\gamma},
\qquad
\vvmn{1}_{\gamma\gamma'}=\vvn{-1}_{\gamma'\gamma};
\label{symprop}
\ee
\item
identity involving the central charge
\be
\sum_{\gamma'}\(\vvpn{1}_{\gamma\gamma'}Z_{\gamma'}-\vvmn{1}_{\gamma\gamma'}\bZ_{\gamma'}\)
=\sum_{\gamma'}\(Z_{\gamma'}\vvn{1}_{\gamma'\gamma}-\bZ_{\gamma'}\vvn{-1}_{\gamma'\gamma}\)=0.
\ee
\end{itemize}

\item
Then we combine the matrices defined above into a new object
\be
\Min_{\gamma\gamma'}=\delta_{\gamma\gamma'}-2\sum_{\gamma''}\cQ_{\gamma\gamma''} \vv_{\gamma''\gamma'}.
\label{defmatA}
\ee
We are really interested in the inverse of this matrix which can always be found by an expansion treating the second term
as a perturbation.
Note that although $\Min_{\gamma\gamma'}$ is not symmetric, the matrix product $(\vv\Min)_{\gamma\gamma'}$ is symmetric.
In fact, the product $(\vv\Min^{-1})_{\gamma\gamma'}$ is also symmetric
which can be shown by expanding $\Min^{-1}$ so that
\be
(\vv\Min^{-1})_{\gamma\gamma'}=\vv_{\gamma\gamma'}+2(\vv\cQ\vv)_{\gamma\gamma'}+4(\vv\cQ\vv\cQ\vv)_{\gamma\gamma'}+\cdots.
\ee
Analogously, $(\Min^{-1}\cQ)_{\gamma\gamma'}$ is symmetric as well.
Using this property, one can also find the following useful identity
\be
2\vv \Min^{-1}\cQ=2(\vv\Min^{-1})^T\cQ=2\Min^{-T}\vv\cQ=\Min^{-T}\(\delta-\Min^T\)=\Min^{-T}-\delta.
\label{identMT}
\ee

\item
It is convenient also to introduce two vectors
\be
\begin{split}
\zzb_\gamma=&\, \sum_{\gamma'}\(Z_{\gamma'}\vvn{1}_{\gamma'\gamma}+\bZ_{\gamma'}\vvn{-1}_{\gamma'\gamma}\),
\\
\cW_{\gamma}=&\, 4\pi \sum_{\gamma'}\[\bZ_{\gamma'}\vvp_{\gamma'\gamma}
-\sum_{\tgam,\tgam'}\zzb_{\tgam}\Min^{-1}_{\tgam\tgam'}\cQ_{\tgam'\gamma'}\vvpn{1}_{\gamma'\gamma} \]
\end{split}
\label{def-vectors}
\ee
and a potential
\be
\begin{split}
\Uin=&\,
e^{-\cK}-2\sum_{\gamma,\gamma'} \vvp_{\gamma\gamma'}\bZ_\gamma Z_{\gamma'}
+\sum_{\gamma,\gamma'} \zzb_\gamma(\Min^{-1}\cQ)_{\gamma\gamma'}\zzb_{\gamma'}
,
\label{defUin}
\end{split}
\ee
which is a real function due to the property \eqref{symprop}.

\item
Besides, we define several 1-forms. The first one is a certain linear combination of the differentials of the RR-fields
\be
\cCf_\gamma= N^{\Lambda\Sigma}\(q_\Lambda-\Re F_{\Lambda\Xi}p^\Xi\)\(\de\tzeta_\Sigma-\Re F_{\Sigma\Theta}\de\zeta^\Theta\)
+\frac14\, N_{\Lambda\Sigma}\,p^\Lambda\,\de\zeta^\Sigma,
\label{connC}
\ee
which is built in the way analogous to $\cQ_{\gamma\gamma'}$ \eqref{defmatQ}.
The second, which we call $\cV$, appears explicitly in the HM metric \eqref{metric} as the quantum part of the connection
on the circle bundle parametrized by the NS-axion $\sigma$.
In terms of $\cCf_\gamma$ and the other quantities introduced above, it reads
\be
\begin{split}\hspace{-0.6cm}
\cV=&\,
\frac{\tau_2^2}{2}\, e^{-\cK}\(1-\frac{16r}{\tau_2^2\Uin}\) \cA_K
+\frac{16r}{\tau_2\Uin}\sum_{\gamma,\gamma'}\zzb_{\gamma'}\Min^{-1}_{\gamma'\gamma}
\(\cCf_\gamma-\frac{\I}{16\pi^2}\sum_{\gamma''}\Om{\gamma''}\langle\gamma,\gamma''\rangle \de\Igam{\gamma''}\)
\\
&\, +\frac{2r }{\pi\I\Uin}\sum_\gamma  \[\(\cW_\gamma+\frac{\tau_2\Uin}{16\pi\I r}\,\Om{\gamma}\Igp\)\de Z_\gamma
-\(\bar\cW_\gamma+\frac{\tau_2\Uin}{16\pi\I r}\,\Om{\gamma}\Igm\)\de \bZ_\gamma  \].
\end{split}
\label{conn}
\ee
Finally, we introduce
\bea
\cY_\gamma&=&\I N^{\Lambda\Sigma}\, \bV_{\gamma\Lambda}\, \cY_\Sigma
\nn\\
&=& \I \cCf_\gamma-\hf\, \de\Theta_\gamma-\frac{\I}{8\pi^2}\sum_{\gamma'}
\Om{\gamma'}\(\cQ_{\gamma\gamma'}+\frac{\I}{2}\,\langle\gamma,\gamma'\rangle\)\de\Igam{\gamma'},
\label{defUgam}
\\
\hat\Sigma&=& 2 \(1-\frac{8r}{\tau_2^2\Uin}\)\de r
+\frac{\I}{4}\(\de \sigma+\tzeta_\Lambda\de \zeta^\Lambda-\zeta^\Lambda\de \tzeta_\Lambda
+\frac{1}{64\pi^4}\sum\limits_{\gamma,\gamma'}\Om{\gamma} \Om{\gamma'}\langle\gamma,\gamma'\rangle \Igg{}\de\Igam{\gamma'}+\cV\) .
\nn
\eea
Both these 1-forms are of (1,0) Dolbeault type. Whereas this is evident for $\cY_\gamma$, for $\hat\Sigma$ this follows from
the following representation
\be
\hat{\Sigma} = \Sigma + f_\Lambda \de z^\Lambda + g^\Lambda \cY_\Lambda
\ee
with
\be
\begin{split}
f_\Lambda =&\,\(\frac{\tau_2^2}{8}-\frac{2r}{\Uin}\)\, N_{\Lambda\Sigma}\bz^\Sigma
+\frac{1}{\pi}\sum_\gamma
\( \frac{\tau_2}{16\pi\I}\, \Om{\gamma}\Igp +\frac{r}{\Uin}\, \cW_{\gamma}\) V_{\gamma\Lambda},
\\
g^\Lambda=&\,  \frac{4\I r}{\tau_2\Uin}\, N^{\Lambda\Sigma}\sum_{\gamma,\gamma'} \zzb_{\gamma}\Min^{-1}_{\gamma\gamma'}\bV_{\gamma'\Sigma}.
\end{split}
\ee

\end{itemize}

In terms of all these notations, one can show that the quaternionic 2-form \eqref{om3-start}
can be rewritten as
\bea
\omega_3 &= &  \frac{\I\,\hat\Sigma\wedge \bar{\hat\Sigma}}{4 r^2\(1-\frac{8r}{\tau_2^2\Uin}\)}
-\frac{\I}{2r}\(N^{\Lambda\Sigma}-\frac{\tau_2^2}{8r}\,z^\Lambda\bz^\Sigma\)\cY_\Lambda\wedge\bar\cY_\Sigma
-\frac{\I}{r}\sum_{\gamma,\gamma'}(\vv\Min^{-1})_{\gamma\gamma'}\cY_{\gamma}\wedge \bar\cY_{\gamma'}
\nn\\
&&
+\frac{\I}{2r\Uin}\sum_{\gamma}\((\zzb\Min^{-1})_\gamma \cY_\gamma+\frac{\tau_2}{4\pi}\,\cW_{\gamma}\de Z_{\gamma}\)
\wedge
\sum_{\gamma'}\((\zzb\Min^{-1})_{\gamma'}\bar\cY_{\gamma'}+\frac{\tau_2}{4\pi}\, \bar\cW_{\gamma'}\de\bZ_{\gamma'}\)
\nn\\
&&
+\frac{\I\tau_2}{2r}\sum_{\gamma,\gamma',\gamma''}\Min^{-1}_{\gamma\gamma'}
\[\vvpn{1}_{\gamma\gamma''}\(\de Z_{\gamma''}-\Uin^{-1}Z_{\gamma''}\p e^{-\cK}\)\wedge \bar\cY_{\gamma'}
+\cY_{\gamma'}\wedge\vvmn{1}_{\gamma\gamma''}\(\de\bZ_{\gamma''}-\Uin^{-1}\bZ_{\gamma''}\bar\p e^{-\cK}\)\]
\nn\\
&&
+\frac{\I\tau_2^2}{8r}\[ \Uin^{-1}\p e^{-\cK}\wedge\bar\p e^{-\cK}-N_{\Lambda\Sigma}\de z^\Lambda \wedge\de\bz^\Sigma
-\frac{1}{2\pi\Uin}\sum_{\gamma}\Bigl(\cW_{\gamma}\de Z_{\gamma}\wedge \bar\p e^{-\cK}+\p e^{-\cK}\wedge \bar\cW_{\gamma}\de\bZ_{\gamma}\Bigr) \]
\nn\\
&&
+\frac{\I\tau_2^2}{4r}\sum_{\gamma,\gamma'}\vvp_{\gamma\gamma'}\de Z_{\gamma'}\wedge\de \bZ_{\gamma}
-\frac{\I\tau_2^2}{2r}\sum_{\gamma,\gamma'}(\Min^{-1}\cQ)_{\gamma\gamma'}\sum_{\tgam}\vvpn{1}_{\gamma\tgam}\de Z_{\tgam}
\wedge
\sum_{\gamma''}\vvmn{1}_{\gamma'\gamma''}\de \bZ_{\gamma''}.
\label{om3}
\eea
All terms appearing in this representation are explicitly of (1,1) Dolbeault type. Therefore, one can apply the rule \eqref{metom}
which immediately produces the metric \eqref{metric} given in the main text.

\section{Metric on $\cM'_H$}
\label{ap-GMN}

In this appendix we derive the rigid limit of the D-instanton corrected HM metric \eqref{metric}
and uncover its geometric structure.
As discussed in \S\ref{subsubsec-quantum}, in this limit
some of worldsheet and D-instantons actually decouple and therefore we should restrict our attention
only to the charges $\gamma$ belonging to the lattice $\Geff=\{ \gamma=(0,p^I,q_{\hI},q_0)\}$.
This means that in all sums over charges appearing in \eqref{metric}
the condition $\gamma\in\Geff$ should be inserted.

\subsection{Scaling behavior}
\label{subap-scale}

Let us first find the scaling behavior of various quantities entering the metric.
Noticing that the quantum part of the prepotential \eqref{lve} remains finite,
for the real and imaginary parts of its second derivative one obtains
\be
\Re F_{\Lambda\Sigma}\sim\(\begin{array}{ccc}
\scl{2}  & \ \scl{2}\  & \scc
\\
\scl{2} & \scc & \scc
\\
\scc & \scc & \scc
\end{array}\),
\qquad
N_{\Lambda\Sigma}\sim\(\begin{array}{ccc}
\scl{3}  & \ \scl{}\  & \scc
\\
\scl{} & \scl{} & \scc
\\
\scc & \scc & \scc
\end{array}\),
\ee
where the rows and columns correspond to the splitting of the index $\Lambda=(0,\hA ,I)$.
To get the scaling of the inverse matrix $N^{\Lambda\Sigma}$, one can split $N_{\Lambda\Sigma}$ into its classical part $N^{\rm cl}$ given in \eqref{Ncl}
and the part $N^{\rm q}$ encoding the quantum corrections. After that the expansion
\be
N^{-1}=(N^{\rm cl})^{-1}-(N^{\rm cl})^{-1}N^{\rm q}(N^{\rm cl})^{-1}+\cdots
\ee
together with the explicit expression for the $(N^{\rm cl})^{-1}$ \eqref{Nclinv} and the scaling \eqref{sc-invkappa}, result in
\be
N^{\Lambda\Sigma}\sim\(\begin{array}{ccc}
\scl{-3}  & \ \scl{-3}\  & \scl{-3}
\\
\scl{-3} & \scl{-1} & \scl{-1}
\\
\scl{-3} & \scl{-1} & \scc
\end{array}\).
\ee
Similarly, for the gauge coupling matrix one obtains
\be
\Re \cN_{\Lambda\Sigma}\escc,
\qquad
\Im\cN_{\Lambda\Sigma}\sim\(\begin{array}{ccc}
\scl{3}  & \ \scl{}\  & \scc
\\
\scl{} & \scl{} & \scc
\\
\scc & \scc & \scc
\end{array}\),
\qquad
\Im\cN^{\Lambda\Sigma}\sim\(\begin{array}{ccc}
\scl{-3}  & \ \scl{-3}\  & \scl{-3}
\\
\scl{-3} & \scl{-1} & \scl{-1}
\\
\scl{-3} & \scl{-1} & \scc
\end{array}\).
\ee
In particular, the scaling of the inverse matrices implies that
\be
N^{IJ}=-\hf\, g^{IJ}+O(\Lambda^{-1}),
\qquad
\cI^{IJ}=-g^{IJ}+O(\Lambda^{-1}),
\label{invIg}
\ee
where $g^{IJ}$ is the inverse of $g_{IJ}=-\hf\, N_{IJ}$.

Combining the condition $\gamma\in\Geff$ with these results, it is easy to see that
the central charge $Z_\gamma$ remains finite and unaffected by the limit so as the vectors $V_{\gamma\Lambda}$, $\zzb_\gamma$
and the matrices $\vv_{\gamma\gamma'}$, $\vvpm_{\gamma\gamma'}$, $\vvpmn{n}_{\gamma\gamma'}$, $\Min_{\gamma\gamma'}$.
The matrix $\cQ_{\gamma\gamma'}$ also remains finite, but simplifies because some components of the matrix $N^{\Lambda\Sigma}$ vanish.
\be
\cQ_{\gamma\gamma'} = \frac14\, N_{IJ} p^I p'^J + N^{IJ} (q_I -\Re F_{IK} p^L) (q'_J-\Re F_{JL} p'^L)+O(\Lambda^{-1}).
\ee
The four-dimensional dilaton (coinciding with the contact potential) $r$ and the potential $\Uin$ \eqref{defUin}
have the leading contributions scaling as $\Lambda^3$.

An important role is played by the one-form $\cY_\Lambda$ \eqref{holforms}. However, its components have too different scaling
and, instead of working with them, it turns out to be more convenient to introduce two real one-forms, $y_\Lambda$ and $w^\Lambda$, defined by
\be
\cY_\Lambda=y_\Lambda-F_{\Lambda\Sigma}w^\Sigma.
\label{relYyw}
\ee
Their explicit expressions are
\be
\begin{split}
y_\Lambda=&\,\de\tzeta_\Lambda-\frac{\I}{4\pi^2}\sum_{\gamma\in\Geff}\Om{\gamma}
\(\frac14\, N_{\Lambda\Sigma}p^\Sigma-\Re F_{\Lambda\Lambda'}N^{\Lambda'\Sigma}\Re V_{\gamma \Sigma}\)
\de\Igam{\gamma},
\\
w^\Lambda=&\, \de\zeta^\Lambda+\frac{\I}{4\pi^2}\sum_{\gamma\in\Geff}\Om{\gamma}
N^{\Lambda\Sigma}\Re V_{\gamma\Sigma}\de\Igam{\gamma},
\end{split}
\label{comp-cY}
\ee
and it is easy to check that in the limit they both remain finite except one contribution in $y_0$ which scales as $\Lambda$.
It turns out that the one-forms $\cCf_\gamma$ \eqref{connC} and $\cY_\gamma$ \eqref{defUgam} also have a divergent piece
which is determined by the same quantity.
Thus, the three one-forms can be written as
\be
y_0=-\frac{\I}{4\pi^2}\sum_{\gamma\in\Geff}\Om{\gamma}\cCd_\gamma\de\Igam{\gamma}+y_0^{(0)},
\qquad
\cCf_\gamma=\cCd_\gamma\de\zeta^0+\cCf^{(0)}_\gamma,
\qquad
\cY_\gamma=\I\cCd_\gamma w^0+\cY^{(0)}_\gamma,
\ee
where
\be
\cCd_\gamma=- \Re F_{0\hA}N^{\hA\Lambda}\(q_{\Lambda}-\Re F_{\Lambda\Sigma}p^\Sigma\)
\label{diverg}
\ee
diverges\footnote{Note that not all terms in $\cCd_\gamma$ scale as $\Lambda$. Some of them stay finite or even decay, but
we find convenient to combine all of them into one expression. What is important is the behavior of the leading contribution.},
whereas $y_0^{(0)}$, $\cCf^{(0)}_\gamma$ and $\cY^{(0)}_\gamma$ are all finite.

\subsection{Evaluation of the limit}
\label{subap-limit}

Before we apply the scaling results from the previous subsection to the metric \eqref{metric},
it is convenient to rewrite the second term on the first line using \eqref{relYyw}.
Then it becomes
\bea
-\frac{1}{r}\(N^{\Lambda\Sigma}-\frac{\tau_2^2}{8r}\,z^\Lambda\bz^\Sigma\)\cY_\Lambda\bar\cY_\Sigma
&=&-\frac{1}{2r}\, \Im\cN^{\Lambda\Sigma}\(y_\Lambda-\Re\cN_{\Lambda\Lambda'}w^{\Lambda'}\)\(y_\Lambda-\Re\cN_{\Lambda\Lambda'}w^{\Lambda'}\)
\nn\\
&&-\frac{1}{2r}\, \Im\cN_{\Lambda\Sigma}w^\Lambda w^\Sigma
+\frac{\tau_2^2}{8r^2}\(1-\frac{16r}{\tau_2^2}\, e^\cK\)\left|z^\Lambda\cY_\Lambda\right|^2.
\label{RR-terms}
\eea
Note that the coefficient of the last term is given by
\be
\begin{split}
1-\frac{16r}{\tau_2^2}\, e^\cK =&\, \frac{e^\cK}{\tau_2^2}\[\frac{\chi_\CY}{12\pi}
+\frac{ \I\tau_2}{4\pi^2}\sum\limits_{\gamma\in\Geff} \Om{\gamma}
\(\Zg{}\Igp+\bZg{}\Igm\)\]
\end{split}
\label{coef-c}
\ee
and scales as $\Lambda^{-3}$.

The Lagrangian based on the metric \eqref{metric} can be represented as in \eqref{fullL}\footnote{Again the factor
$\sqrt{\tau_2}$ is included to preserve the modular symmetry.}
\be
\cL_{\rm bos} =-\frac{\sqrt{-g}\sqrt{\tau_2}}{4\kappa^2 r}\(\cL_+ +\cL_0+\cL_-\),
\label{fullLq}
\ee
where the three terms in the brackets correspond to divergent, finite and vanishing contributions, respectively.
Taking the gravitation coupling $\kappa^2$ to scale as $\Lambda^{-3}$, one ensures that the prefactor is constant.
Then the contribution $\cL_-$ drops out and we will not specify its form since it is completely irrelevant.
On the other hand, the divergent part $\cL_+$ imposes strongly classical equations of motion which lead to the freezing of some fields.
To compute its effect, let us first denote
\bea
\cAd &=& - \cI_{00} - 4 \sum_{\gamma,\gamma'\in\Geff} (\vv\Min^{-1})_{\gamma\gamma'}\cCd_\gamma \cCd_{\gamma'},
\\ \nn
\cBd &=& \cI_{0\hA} w^{\hA} + \I\tau_2 \sum_{\gamma,\gamma',\gamma''\in\Geff}\Min^{-1}_{\gamma\gamma'}
\[\vvpn{1}_{\gamma\gamma''}\de Z_{\gamma''}-\vvmn{1}_{\gamma\gamma''}\de\bZ_{\gamma''}\]\cCd_{\gamma'}
+ 4\sum_{\gamma,\gamma'\in\Geff} (\vv\Min^{-1})_{\gamma\gamma'}\cCd_\gamma \Im\cY^{(0)}_{\gamma'}
\eea
and redefine $w^0$ as
\be
\hw^0=w^0-\cAd^{-1}\cBd.
\ee
One can show that in terms of these notations, $\cL_+$ can be represented as
\be
\sqrt{\tau_2}\cL_+ = 2r\(1-\frac{8r}{\tau_2^2\Uin}\)(\p_\mu \log r)^2 +\frac{\cAd}{2}\,(\hw_\mu^0)^2- \frac12\,\cI_{\hA\hB} w_\mu^{\hA} w^{\hB\mu}
- \frac{\tau_2^2}{4}\, N_{\hA\hB}\p_\mu z^{\hA}\p^\mu\bz^{\hB}
\label{modLp}
\ee
and leads to very simple equations of motion
\be
\begin{split}
\p_\mu\log r=&\, 2\p_\mu\log\tau_2+O(\Lambda^{-1})=O(\Lambda^{-3}),
\\
\hw_\mu^0=&\,\p_\mu\zeta^0+O(\Lambda^{-2})=O(\Lambda^{-3}),
\\
w_\mu^{\hA}=&\, \p_\mu\zeta^{\hA}+O(\Lambda^{-1})=O(\Lambda^{-1}).
\\
\p_\mu z^{\hA}=&\, O(\Lambda^{-1}).
\end{split}
\label{freez}
\ee
The non-vanishing r.h.s. of these equations correspond to the omitted contributions coming from $\cL_0+\cL_-$, and the power of $\Lambda$
is determined by the growth rate of the coefficients in \eqref{modLp}, which follows from the results of the previous subsection
and that $\cAd\escl{3}$, $\cBd\escl{}$.
Thus, in the leading approximation the fields $\tau_2$, $\zeta^0=\tau_1$, $\zeta^{\hA}$ and $z^{\hA}$
become frozen and have vanishing variations.
Furthermore, substituting \eqref{freez} back into $\cL_+$ \eqref{modLp}, one finds that it behaves as $O(\Lambda^{-1})$
and thus does not contribute to $\cL_0$.

Finally, let us turn to the finite part of the Lagrangian.
Imposing the equations \eqref{freez} and taking into account that $\Re\cN_{IJ}=\Re F_{IJ}+O(\Lambda^{-1})$,
the metric corresponding to $\cL_0$ reduces to
\bea
&&
\sqrt{\tau_2}\de s^2_{\cM'_H} =
\frac{\tau_2^2}{2} \,g_{IJ} \de z^I\de\bz^J +\frac12\, g_{IJ} w'^I w'^J
+\frac12\, g^{IJ} \(y'_I - \Re F_{IK} w'^K\)\(y'_J - \Re F_{JL} w'^L\)
\nn\\
&&\quad
+ \tau_2\sum_{\gamma,\gamma',\gamma''\in\Geff}\Min^{-1}_{\gamma\gamma'}
\[\vvpn{1}_{\gamma\gamma''} \de' Z_{\gamma''}\cY'_{\gamma'} + \vvmn{1}_{\gamma\gamma''} \de' \bZ_{\gamma''} \cY'_{\gamma'}\]
- 2 \sum_{\gamma,\gamma'\in\Geff} (vM^{-1})_{\gamma\gamma'} \cY'_\gamma \bar\cY'_{\gamma'}
\label{finLag0}\\
&&\quad
+\frac{\tau_2^2}{2} \sum_{\gamma,\gamma'\in\Geff} \vvp_{\gamma\gamma'} \de' Z_{\gamma'} \de' \bZ_{\gamma}
-\tau_2^2 \sum_{\gamma,\gamma'\in\Geff}(\Min^{-1}\cQ)_{\gamma\gamma'}\sum_{\tgam\in\Geff}\vvpn{1}_{\gamma\tgam}\de' Z_{\tgam}
\sum_{\tgam'\in\Geff}\vvmn{1}_{\gamma'\tgam'}\de' \bZ_{\tgam'}.
\nn
\eea
Here the differential $\de'$ acts only on the fields $z^I$, $\zeta^I$ and $\tzeta_I$, and we defined
\be
\begin{split}
y'_I=&\,\de\tzeta_I+\frac{\I}{8\pi^2}\sum_{\gamma\in\Geff}\Om{\gamma}
\(g_{IJ}p^J-\Re F_{IJ}\,g^{JK}\(q_{K}-\Re F_{KL}p^L\)\)
\de'\Igam{\gamma},
\\
w'^I=&\, \de\zeta^I-\frac{\I}{8\pi^2}\sum_{\gamma\in\Geff}\Om{\gamma}
g^{IJ}\(q_{J}-\Re F_{JK}p^K\)\de'\Igam{\gamma},
\\
\cY'_\gamma =&\, -\frac{\I}{2}\, g^{IJ}(q_I - F_{IK} p^K) \cY'_J\, .
\end{split}
\label{comp-cYI}
\ee
where, similarly to \eqref{relYyw}, we have
\be
\begin{split}
\cY'_I=&\, y'_I-F_{IJ}w'^J
\\
=&\, \de\tzeta_I - F_{IJ} \de\zeta^J -\frac{1}{8\pi^2} \sum_{\gamma\in\Geff} \Om{\gamma} (q_I - F_{IJ} p^J) \de' \cJ_\gamma^{(1)}.
\end{split}
\label{relYywI}
\ee
Then, using the identity
\be
g_{IJ} w'^I w'^J+ g^{IJ} (y'_I - \Re F_{IK} w'^K) (y'_J - \Re F_{JL} w'^L )
= g^{IJ} \cY'_I \bar\cY'_J\, ,
\ee
which coincides with the finite part of \eqref{RR-terms},
one can rewrite the metric \eqref{finLag0} precisely as in \eqref{finLag}.

\subsection{HK structure}
\label{subap-HK}

Now we want to prove that the limiting space $\cM'_H$ is an HK manifold. For this purpose, it is enough to show that
it carries a holomorphic symplectic structure, which in turn can be achieved by constructing
a globally defined holomorphic symplectic form on the trivial $\CP$ bundle over $\cM'_H$, which gets interpretation of the twistor space.
Such symplectic form has a representation
\be
\Omega=\I t^{-1}\omega'_+ +\omega'_3+\I t\,\omega'_-,
\label{Omega}
\ee
where $t$ is the stereographic coordinate on $\CP$.
Then the metric \eqref{finLag} must be such that $\omega'_3$ is the K\"ahler form in the complex structure in which $\omega'_+$ is holomorphic.

To find such $\Omega$, note that locally it can always be trivialized by the choice of Darboux coordinates
\be
\Omega=\de\etai{i}^I\wedge\de\mui{i}_I,
\label{OmHK}
\ee
where, as usual, the index $\scriptstyle{[i]}$ labels open patches of an atlas on the twistor space.
Thus, what we need is to specify a consistent set of Darboux coordinates.
We claim that one recovers the metric \eqref{finLag} if one identifies these Darboux coordinates away form the poles of $\CP$
with the corresponding Darboux coordinates on the twistor space
of the initial QK manifold $\cM_H$, i.e. one takes
\be
\eta^I(t)=\xi^I(t),
\qquad
\mu_I(t)=\txi_I(t),
\label{ident-Dc}
\ee
whereas the Darboux coordinates around $t=0$ are obtained by applying a holomorphic symplectic transformation
(c.f. \eqref{QKgluing}) with the generating function given by
\be
\label{gensymp-HK}
\Hij{+}= \frac{\tau_2^2}{4t^2}\, f\(\frac{2t}{\tau_2}\, \etai{+}\) +\cG(\etai{+},\mu),
\ee
where the prepotential $f(\uz^I)$ is defined in \eqref{ffull}.
This identification is possible because, under the restriction $\gamma\in\Geff$,
the integral equations \eqref{eqTBA} fixing the Darboux coordinates on $\cZ_\cM$ become a closed system for $\xi^I$ and $\txi_I$.
They also involve $\xi^0$ and $\xi^X$, but these Darboux coordinates are fixed in terms of the frozen fields and do not receive
any quantum corrections (this happens because the components $p^0$ and $p^X$ of the magnetic charge are taken to vanish)
\be
\xi^0=\tau_1+\frac{\tau_2}{2}\(t^{-1}-t\),
\qquad
\xi^X=\zeta^X+\frac{\tau_2}{2}\( \uz^X t^{-1}-\buz^X t\).
\label{simpleDc}
\ee

Besides, it is important to note that $F_I(z)=f_I(z)$ due to
the condition on the intersection numbers \eqref{restr-kap} and the restriction on the charges of worldsheet instantons.
This makes it possible to replace \eqref{eqTBA} by
\be
\begin{split}
\cX'_\gamma(t) =& \exp\Biggl[-2\pi\I\(\Thkl'+\frac{\tau_2}{2}\(Z'_\gamma \,t^{-1}-\bZ'_\gamma\,t\)\)
\Biggr.
\\
&\, \Biggl.\qquad
+\frac{1}{4\pi\I}\sum_{\gamma'\in\Geff} \Om{\gamma'}\, \langle \gamma ,\gamma'\rangle
\int_{\ell_{\gamma'} }\frac{\de t'}{t'}\, \frac{t+t'}{t-t'}\,
\log\(1-\cX'_{\gamma'}(t')\)\Biggr],
\end{split}
\label{eqTBA-HK}
\ee
where
\be
\begin{split}
\cX'_\gamma =&\, \qr\, e^{-2\pi \I \(q_0\xi^0+q_X\xi^X+q_{I} \eta^{I}-p^I\mu_I\)},
\\
\Thkl'=&\, q_0\tau_1+q_X\zeta^X+q_{I} \zeta^{I} - p^I\tzeta_I,
\\
Z'_\gamma =&\, q_0+q_X z^X+q_{I} z^{I}- p^I f_I(z).
\end{split}
\ee
The resulting system of integral equations coincides with the equations for Darboux coordinates on the twistor space of
the HK moduli space of a 4d $N=2$ gauge theory compactified on a circle \cite{Gaiotto:2008cd}, which has flavor charges $q_0$ and $q_X$
and is characterized by the holomorphic prepotential $f(z^I)$.

Let us finally show that this twistorial construction indeed leads to the metric \eqref{finLag}.
To this end, we first perform the symplectic transformation generated by \eqref{gensymp-HK}.
Although it appears to be similar to the canonical transformation generated by \eqref{gensymp} which we encountered in
the computation of the D-instanton corrected HM metric, they are not identical because the prepotential $f$ is not homogeneous in contrast to $F$.
Keeping this difference in mind, computing the expansion coefficients of $\etai{+}^I$ and $\mui{+}_I$ around $t=0$ and substituting them into
\be
\begin{split}
\omega'_+=&\, -\I\de\eta_{[+]}^{I,0}\wedge\de\mu_{I,0}^{[+]},
\\
\omega'_3=&\,\de\eta^{I,0}_{[+]}\wedge\de\mu^{[+]}_{I,0}+
\de\eta^{I,-1}_{[+]}\wedge\de\mu^{[+]}_{I,1},
\end{split}
\ee
which follows from a combination of \eqref{Omega} and \eqref{OmHK}, one finds that the basis of (1,0) forms encoding
the complex structure $J'_3$ consists of $\de z^I$ and $\cY'_I$ \eqref{relYywI}, whereas the K\"ahler form $\omega'_3$ is given by
\be
\begin{split}
\omega'_3=&\,
\de\zeta^I\wedge\de\tzeta_I
+\frac{\I\tau_2^2}{2}\, g_{IJ}\de z^I\wedge\de\bz^J
-\frac{1}{128\pi^4} \sum_{\gamma,\gamma'\in\Geff} \Om{\gamma}\Om{\gamma'} \langle\gamma,\gamma'\rangle\,  \de'\cJ_{\gamma}^{(1)} \wedge \de'\cJ_{\gamma'}^{(1)}
\\
&\, -\frac{1}{8\pi^2}\sum_{\gamma\in\Geff} \Om{\gamma}\Bigl[(q_I\de\zeta^I-p^I\de\tzeta_I)\wedge \de'\Igg{}
+\tau_2\de' Z_\gamma\wedge \de' \Igp\Bigr].
\end{split}
\label{om3HK1}
\ee

Next we observe that the last term in \eqref{om3HK1} can be rewritten as
\be
\begin{split}
&\, \frac{\tau_2}{16\pi^2}\sum_{\gamma\in\Geff} \Om{\gamma}\(\de' \Igp\wedge\de' Z_\gamma - \de' \Igm\wedge\de' \bZ_\gamma\)
\\
=&\, \frac{\I\tau_2}{2} \sum_{\gamma,\gamma'\in\Geff}
\Bigl(\de'\Theta_\gamma\wedge \(\vvpn{1}_{\gamma\gamma'} \de'Z_{\gamma'}-\vvmn{1}_{\gamma\gamma'} \de'\bZ_{\gamma'}\)
+ \tau_2 \vvp_{\gamma\gamma'} \de' Z_{\gamma'} \wedge \de' \bZ_{\gamma} \Bigr).
\end{split}
\ee
Besides, one has
\bea
\nn
&& \de\zeta^I\wedge\de\tzeta_I -\frac{1}{128\pi^4} \sum_{\gamma,\gamma'\in\Geff} \Om{\gamma}\Om{\gamma'} \langle\gamma,\gamma'\rangle\,
\de'\cJ_{\gamma}^{(1)} \wedge \de'\cJ_{\gamma'}^{(1)}
\\
&& \qquad\qquad
=\frac{\I}{2}\, g^{IJ} \cY'_I \wedge \bar\cY'_J - 2\I \sum_{\gamma,\gamma'\in\Geff} (\vv \Min^{-1})_{\gamma\gamma'} \cY'_\gamma\wedge\bar\cY'_{\gamma'}
\\ && \nn
- \tau_2 \sum_{\gamma,\gamma'\in\Geff} \Min^{-1}_{\gamma\gamma'}
\[\cCf'_\gamma-\frac{\I}{16\pi^2}\sum_{\tgam\in\Geff}\Om{\tgam}\langle\gamma,\tgam\rangle \de'\Igam{\tgam}\]
\wedge
\sum_{\tgam'\in\Geff}\(\vvpn{1}_{\gamma'\tgam'}\de' Z_{\tgam'}+\vvmn{1}_{\gamma'\tgam'}\de' \bZ_{\tgam'}\),
\eea
where
\be
\cCf'_\gamma= -2g^{IJ}\(q_I-\Re F_{IK}p^K\)\(\de\tzeta_J-\Re F_{JL}\de\zeta^L\)
-\frac12\, g_{IJ}\,p^I\,\de\zeta^J.
\label{cCfp}
\ee
Finally, we need also the following identity
\bea
&&\I\sum_{\gamma,\gamma',\gamma''\in\Geff}\Min^{-1}_{\gamma\gamma'}\(\vvpn{1}_{\gamma\gamma''}\de' Z_{\gamma''}\wedge \bar\cY'_{\gamma'}
-\vvmn{1}_{\gamma\gamma''}\de'\bZ_{\gamma''}\wedge \cY'_{\gamma'}\)
\nn\\
&=&-\sum_{\gamma,\gamma',\tgam'\in\Geff}\Min^{-1}_{\gamma\gamma'}\(\cCf'_{\gamma}
-\frac{\I}{16\pi^2}\sum_{\tgam\in\Geff}\Om{\tgam}\langle\gamma,\tgam'\rangle \de'\Igam{\tgam}\)
\wedge \(\vvpn{1}_{\gamma'\tgam'}\de' Z_{\tgam'}+\vvmn{1}_{\gamma'\tgam'}\de' \bZ_{\tgam'}\)
\nn\\
&&
-\frac{\I}{2}\sum_{\gamma,\gamma'\in\Geff}\(\vvpn{1}_{\gamma\gamma'}\de' Z_{\gamma'}-\vvmn{1}_{\gamma\gamma'}\de' \bZ_{\gamma'}\)
\wedge\Biggl[\de'\Theta_{\gamma}
\Biggr.
\\
&&\Biggl.\qquad
-\tau_2\sum_{\gamma''\in\Geff}(\Min^{-1}\cQ)_{\gamma\gamma''}
\(\sum_{\tgam\in\Geff}\(\vvpn{1}_{\gamma''\tgam}\de' Z_{\tgam}+\vvmn{1}_{\gamma''\tgam}\de' \bZ_{\tgam}\)\)\Biggr],
\nn
\eea
which allows express terms with $\de'\Theta_\gamma$ and $\cC'_\gamma$ in terms of (1,0)-forms and their conjugate.
Collecting all these relations together and suing them in \eqref{om3HK1}, one arrives at the following expression for the K\"ahler form
\bea
\omega'_3 &=&  \frac{\I\tau_2^2}{2}\, g_{IJ}\de z^I\wedge\de\bz^J +\frac{\I}{2}\, g^{IJ} \cY'_I \wedge \bar\cY'_J
- 2\I \sum_{\gamma,\gamma'\in\Geff} (\vv M^{-1})_{\gamma\gamma'} \cY'_\gamma\wedge\bar\cY'_{\gamma'}
\nn\\ &&
+\I\tau_2\sum_{\gamma,\gamma',\gamma''\in\Geff}\Min^{-1}_{\gamma\gamma'}
\[\vvpn{1}_{\gamma\gamma''}\de' Z_{\gamma''} \wedge \bar\cY'_{\gamma'}
+\cY'_{\gamma'} \wedge\vvmn{1}_{\gamma\gamma''} \de'\bZ_{\gamma''}\]
\\ &&
-\I\tau_2^2\sum_{\gamma,\gamma'\in\Geff}(\Min^{-1}\cQ)_{\gamma\gamma'}\sum_{\tgam\in\Geff}\vvpn{1}_{\gamma\tgam}\de' Z_{\tgam}
\wedge
\sum_{\tgam'\in\Geff}\vvmn{1}_{\gamma'\tgam'}\de' \bZ_{\tgam'}
+\frac{\I\tau_2^2}{2}\sum_{\gamma,\gamma'\in\Geff} \vvp_{\gamma\gamma'}  \de' Z_{\gamma'} \wedge \de' \bZ_{\gamma}.
\nn
\eea
It precisely corresponds to the metric \eqref{finLag}.

\section{Torus reduction of 5d gauge theory}
\label{ap-torus}

In this appendix we perform compactification of the five-dimensional action \eqref{bosL5d} on the torus.
The spacetime metric $g_{\hmu\hnu}$, whose signature in our conventions is $(-,+,+,+,+)$, is taken to be as in \eqref{metric5d}
where coordinates $x^3$ and $x^4$ parametrize the torus directions.
The periodicity of holonomies under large gauge transformations is set to be $2\pi$,
which makes natural to define the variables
\be
\vth_1^I = \frac{1}{2\pi}\oint_{S^1_3} A_3^I \, \de x^3,
\qquad
\vth_2^I = \frac{1}{2\pi}\oint_{S^1_4} A_4^I \, \de x^4
\label{eq:holonomies-norm-tilde}
\ee
with period $1$.
Introducing
\be
\thtau^I \equiv \vth_1^I - \tau \vth_2^I
\label{defvart}
\ee
and assuming the independence of all fields on the torus coordinates, one finds that
the kinetic term for vectors and the Chern-Simons term give, respectively,
\bea
	\int_{T^2} \de x^3\de x^4 \sqrt{\gT}\,\Fg_{IJ} F^I_{\hmu\hnu} F^{J\hmu\hnu}
&=&  \cV \Fg_{IJ} F^I_{\mu\nu} F^{J\mu\nu}
+\frac{ 8 \pi^2}{\tau_2}\, \Fg_{IJ}\p_\mu \thtau^I  \p^\mu \bthtau^J,
\\
\int_{T^2} \de x^3\de x^4\cF_{IJK}\epsilon^{\hmu\hnu\hlambda\hrho\hsigma} A_{\hmu}^I F_{\hnu\hlambda}^J F_{\hrho\hsigma}^K
&=&
16\pi^2\cF_{IJK} \epsilon^{\mu\nu\lambda}\(  F_{\mu\nu}^I \(\vth_2^J\p_{\lambda}  \vth_1^K - \vth_1^J  \p_{\lambda}  \vth_2^K\)
- 2  A_\mu^I  \p_\nu \vth_1^J \p_\lambda \vth_2^K \).	
\nn
\eea
Note that the CS term does not produce the factor of volume because the integrand does not contain the factor $\sqrt{\gT}$.
As a result, integrating by parts and using the Bianchi identity $\epsilon^{\mu\nu\lambda}\p_\mu F^I_{\nu\lambda}\equiv 0$,
the reduced action can be brought to the following form
\be
\begin{split}
S_{\rm bos}^{5\to 3d}
= &\, -\int\de^3 x\[\Fg_{IJ}\( \frac{\cV }{8\pi}\, F^I_{\mu\nu} F^{J\mu\nu}
+ \frac{\pi}{\tau_2}\, \p_\mu \thtau^I  \p^\mu \bthtau^J
+\frac{\cV}{4\pi}\, \p_\mu\vph^I\, \p^\mu\vph^J\)
\right.
\\
&\,\qquad
\biggl. + \frac{\pi}{2}\,\Fg_{IJK} \epsilon^{\mu\nu\lambda} F^I_{\mu\nu} \(\vth_2^J\p_{\lambda}\vth_1^K - \vth_1^J  \p_{\lambda}\vth_2^K\)\biggr].
\end{split}
\label{red-S3d}
\ee

The action \eqref{red-S3d} contains the vector fields $A^I_\mu$ and therefore describes a coupling of three-dimensional tensor multiplets.
It can be turned into a nonlinear $\sigma$-model for hypermultiplets by dualizing the vector fields into scalars.
This is done by adding to the action the term
\be
\Delta S =\pi\int\de^3 x\, \lambda_I\, \epsilon^{\mu \nu\lambda}\p_\mu F^I_{\nu\lambda}
\ee
such that the variation with respect to the Lagrange multipliers $\lambda_I$ induces equations of motion which are simply the Bianchi identity.
Integrating this term by parts and varying the total action $S_{\rm bos}^{5\to 3d}+\Delta S$ with respect to $F^I_{\mu\nu}$, one finds instead
\be
F^I_{\mu\nu} = -\frac{4\pi^2}{\cV}\, \Fg^{IJ} {\epsilon_{\mu\nu}}^{\lambda}
\(\p_\lambda\lambda_J+ \hf\,\Fg_{JKL} \(\vth_2^K\p_{\lambda}  \vth_1^L - \vth_1^K \p_{\lambda}\vth_2^L\)\),
\ee
where $\Fg^{IJ}$ is the inverse of $\Fg_{IJ}$. Substituting this field strength back into the action, one obtains\footnote{One should remember
that due to our choice of signature the contraction of two Levi-Civita symbols gives $\eps^{\mu\rho\sigma}\eps_{\nu\rho\sigma}=-2\delta^\mu_\nu$.}
a $\sigma$-model for complex fields $\thtau^I$ and real fields $\vph^I$ and $\lambda_I$:
\bea
&& S_{\rm bos}^{3d}
= -\int\de^3 x\[\Fg_{IJ}\(  \frac{ \pi}{\tau_2}\, \p_\mu \thtau^I  \p^\mu \bthtau^J
+\frac{\cV}{4\pi}\, \p_\mu\vph^I\, \p^\mu\vph^J\)
\right.
\label{S3d}\\
&&
\biggl. +
\frac{4\pi^3}{\cV}\, \Fg^{IJ} \(\p_\mu\lambda_I+\hf\,\Fg_{IKL} \(\vth_2^K\p_{\mu}\vth_1^L - \vth_1^K \p_{\mu}\vth_2^L\)\)
\(\p^\mu\lambda_J+\hf\,\Fg_{JMN} \(\vth_2^M\p^\mu  \vth_1^N - \vth_1^M \p^\mu\vth_2^N\)\)
\biggr].
\nn
\eea

\section{Toric data}
\label{ap-toricdata}

Let  $\CY$ be a smooth compact threefold described by a homogeneous polynomial equation in some coordinate patch
of a desingularized ambient four-dimensional toric Fano variety $\widetilde \cA$.
The relevant topological information about $\CY$ can be encoded by two sets of data.

First, one should specify a reflexive polytope $\Delta$ with vertices belonging to a lattice $\tM\simeq \IZ^4$.
We will denote the dual polytope, defined within the dual lattice $\tN \simeq {\rm{Hom}}(\tM,\IZ)$, by $\Delta^*$.
Both of them can be represented by matrices with 4 columns whose rows correspond to points in $\tM$ (resp. $\tN$)
defining their vertices.
A crucial property of $\Delta^*$ is that it contains a single interior point, which is the origin of $\tN$,
while all other lattice points contained in $\Delta^*$ lie on its boundary (including the vertices).
Moreover, the boundary lattice points have an important geometric interpretation corresponding to
the toric divisors $\tD_i$ of the ambient space. Due to this, it will be convenient to introduce a matrix $\bar\Delta^*$
whose rows describe {\it all} such boundary points.

However, these polytopes define in general a singular variety.
The desingularization of the ambient space is encoded by the second data, a simplicial triangulation of $\Delta^*$
by unit-volume simplices with a vertex at the origin, or more precisely a \emph{fine, star, regular} triangulation.
It is specified by describing
each four-simplex by a set of 4 toric divisors corresponding to its edges.
Thus, it can be encoded by a matrix also with 4 columns and the number of rows equal to the number of simplices used
in the triangulation. The entries of the matrix specify rows of $\Delta^*$.

As a side remark, let us recall that different triangulations of $\Delta^*$ may
(or may not) correspond to different Calabi-Yau geometries.
When more than one triangulation describes the same Calabi-Yau, these represent
different chambers of the K\"ahler cone of $\CY$, also known as \emph{phases} of the geometry.

Below we specify the toric data for the two Calabi-Yau manifolds used in section \ref{sec-examples}.
We call them by their number in the database \cite{Altman:2014bfa}.

\subsubsection*{Polytope 337 (geometry 1)}

The polytopes corresponding to this example are
\be
\Delta =
\left(\scriptsize
\begin{array}{cccc}
 1 & 0 & 0 & 0 \\
 1 & 2 & 0 & 0 \\
 0 & 0 & 1 & 0 \\
 0 & 3 & 1 & 0 \\
 0 & 0 & 0 & 1 \\
 0 & 2 & 0 & 1 \\
 -1 & 1 & 2 & -1 \\
 3 & -2 & -2 & -2 \\
 -2 & -2 & -2 & 3 \\
 -3 & -3 & -2 & 3 \\
 3 & -3 & -2 & -3
\end{array}
\right),
\qquad
\bar\Delta^* =
\left(\scriptsize
\begin{array}{cccc}
 -1 & 0 & -1 & -1 \\
 -1 & 0 & -1 & 0 \\
 -1 & 0 & 0 & -1 \\
 -1 & 1 & -1 & -1 \\
 0 & 0 & -1 & -1 \\
 0 & 1 & -1 & 0 \\
 1 & -1 & 2 & 1 \\
 2 & -1 & 2 & 2
\end{array}
\right)\,.
\label{polytop1}
\ee
They define a singular variety. We are interested in the Calabi-Yau manifold
defined by the desingularization corresponding to geometry 1 in the list of \cite{Altman:2014bfa}.
Such desingularization admits three phases, described by the following three triangulations
\be
T_1 =
\left(\scriptsize
\begin{array}{cccc}
 0 & 1 & 2 & 3 \\
 0 & 1 & 2 & 6 \\
 0 & 1 & 3 & 4 \\
 0 & 1 & 4 & 6 \\
 0 & 2 & 3 & 4 \\
 0 & 2 & 4 & 6 \\
 1 & 2 & 3 & 6 \\
 1 & 3 & 4 & 5 \\
 1 & 3 & 5 & 6 \\
 1 & 4 & 5 & 7 \\
 1 & 4 & 6 & 7 \\
 1 & 5 & 6 & 7 \\
 2 & 3 & 4 & 6 \\
 3 & 4 & 5 & 6 \\
 4 & 5 & 6 & 7
\end{array}
\right),
\qquad
T_2 =
\left(\scriptsize
\begin{array}{cccc}
 0 & 1 & 2 & 3 \\
 0 & 1 & 2 & 6 \\
 0 & 1 & 3 & 5 \\
 0 & 1 & 4 & 5 \\
 0 & 1 & 4 & 6 \\
 0 & 2 & 3 & 4 \\
 0 & 2 & 4 & 6 \\
 0 & 3 & 4 & 5 \\
 1 & 2 & 3 & 6 \\
 1 & 3 & 5 & 6 \\
 1 & 4 & 5 & 7 \\
 1 & 4 & 6 & 7 \\
 1 & 5 & 6 & 7 \\
 2 & 3 & 4 & 6 \\
 3 & 4 & 5 & 6 \\
 4 & 5 & 6 & 7
\end{array}
\right),
\qquad
T_3 =
\left(\scriptsize
\begin{array}{cccc}
 0 & 1 & 2 & 3 \\
 0 & 1 & 2 & 6 \\
 0 & 1 & 3 & 5 \\
 0 & 1 & 4 & 5 \\
 0 & 1 & 4 & 7 \\
 0 & 1 & 6 & 7 \\
 0 & 2 & 3 & 4 \\
 0 & 2 & 4 & 6 \\
 0 & 3 & 4 & 5 \\
 0 & 4 & 6 & 7 \\
 1 & 2 & 3 & 6 \\
 1 & 3 & 5 & 6 \\
 1 & 4 & 5 & 7 \\
 1 & 5 & 6 & 7 \\
 2 & 3 & 4 & 6 \\
 3 & 4 & 5 & 6 \\
 4 & 5 & 6 & 7
\end{array}
\right).
\label{triang1}
\ee
We recall that the $i$-th row of a triangulation $T$ has components $T_{ij} = (n_j)_{j=1\dots4}$,
encoding the 4-simplex delimited by the four toric divisors $\{\tD_{n_j+1}\}_{j=1\dots4}$
where $\tD_n$ is the divisor specified by the $n$-th row of $\bar\Delta^\star$.

\subsubsection*{Polytope 1439 (geometry 2)}

In this example the two polytopes are encoded by the matrices
\be
\Delta =
\left(\scriptsize
\begin{array}{cccc}
 1 & 0 & 0 & 0 \\
 0 & 1 & 0 & 0 \\
 2 & 3 & 4 & 0 \\
 2 & 3 & 0 & 4 \\
 -6 & -5 & 4 & -8 \\
 -6 & -5 & -8 & 4
\end{array}
\right),
\qquad
\bar\Delta^* =
\left(\scriptsize
\begin{array}{cccc}
 -1 & -1 & 1 & 1 \\
 -1 & -1 & 1 & 2 \\
 -1 & -1 & 2 & 1 \\
 -1 & -1 & 3 & 3 \\
 -1 & 3 & -2 & -2 \\
 1 & -1 & 0 & 0 \\
 -1 & -1 & 2 & 2 \\
 0 & 1 & -1 & -1
\end{array}
\right).
\label{polytop2}
\ee
They again define a singular manifold. This time we take the desingularization corresponding to geometry 2
in the classification of \cite{Altman:2014bfa}, which is described by the following triangulation matrix
\be
T =
\left(\scriptsize
\begin{array}{cccc}
 0 & 1 & 4 & 6 \\
 0 & 1 & 4 & 7 \\
 0 & 1 & 5 & 6 \\
 0 & 1 & 5 & 7 \\
 0 & 2 & 4 & 6 \\
 0 & 2 & 4 & 7 \\
 0 & 2 & 5 & 6 \\
 0 & 2 & 5 & 7 \\
 1 & 3 & 4 & 6 \\
 1 & 3 & 4 & 7 \\
 1 & 3 & 5 & 6 \\
 1 & 3 & 5 & 7 \\
 2 & 3 & 4 & 6 \\
 2 & 3 & 4 & 7 \\
 2 & 3 & 5 & 6 \\
 2 & 3 & 5 & 7
\end{array}
\right).
\label{triang2}
\ee

\providecommand{\href}[2]{#2}\begingroup\raggedright\endgroup


\end{document}